\pdfoutput=1
\documentclass[letterpaper,aps,pre,showpacs,preprintnumbers,superscriptaddress]{revtex4-1}

\usepackage{graphicx}
\usepackage{dcolumn}
\usepackage{bm}
\usepackage{color}
\usepackage{mathtools}
\usepackage{amssymb}
\usepackage{epsfig}
\usepackage{afterpage}
\usepackage{epstopdf}
\usepackage{tikz}
\usetikzlibrary{"arrows","shapes.geometric"}
\usetikzlibrary{"calc"}

\begin{document}

\title{Strong anisotropy in two-dimensional surfaces with generic scale
invariance: Non-linear effects}

\author{Edoardo Vivo}
\affiliation{Departamento de
Matem\'{a}ticas and Grupo Interdisciplinar de Sistemas Complejos
(GISC), Universidad Carlos III de Madrid, Avenida de la
Universidad 30, E-28911 Legan\'{e}s, Spain}
\author{Matteo Nicoli}
\affiliation{Center for Interdisciplinary Research on Complex Systems, Department of Physics,
Northeastern University, Boston, MA 02115, USA}
\author{Rodolfo Cuerno}
\affiliation{Departamento de
Matem\'{a}ticas and Grupo Interdisciplinar de Sistemas Complejos
(GISC), Universidad Carlos III de Madrid, Avenida de la
Universidad 30, E-28911 Legan\'{e}s, Spain}

\date{\today}

\begin{abstract}
We expand a previous study [Phys.\ Rev.\ E {\bf 86}, 051611 (2012)]
on the conditions for occurrence of strong anisotropy (SA) in the scaling
properties of two-dimensional surfaces displaying generic scale invariance. There,
a natural Ansatz was proposed for SA, which arises naturally when analyzing data from e.g.\
thin-film production experiments. The Ansatz was tested in Gaussian
(linear) models of surface dynamics and in non-linear models, like the Hwa-Kardar
(HK) equation [Phys.\ Rev.\ Lett.\ {\bf 62}, 1813 (1989)], which are susceptible
of accurate approximations through the former. In contrast, here we analyze non-linear
equations for which such type of approximations fail. Working within
generically-scale-invariant situations, and as representative case studies,
we formulate and study a generalization of the HK equation for conserved dynamics,
and reconsider well-known systems, such as the conserved and the non-conserved anisotropic Kardar-Parisi-Zhang
equations. Through the combined use of Dynamic Renormalization Group analysis and direct numerical
simulations, we conclude that the occurrence of SA in two-dimensional surfaces requires
dynamics to be conserved. We find that, moreover, SA is not generic in parameter space but requires,
rather, specific shapes from the terms appearing in the equation of motion, whose justification
needs detailed information on the dynamical process that is being modeled in
each particular case.
\end{abstract}



\maketitle

\section{Introduction}

Scale invariant, two-dimensional surfaces that are anisotropic in space abound in Science and Technology,
for systems spanning many orders of magnitude in length scales. Examples range from epitaxial thin films in nanoscience
\cite{misbah:2010} to micro and macroscopic crack formation in solids \cite{alava:2006,bonamy:2011}, to geological systems,
such as landscape evolution induced by rivers \cite{Romo,rodriguez-iturbe:2001}. Mathematically, the surfaces that occur in these and many other systems are self-affine fractals \cite{Barabasi}, whose fractal dimension (or, equivalently, roughness exponent) differs, depending on the direction along which it is measured.

Due to the lack of characteristic distances, the scaling behavior just described is a form of anisotropic {\em critical} behavior \cite{henkel:book_v2,tauber:book}, which moreover often occurs without the need of parameter fine-tuning that adjusts the system to a critical point. These are thus examples of so-called generic scale invariance (GSI) \cite{grinstein:1990,grinstein:1991,grinstein:1995}. Additional well-known instances of space anisotropies ensuing under GSI conditions are driven diffusive systems (DDS) \cite{schmittmann:book} and self-organized criticality (SOC) \cite{christensen:book}. Still another context for this type of behavior, which has remained relatively less studied, is that of surface kinetic roughening \cite{Barabasi}. While differing from DDS in the fact that dynamics and noise are not necessarily conserved, kinetic roughening systems also differ from SOC systems in the fact that their typical time scales for response are not separated from those characterizing the external driving \cite{grinstein:1991,grinstein:1995}.

In this work we pursue a continuum description of GSI systems through stochastic partial differential equations \cite{kardar:book}. Within such a framework, our cases of interest will be those conditions that lead to GSI while applying to the most important universality classes in surface kinetic roughening. Namely \cite{grinstein:1991,grinstein:1995}, systems with non-conserved dynamics, like the celebrated Kardar-Parisi-Zhang (KPZ) equation \cite{kardar:1986}, or else systems with conserved dynamics and non-conserved noise, like e.g.\ the so-called conserved KPZ (cKPZ) equation \cite{lai:1991}. Both equations have been shown to be directly relevant to the growth of two-dimensional interfaces. For instance, the KPZ equation does describe the asymptotic behavior of many thin films whose interfacial dynamics is not constrained by conservation laws \cite{Barabasi}, as e.g.\ for silica films grown by chemical vapor deposition \cite{ojeda:2000}, while the cKPZ equation plays a cental role in the dynamics of
epitaxial surfaces grown by Molecular Beam Epitaxy (MBE), in which adatom desorption is typically suppressed, inducing interfacial conservation laws \cite{misbah:2010}.

Remarkably, the anisotropic generalizations of the two previous equations, namely, the so-called anisotropic KPZ (aKPZ) \cite{Wolf} and conserved anisotropic KPZ (caKPZ) equations \cite{kallabis:1998,kallabis_thesis}, {\em do not} lead asymptotically to anisotropic behavior (strong anisotropy, SA). Rather, in spite of being nominally anisotropic, they lead to isotropic asymptotics (weak anisotropy, WA), in universality classes that depend on parameter conditions. This fact contrasts strikingly with the unambiguous observation of SA in experiments on surface kinetic roughening for two-dimensional interfaces, see \cite{us_exp} and references therein. To cite a few, anisotropic behavior occurs for growth by Molecular Beam Epitaxy (MBE), both under morphologically unstable conditions, as for growth on Si(001) \cite{schelling:1999,uwaha:1999,verga:2009_12}, or for morphologically stable ones, as for growth of GaAs films \cite{ballestad:2001,ballestad:2002}. Also erosion, rather than growth, of thin films by ion-
beam sputtering (IBS) induces space anisotropies related with the different roles played by the direction on the target that lies along the projection of the ion beam, and the direction perpendicular to it, leading to SA regardless of the morphological stability conditions of the experiment \cite{Luis,Michely}. Macroscopically, fracture of solids provides still another instance for the occurrence of space anisotropies, in this case between the crack propagation and crack front directions \cite{alava:2006,bonamy:2011}, and concomitant SA properties.

\subsection{Ans\"atze for anisotropic kinetic roughening}

The previous facts underscore the lack of a sufficient understanding of SA kinetic roughening for the case of two-dimensional interfaces, with particular experimental relevance. Actually, the interface equations that can potentially describe a range of anisotropic systems still remain to be identified, as is the case e.g.\ for IBS systems \cite{cuerno:2011}. This motivated us to perform a systematic study of the phenomenon, with the aim to identify general conditions on the occurrence of isotropic vs anisotropic behavior. Our program started with the formulation of a scaling Ansatz for SA \cite{us}, as encoded in the asymptotic behavior of the surface structure factor, that could be readily applied to analyze experimental data on, say, surface dynamics of thin films \cite{us_exp}. Specifically, suppose the scalar field $h(\mathbf{r})$ describes the height of a surface above point $\mathbf{r}=(x,y)$ on a reference plane. A convenient characterization of its fluctuations can be performed through the power
spectral density (PSD) or height structure factor \cite{Barabasi},
\begin{equation}
S(\mathbf{k}) = \langle |h_{\mathbf{k}}|^2 \rangle ,
\label{psd}
\end{equation}
where $\mathbf{k}=(k_x,k_y)$ is wave-vector, $h_{\mathbf{k}}$ is the space Fourier transform of $h(\mathbf{r})$, and brackets denote averages over the noise distribution. For a system displaying SA, we postulate \cite{us} that the stationary PSD scales with wave-vector components $k_x$ and $k_y$ as
\begin{equation}
\label{psd2d}
S(k_x,k_y) \sim \frac{1}{k_x^{2\tilde{\alpha}_x} + \nu k_y^{2\tilde{\alpha}_y}},
\end{equation}
where we refer to $\tilde{\alpha}_x$ and $\tilde{\alpha}_y$ as roughness exponents in momentum space, and $\nu$ is a mere constant. It is convenient to make also contact with observables in real space, such as the 2D height-difference correlation function
\begin{equation}
G(r) = \langle [h(\mathbf{r}+\mathbf{r}_0)-h(\mathbf{r}_0)]^2 \rangle ,
\label{G}
\end{equation}
where $\mathbf{r}_0=(x_0,y_0)$ is an arbitrary position on the substrate plane and $r = |\mathbf{r}|$. Indeed, a natural, equivalent definition of SA is that the value of the roughness exponent changes with the direction along the latter.
Namely, by defining 1D versions of the height-difference correlation function along the two substrate directions,
\begin{eqnarray}
G_x(x) & = & \langle \left[h(x_0+x,y_0) - h(x_0,y_0)\right]^2\rangle, \label{Gx} \\
G_y(y) & = & \langle \left[h(x_0,y_0+y) - h(x_0,y_0)\right]^2\rangle, \label{Gy}
\end{eqnarray}
and under kinetic roughening conditions, scaling behavior ensues \cite{Barabasi}, $G_x(x)\sim x^{2\alpha_x}$ and $G_y(y)\sim y^{2\alpha_y}$, where $\alpha_{x,y}$ are two potentially different roughness exponents. The system is said to display SA if indeed $\alpha_x \neq \alpha_y$, whereas WA occurs when the steady state of the system is actually \emph{isotropic}, so that $\alpha_x = \alpha_y=\alpha$, with $G(r) \sim r^{2\alpha}$ in such a case. In Ref.\ \cite{us} we proved that, indeed, Eq.\ \eqref{psd2d} is equivalent to SA for the correlation functions \eqref{Gx}, \eqref{Gy}, provided exponents are related as
\begin{eqnarray}
&& 2\alpha_x = 2\tilde \alpha_x - \zeta -1, \label{eq_relalphax} \\[5pt]
&& 2\alpha_y = 2\tilde \alpha_y - 1/\zeta -1 , \label{eq_relalphay}
\end{eqnarray}
where we have introduced the anisotropy exponent
\begin{equation}
\zeta = \dfrac{\tilde{\alpha}_x}{\tilde{\alpha}_y} =  \dfrac{\alpha_x}{\alpha_y},
\end{equation}
the second equality being a consequence of Eqs.\ \eqref{eq_relalphax} and \eqref{eq_relalphay}. Thus, SA is simply stated as $\zeta\neq 1$. Conversely, WA implies $\zeta= 1$, so that $\alpha_x=\alpha_y=\alpha$ and Eq.\ \eqref{psd2d} is asymptotically equivalent to the isotropic behavior of the 2D PSD function \cite{Barabasi} $S(\mathbf{k}) \sim k^{-(2\alpha + 2)}$, with $k=|\mathbf{k}|$.

In turn, as an alternative to the 1D correlation functions \eqref{Gx}, \eqref{Gy}, one may consider the power spectral densities $S_x(k_x)$ and $S_y(k_y)$ of 1D cuts of the 2D interface along the $x$ and $y$ directions. In many experimental or numerically simulated systems, this is a way to improve over the signal-to-noise ratio of the 2D PSD. Thus for instance, considering a fixed value $y=y_0$, one defines
\begin{equation}
S_x(k_x) = \langle h^{(y_0)}_{k_x} h^{(y_0)}_{-k_x} \rangle ,
\label{def_Sx}
\end{equation}
where $h^{(y_0)}_{k_x}$ is the Fourier transform of the corresponding 1D profile $h(x,y_0)$. Analogously one can define $S_y(k_y)$ for a cut along the $y$ direction at a fixed $x=x_0$ value. As a consequence of Eq.\ \eqref{psd2d}, these functions scale as \cite{keller:2009,us}
\begin{eqnarray}
\label{eq:psd1d_scaling}
&&S_x(k_x)\sim k_x^{-(2\tilde{\alpha}_x-\zeta)} =  k_x^{-(2\alpha_x+1)} , \\[5pt]
&&S_y(k_y)\sim k_y^{-(2\tilde{\alpha}_y-1/\zeta)} = k_y^{-(2\alpha_y+1)}, \label{eq:psd1d_scaling2}
\end{eqnarray}
that provide the natural generalization to the SA case of the scaling behavior of the PSD of 1D cuts of the surface in the isotropic case, in which $\alpha_x=\alpha_y=\alpha$ and $S_{x,y} \sim k_{x,y}^{-(2\alpha+1)}$ \cite{Barabasi,hansen:2001}.

With respect to the time evolution, the isotropic behavior is encoded in the standard Family-Vicsek (FV) Ansatz for kinetic roughening surfaces, which is typically formulated in terms of the surface roughness $W^2(t) = \langle (h-\bar{h})^2 \rangle = \int S(\mathbf{k}) \; {\rm d}\mathbf{k}$. Thus \cite{Barabasi}, $W \sim t^{\beta}$ for $t \ll t^{1/z}$, while $W \sim L^{\alpha}$ for $t \gg t^{1/z}$, where $z$ is the so-called dynamic exponent, $t^{1/z}$ is proportional to the length-scale below which non-trivial correlations have built up among height values at different substrate positions, $L$ is the lateral system size, and $\beta = \alpha/z$ is usually termed the growth exponent. As shown in \cite{us}, for SA systems, the Ansatz \eqref{psd2d} implies that the behavior of the roughness of 1D line profiles is $W_{x,y} \sim t^{\beta}$ for $t \ll t^{1/z_{x,y}}$, while $W_{x,y} \sim L^{\alpha_{x,y}}$ for $t \gg t^{1/z_{x,y}}$. Namely, there are two dynamic exponents $z_{x,y}$, which are related as $z_y=z_x/\zeta$, and a single growth exponent, since then $\beta_x = \alpha_x/z_x = \beta_y = \alpha_y/z_y = \beta$. Indeed, for $\zeta=1$ one has that $\alpha_x=\alpha_y=\alpha$ and $z_x=z_y=z$, and WA behavior ensues. This SA dynamic behavior has been confirmed in the experiments of Ref.\ \cite{us_exp}. Overall, for a SA system there are then three independent critical exponents, e.g.\ $\alpha_x$, $z_x$, and $\zeta$.

Note that, although anisotropic kinetic roughening was previously encoded into a scaling Ansatz \cite{schmittmann:2006} originating in the study of critical dynamics of equilibrium statistical-mechanical systems \cite{henkel:book_v1}, such a theoretically powerful formulation is not particularly natural for the characterization of actual two-dimensional surfaces. In Refs.\ \cite{us,us_exp} we have clarified the relation between Eq.\ \eqref{psd2d} and the behavior of standard observables employed in the experimental characterization of anisotropic thin films through e.g.\ 1D correlations like $G_{x,y}$, or power spectral densities of 1D profiles $S_{x,y}$. We have moreover shown how all these results can be employed in a consistent characterization of SA for actual experimental data on IBS-sputtered silicon surfaces \cite{us_exp}.

Frequently, e.g.\ in the MBE or IBS systems mentioned above, physical properties and geometric constraints dictate the appropriate choices for the $x$ and $y$ directions. Still, as shown in \cite{us}, under conditions for strong anisotropy any choice of two orthogonal directions will lead to the same set of two different exponents $\tilde{\alpha}_{x,y}$, which guarantees the generality of Ansatz \eqref{psd2d}. In the case of e.g.\ fracture, alternative choices for anisotropic scaling Ans\"atze are also available, in which e.g.\ either an auxiliary dynamics is postulated \cite{bonamy:2011,ponson:2006}, or else expansions of observables over appropriate functional bases are performed that exploit the fact that isotropic materials often have anisotropic fracture surfaces only because of the breaking of isotropy by the initial conditions \cite{bouchbinder:2005,bouchbinder:2006}.

From the theoretical point of view, Ansatz \eqref{psd2d} was motivated by the behavior of exact solutions to linear interface equations displaying SA, and was further validated in Ref.\ \cite{us} against a non-linear system for which SA is also well known to occur, namely, the Hwa-Kardar equation \cite{HK:1989,HK:1992}. This equation was originally put forward to describe the evolution of the surface height for a running sand-pile, a particular instance of a supercritical SOC system. The equation has conserved dynamics, reflecting the conservation in the number of sand grains by the relaxation dynamics, and non-conserved noise, as a reflection of the non-conserved driving field associated with sand-grain addition. Thus, it features GSI, characterized by anisotropic scaling exponents which are believed to be exact \cite{HK:1989,HK:1992}. As it turns out, one can write down an exactly solvable, linear equation with the same exponents \cite{us}, which allowed to elucidate superficial differences between the SA
scaling of the HK equation and Ansatz \eqref{psd2d} as being due to finite size effects \cite{us}.

In this work, we pursue further the study of SA through continuum interface equations by trying to identify conditions that such type of models have to fulfill, in order to display this type of scaling. This characterization might prove useful when invoking universality principles \cite{cuerno:2004,cuerno:2007} in order to put forward a continuum equation for a system featuring SA. To this end, we focus on a number of representative equations, all of which display GSI, and which remained outside the analysis in \cite{us}, due to the unavailability of accurate approximations through linear equations for most of the cases. Thus, we will employ techniques that in principle can tackle strongly non-linear systems, such as the Dynamic Renormalization Group and direct numerical simulations.

In the presence of conserved dynamics, we recall results for the paradigmatic caKPZ equation (which, as mentioned, displays WA), and generalize this equation into a related system which does present SA. Likewise, given that the HK equation has a special shape that does not admit an isotropic limit, we provide a natural generalization of it which does. However, this equation turns out to again display WA. Interestingly, while the caKPZ equation is invariant under a global shift of the height values $h \to h + {\rm const.}$, the (generalized) HK equation does not. Nevertheless, the overall behavior with respect to SA seems common. Thus, as a partial conclusion, we note that SA can appear for conserved dynamics, but it requires a shape from the interface equation that is not generic in parameter space. Actually, as argued with some generality in  \cite{tauber:02}, anisotropies do seem to play a more relevant role in the conserved dynamics case than for non-conserved dynamics. We confirm this in our present
context by revisiting the paradigmatic representative of non-conserved dynamics, namely, the aKPZ equation. Even analyzing some particular limits that had remained unexplored thus far, we confirm the general conclusion on the occurrence of WA throughout parameter space for this model.

The paper is organized as follows. Section II contains a brief reminder on the basic steps of the analytical technique we will employ in order to study the SA properties of the equations just mentioned, namely, the Dynamic Renormalization Group (DRG). This will allow us to establish our notation and some assumptions which are common to all cases discussed. Section III is devoted to equations with conserved dynamics, while the case of non-conserved dynamics is explored in Section IV. We will extract conclusions on modeling of strongly anisotropic systems in the final section V. Finally, we provide two appendices with details on our DRG calculations for the generalized HK equation and for the aKPZ equation. While the former is a new model, the latter has been long studied in the literature \cite{Wolf}, although in a short account. We hope these details can be found useful by the interested reader.

\section{Dynamic Renormalization Group}
\label{sec_cdyn}

In this section we review briefly the main steps taken by the analytical technique which will be extensively used in this work. As mentioned earlier, we will be addressing non-linear equations, trying to extract the scaling exponents $\alpha_{x}$, $\zeta$, and $z_x$ which are predicted in each case. While for linear stochastic equations such as those considered in \cite{us}, values for the latter can be readily extracted from a simple rescaling of coordinates and fields \cite{Barabasi}, in general this is not the case in the presence of nonlinearities. For these, it is a non-trivial balance between the linear and the non-linear operators occurring in the equation which controls its asymptotic behavior. The DRG is a standard perturbative approach to elucidate the interplay among these terms. Originally, the method was developed in the contexts of fluctuating hydrodynamics \cite{McComb} and dynamic critical phenomena \cite{Mazenko}. More recently, it has been successfully applied to understand e.g.\ the
multiscale nature of fluctuating interfaces~\cite{Haselwandter:07}, kinetic roughening in surfaces controlled by unstable nonlocal interactions~\cite{Nicoli:09,nicoli:2011}, or the interplay between noise and morphological instabilities in anisotropic pattern-forming systems \cite{keller:2011}, to cite a few examples.

Here, we sketch the main steps involved in our DRG analysis. These will later applied in a number of cases, computational details being provided in the appendices. For the systems that will be addressed, both linear and non-linear terms share similar structures when written in Fourier-space coordinates $\mathbf{k}$. Specifically, the non-linear evolution equations to be studied in the next sections can be written as
\begin{equation}
\label{gen_eq}
\partial_t h_{\bf k}(t) = \sigma_{\bf k} h_{\bf k}(t) + \lambda
\mathcal{N}[h,\nabla h]_{\bf k} + \eta_{\bf k}(t),
\end{equation}
where in general ${\bf k}$ is wavevector in $d$-dimensional Fourier space, although we will consider $d=2$ in our specific cases. The fluctuating term $\eta$ is taken as a Gaussian white noise with zero mean and variance equal to $2D$. For the sake of generality, at this moment we leave the linear dispersion relation $\sigma_{\bf k}$ unspecified. With respect to the non-linear operator $\mathcal{N}$, it also remains generic, except for the fact that it is a linear combination of {\em quadratic} products of the height field $h$ and its space derivatives ---such as e.g.\ $(\partial_x h)^2, h \partial_{x} h$, etc.---, with $\lambda$ being a representative non-linear coupling constant. The first step of the DRG procedure consists in time-Fourier transforming Eq.\ \eqref{gen_eq},
\begin{equation}
[-\sigma_{\bf k} - i\omega] h_{{\bf k},\omega} = \eta_{{\bf k},\omega} + \lambda
\int_{|{\bf q}|\leq \Lambda} \dfrac{d{\bf q}}{(2\pi)^d}
\int_{-\infty}^{+\infty} \dfrac{d\Omega}{2\pi} \, f_1({\bf q},{\bf k })\,
h_{{\bf q},\Omega} h_{{\bf k}-{\bf q},\omega-\Omega},
\end{equation}
where $\omega$ is time frequency and $\Lambda = \pi/\Delta x$ is the wavenumber cut-off in the system, $\Delta x$
being the lattice spacing in real space. The specific shape of the function $f_1({\bf q},{\bf k })$ depends on that of the non-linear term ${\cal N}$ in the equation. In Fourier space, the noise term still has zero mean $\langle \eta_{{\bf k},\omega}\rangle = 0$
and is delta-correlated, but its variance becomes rescaled as
\begin{equation}
\langle \eta_{{\bf k},\omega} \eta_{{\bf k}',\omega'}\rangle = 2 D
(2\pi)^{d+1}\, \delta_{{\bf k} + {\bf k}'}\, \delta_{\omega+ \omega'}.
\end{equation}
Following the standard Forster-Nelson-Stephen procedure~\cite{McComb}, the height and the noise fields are split into
two types of components, slow modes $h^<_{\bf{k},\omega}$,  $\eta^<_{\bf{k},\omega}$ for $k\in (0,\Lambda/b)$, and
fast modes $h^>_{\bf{k},\omega}$, $\eta^>_{\bf{k},\omega}$ for $k\in [\Lambda/b,\Lambda]$, where $b = e^{\delta l}>1$ is a
rescaling parameter. For infinitesimal $\delta l$, a small amount of fast modes is eliminated by solving perturbatively the equation for the modes $h^>_{\bf{k},\omega}$, substituting this solution into the equation for the slow
modes, and assuming statistical independence between high and low-frequency components. Formally, the small parameter in the perturbative expansion is the strength of the non-linear term, $\lambda$. This procedure leads to an effective equation in which the fast modes are thus integrated out perturbatively,
\begin{equation}
[-\sigma_{\bf k}-\Sigma({\bf k},0) - i\omega] h^<_{{\bf k},\omega} =
\eta^<_{{\bf k},\omega} + \lambda \int_< \dfrac{d{\bf q}}{(2\pi)^d}
\int \dfrac{d\Omega}{2\pi} \, f_1({\bf q},{\bf k })\, h^<_{{\bf q},\Omega}
h^<_{{\bf k}-{\bf q},\omega-\Omega} + {\rm O}(\lambda^3).
\label{-s-S}
\end{equation}
The effect of this coarse-graining procedure (i.e.\ the elimination of the fast
modes) is obtained by solving the integral
\begin{equation}
\label{Sigma}
\Sigma({\bf k},\omega) =  \lambda^2 D \int_> \dfrac{d{\bf q}}{(2\pi)^d}
\int \dfrac{d\Omega}{2\pi} \,
f_2({\bf q},{\bf k})\, G_0({\bf q},\Omega) G_0(-{\bf q},-\Omega) G_0({\bf k} -
{\bf q},\omega - \Omega) ,
\end{equation}
where the function $f_2({\bf q},{\bf k })$ depends on the exact form of the
nonlinearity. To lighten the notation, in the last two expressions we have omitted the
integration limits in the frequency domain, and we have denoted the integrals over the fast (slow)
modes as $\int_>$ ($\int_<$). In Eq.\ \eqref{Sigma} we have introduced the
bare propagator $G_0({\bf k},\omega) = (-\sigma_{\bf k} - i\omega)^{-1}$ whereas on the
left hand side of Eq.\ \eqref{-s-S} the coarse-grained propagator appears, namely,
\begin{equation}
G_0^<({\bf k},\omega) \equiv  \left[-\sigma_{\bf k} - \Sigma({\bf k},0) -
i\omega\right]^{-1}.
\end{equation}
From this last expression it is obvious that only the parameters appearing in the dispersion
relation are affected by the coarse-graining of the propagator.

The second parameter of the system that is renormalized is the variance of the
noise term. From the equation
\begin{equation}
\langle h^<_{{\bf k},\omega} h^<_{-{\bf k},-\omega}\rangle = 2D^< G_0^<({\bf
k},\omega)G_0^<(-{\bf k},-\omega),
\end{equation}
we can easily derive
\begin{equation}
\langle \eta^<_{{\bf k},\omega} \eta^<_{-{\bf k},-\omega}\rangle = 2\left[D^<
+\Phi({\bf k},0)\right]
	(2\pi)^{d+1}\, \, \delta_{{\bf k} + {\bf k}'}\, \delta_{\omega+ \omega'}
+ {\rm O}(\lambda^3),
\end{equation}
where the coarse-grained noise variance is given by
\begin{equation}
\Phi({\bf k},\omega) = \lambda^2 D^2 \int_> \dfrac{d{\bf q}}{(2\pi)^d}
\int \dfrac{d\Omega}{2\pi} \,
f_3({\bf q},{\bf k})\, |G_0({\bf q},\Omega)|^2 |G_0({\bf k} - {\bf q},\omega -
\Omega)|^2.
\end{equation}
As before, $f_3({\bf q},{\bf k})$ depends on the details of the nonlinearity for
each case considered.

The last step of the coarse-graining procedure has to deal with the corrections
to the non-linear coupling $\lambda$, that can be read off from the perturbative expansion that
allows to rewrite Eq.\ \eqref{-s-S} with the same structure as Eq.\ \eqref{gen_eq}, but with modified
parameters. However, as we have already demonstrated in \cite{nicoli:2011}, for a large class of
nonlinearities this parameter does not renormalize. The systems we consider here have exactly this
kind of behavior.

Finally, after coarse-graining of propagator, noise variance, and nonlinearities, the final step in the
DRG method is a rescaling that restores the value of the wavevector cut-off, $\Lambda/b$ after
coarse-graining, to its bare value $\Lambda$. Within the small $\delta l$ approximation, this moreover
allows to write parameter renormalization in a differential form, taking $l$ as the independent variable.
We will perform this rescaling on a case-by-case basis in the next sections.

\section{Conserved dynamics}

%

We start by considering systems for which dynamics are conserved. Recall that, in contrast with non-conserved dynamics, in such a case GSI ensues if noise is non-conserved \cite{grinstein:1991,grinstein:1995}, irrespective of whether the deterministic part of the dynamical equation is or not invariant under arbitrary global changes in the value of the height
$h(\mathbf{r},t) \to h(\mathbf{r},t) + \mbox{const.}$ We thus consider two representative examples, one in which
such {\em shift invariance} occurs and a different one in which it does not.

\subsection{Systems with shift invariance: conserved aKPZ equation}

A conceptually important example of an anisotropic conserved equation with non-conserved noise, which is invariant under arbitrary shifts of the height, is the caKPZ equation. This model has been formulated and studied  \cite{kallabis:1998,kallabis_thesis} in the context of non-equilibium growth of epitaxial thin films, specifically for surfaces which are vicinal to a high symmetry surface \cite{misbah:2010}. Specifically, the caKPZ equation reads
\begin{equation}
\partial_t h = - \nabla^2 \left[ \nu_x \partial_x^2 h + \nu_y \partial_y^2 h + \frac{\lambda_x}{2} (\partial_x h)^2
+ \frac{\lambda_y}{2} (\partial_y h)^2 \right] + \eta,
 \label{eq:cakpz}
\end{equation}
where a linear first-order derivative term has been omitted, which does not affect our discussion and conclusions \cite{kallabis:1998}. In Eq.\ \eqref{eq:cakpz}, $\nu_{x,y} > 0$ and $\lambda_{x,y}$ are constant parameters. Note, dynamics are explicitly conserved while noise is not, and the equation only depends on $h$ through its space or time derivatives, so that the equation does not single out any preferred height value. For generic parameter values, the DRG analysis performed in \cite{kallabis:1998,kallabis_thesis} leads to the conclusion that the system shows WA (i.e., $\zeta=1$), displaying the scaling exponents of the isotropic conserved KPZ equation, which reads \cite{lai:1991}
\begin{equation}
\partial_t h = - \nabla^2 \left[ \nu \nabla^2 h + \frac{\lambda}{2} (\nabla h)^2 \right] + \eta .
 \label{eq:ckpz}
\end{equation}
Namely, for $d=2$ the scaling exponents of the caKPZ equation are thus predicted to be approximately given through a one-loop DRG analysis \cite{lai:1991} (small corrections occur within a two-loop calculation \cite{janssen:1997}) by $\alpha \simeq 2/3$ and $z \simeq 10/3$. In particular, the change of universality class that occurs in the non-conserved anisotropic KPZ equation when changing the relative signs of the nonlinearities, from non-linear behavior for $\lambda_x \lambda_y > 0$ to linear behavior for $\lambda_x \lambda_y < 0$, does not occur for the caKPZ equation \cite{kallabis:1998,kallabis_thesis}.

\begin{figure*}[t!]
\centering
\begin{minipage}[t]{0.45\textwidth}
 \includegraphics[width=\textwidth]{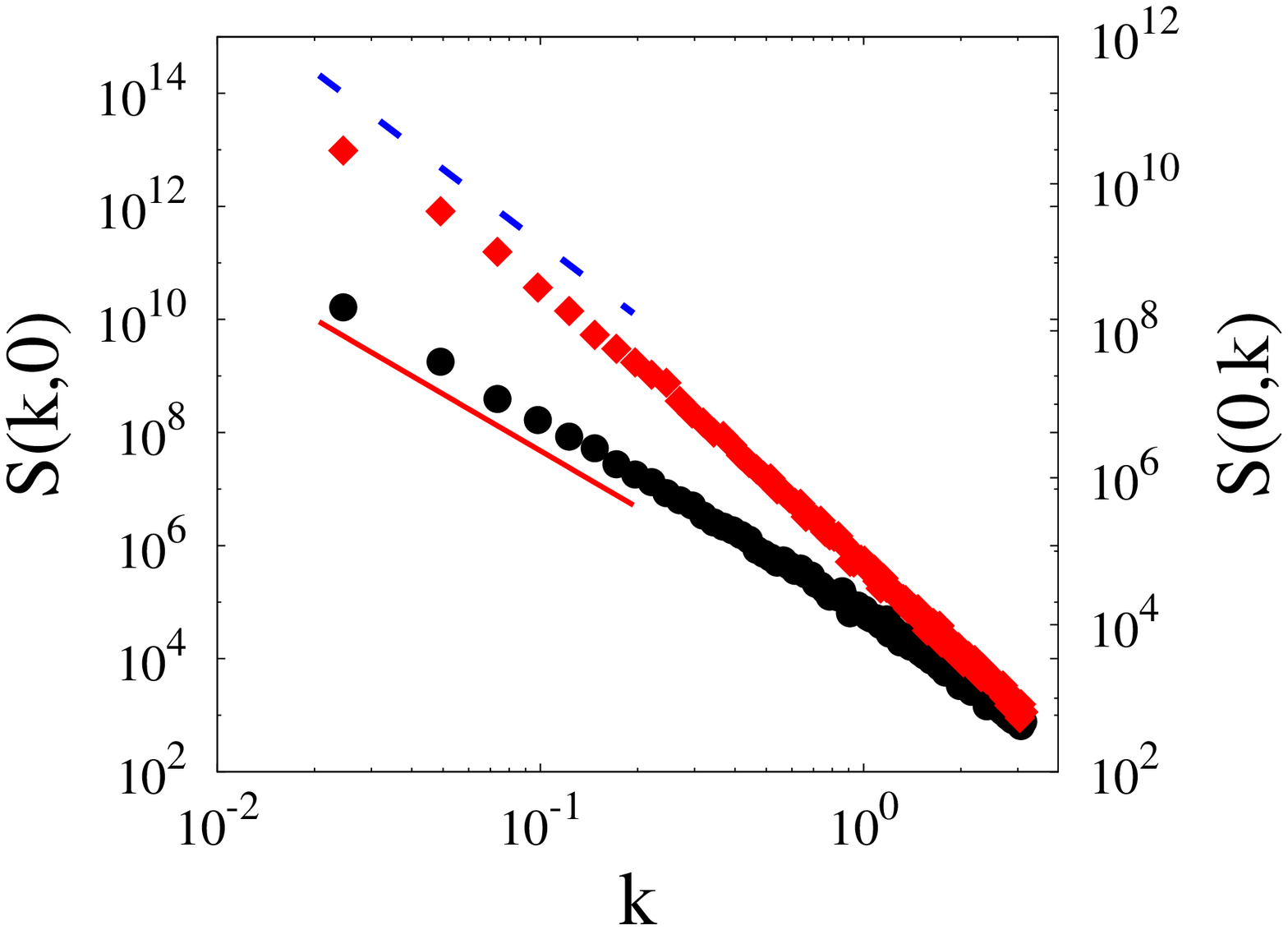}
\end{minipage}
\quad
\begin{minipage}[t]{0.45\textwidth}
\includegraphics[width=\textwidth]{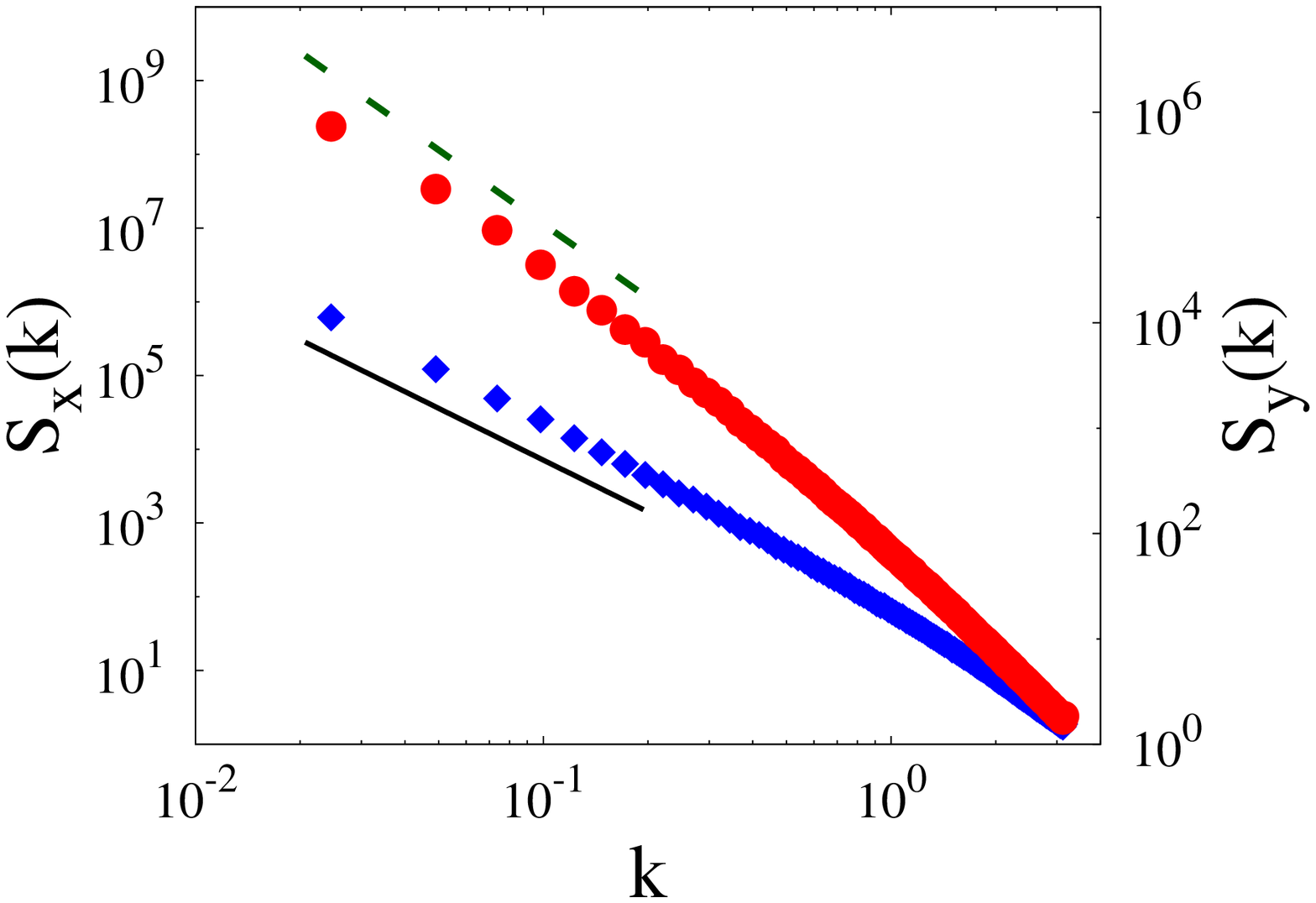}
\end{minipage}
\caption{(Color online) Numerical integrations of Eq.\ \eqref{eq:cakpz_2} for parameters $\nu_x = \nu_y = 1$, $D=1$, $\lambda_x=-3$, $\lambda_y=-1$, $L=256$, $\Delta x = 1$, and $\Delta t = 0.05$. Left panel: One-dimensional projections $S(k,0)$ (black circles, left axis) and $S(0,k)$ (red diamonds, right axis) of the two-dimensional PSD, averaged over 100 different noise realizations. Both the solid red line and the dashed blue line are guides for the eye with slope $-10/3$. Right panel: PSD of one-dimensional cuts $S_x(k)$ (blue squares, left axis) and $S_y(k)$ (red circles, right axis). Both the solid black line and the dashed green line are guides for the eye with slope $-7/3$. All units are arbitrary.}
\label{fig_psd_cakpz_positive}
\end{figure*}

In order to further discuss the scaling properties of the caKPZ equation, we first consider a similar model that shares with it the behavior just described. Thus, consider the equation
\begin{equation}
 \partial_t h = -\nu_x \partial_x^4 h - \nu_y \partial_y^4 h - \frac{\lambda_x}{2}\partial_x^2 (\partial_x h)^2 - \frac{\lambda_y}{2}\partial_y^2 (\partial_y h)^2 + \eta .
\label{eq:cakpz_2}
 \end{equation}
The main difference between Eqs.\ \eqref{eq:cakpz} and \eqref{eq:cakpz_2} is that each term in the latter, e.g.\ $-\frac{\lambda_x}{2}\partial_x^2 (\partial_x h)^2$, is affected by an overall second-order derivative operator with a reduced symmetry as compared to its counterpart in the former, e.g.\ $-\frac{\lambda_x}{2}\nabla^2 (\partial_x h)^2$. Nevertheless, the scaling behavior is not modified, as we have verified by numerical simulations. Specifically, we have integrated Eq.\ \eqref{eq:cakpz_2} by means of a pseudo-spectral integration algorithm as described in \cite{nicoli:2008} and references therein. The results of the simulations are presented in Figs.\ \ref{fig_psd_cakpz_positive} and \ref{fig_psd_cakpz_negative}.

\afterpage{
\begin{figure*}[t]
\centering
\begin{minipage}[b]{0.45\textwidth}
\includegraphics[width=\textwidth]{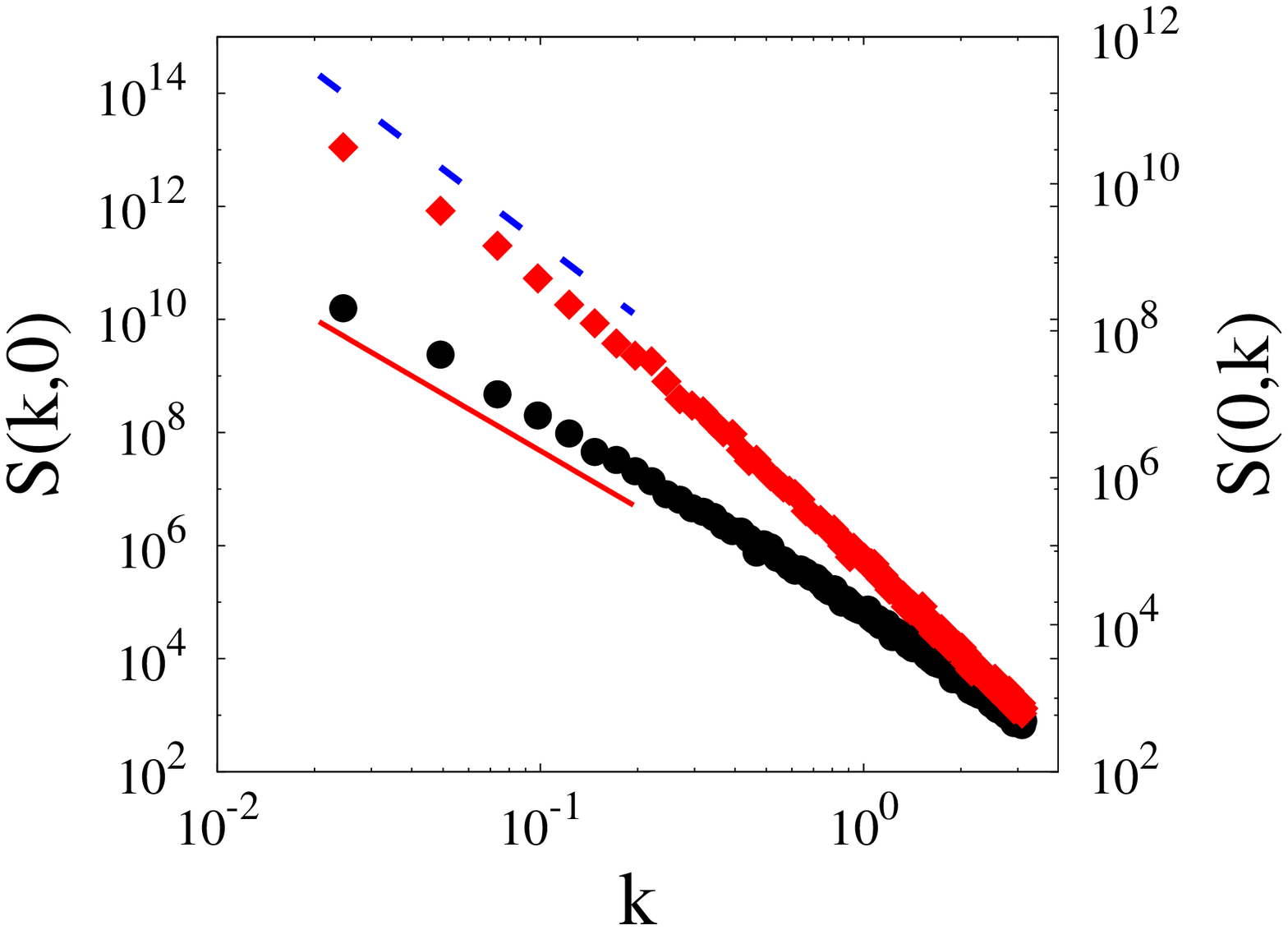}
\end{minipage}
\quad
\begin{minipage}[b]{0.45\textwidth}
\includegraphics[width=\textwidth]{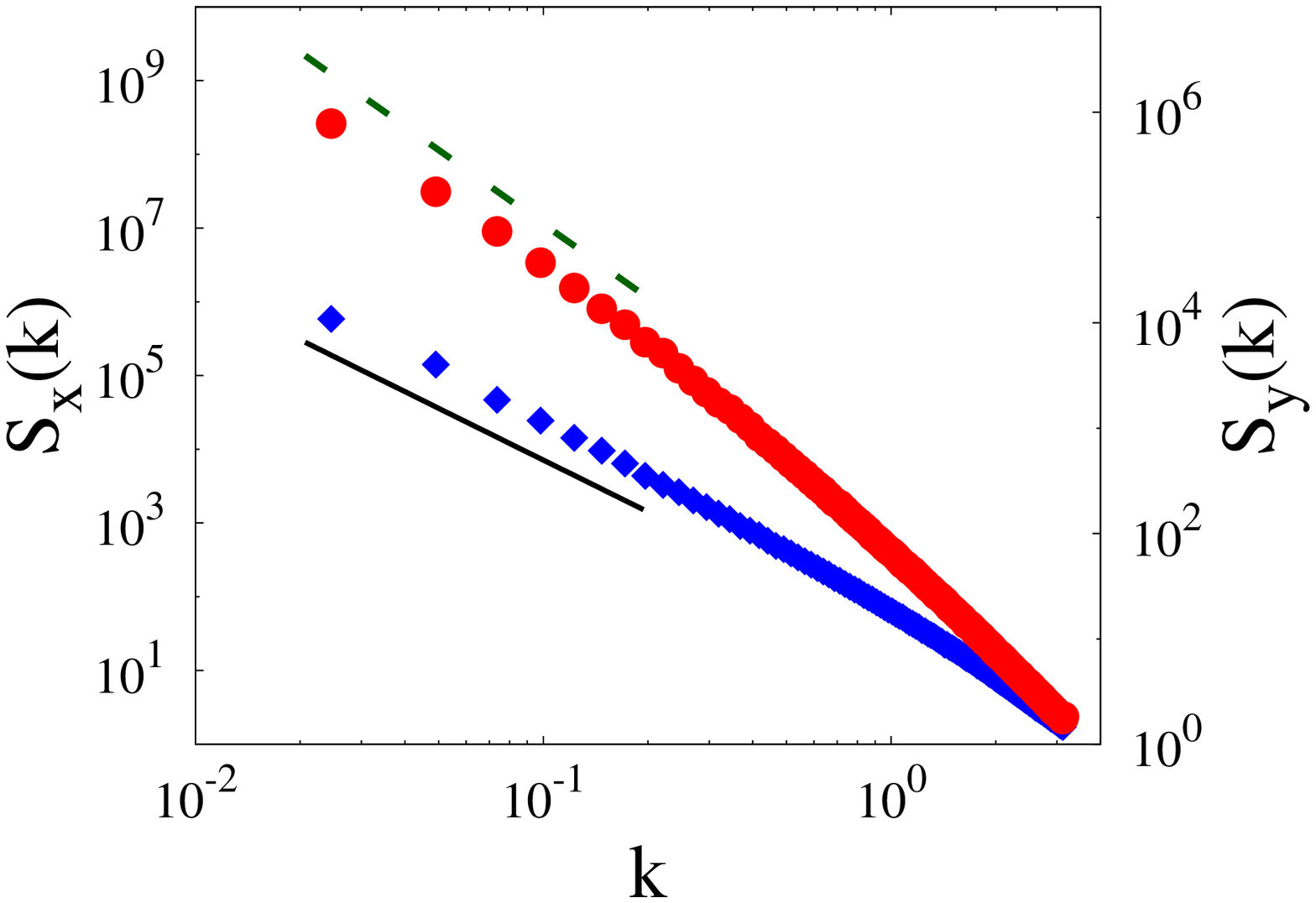}
\end{minipage}
\caption{Numerical integrations of Eq.\ \eqref{eq:cakpz_2} for parameters $\nu_x = \nu_y = 1$, $D=1$, $\lambda_x=-3$, $\lambda_y=1$, $L=256$, $\Delta x = 1$, and $\Delta t = 0.05$. Left panel: (Color online). One-dimensional projections $S(k,0)$ (black circles, left axis) and $S(0,k)$ (red diamonds, right axis) of the two-dimensional PSD, averaged over 100 different noise realizations. Both the solid red line and the dashed blue line are guides for the eye with slope $-10/3$. Right panel: PSD of one-dimensional cuts $S_x(k)$ (blue squares, left axis) and $S_y(k)$
(red circles, right axis). The numerical integrations were performed for the same parameters as in the left panel. Both the solid black line and the dashed green line are guides for the eye with slope $-7/3$. All units are arbitrary.}
\label{fig_psd_cakpz_negative}
\end{figure*}
}
Thus, the left and right panels of Fig.\ \ref{fig_psd_cakpz_positive}
show, respectively, cuts of the 2D PSD function $S(\mathbf{k})$ along the coordinate axes in $\mathbf{k}$-space and 1D PSD functions for cuts of the surface along the $x$ and $y$ directions, $S_{x,y}(k_{x,y})$, for a condition of Eq.\ \eqref{eq:cakpz_2} in which $\lambda_x \lambda_y > 0$. Agreement with asymptotic scaling behavior as in Eqs.\ \eqref{psd2d}, \eqref{eq_relalphax}, \eqref{eq_relalphay}, \eqref{eq:psd1d_scaling}, and \eqref{eq:psd1d_scaling2} for a WA case is very good, using the expected exponents for the isotropic cKPZ equation, $\alpha_x=\alpha_y \simeq 2/3$. Similar agreement is obtained in Fig.\ \ref{fig_psd_cakpz_negative}
for a condition of Eq.\ \eqref{eq:cakpz_2} in which $\lambda_x \lambda_y < 0$. Therefore, the scaling exponents correspond to those of the isotropic cKPZ equation,  irrespective of the relative signs of the nonlinearities, so we can safely say that Eq.\ \eqref{eq:cakpz_2} is in the same universality class as the caKPZ equation, Eq.\ \eqref{eq:cakpz}.

\begin{figure}[!t]
\centering
\begin{minipage}[t]{0.45\textwidth}
\includegraphics[width=\textwidth]{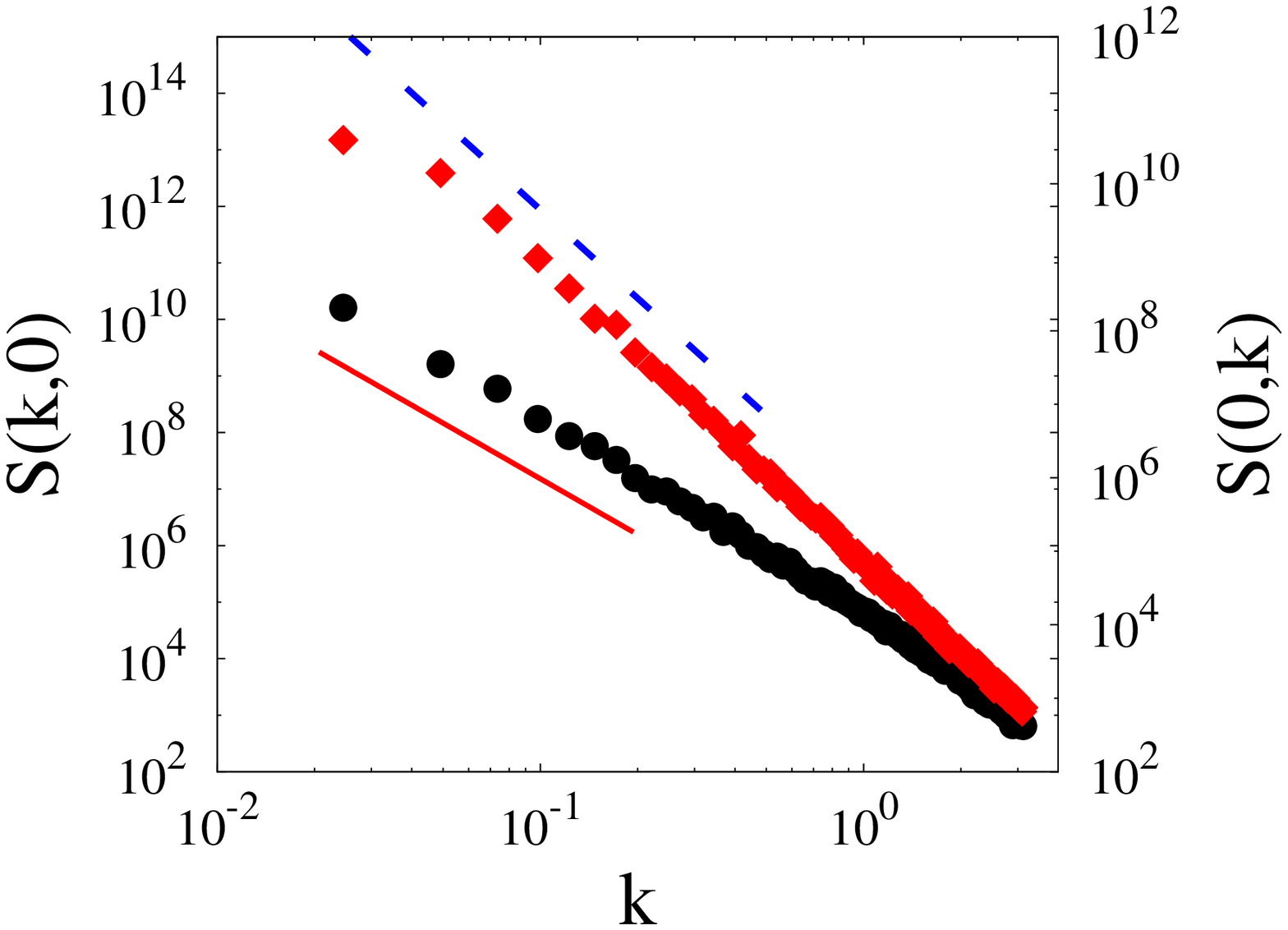}
\end{minipage}
\quad
\begin{minipage}[t]{0.45\textwidth}
\includegraphics[width=\textwidth]{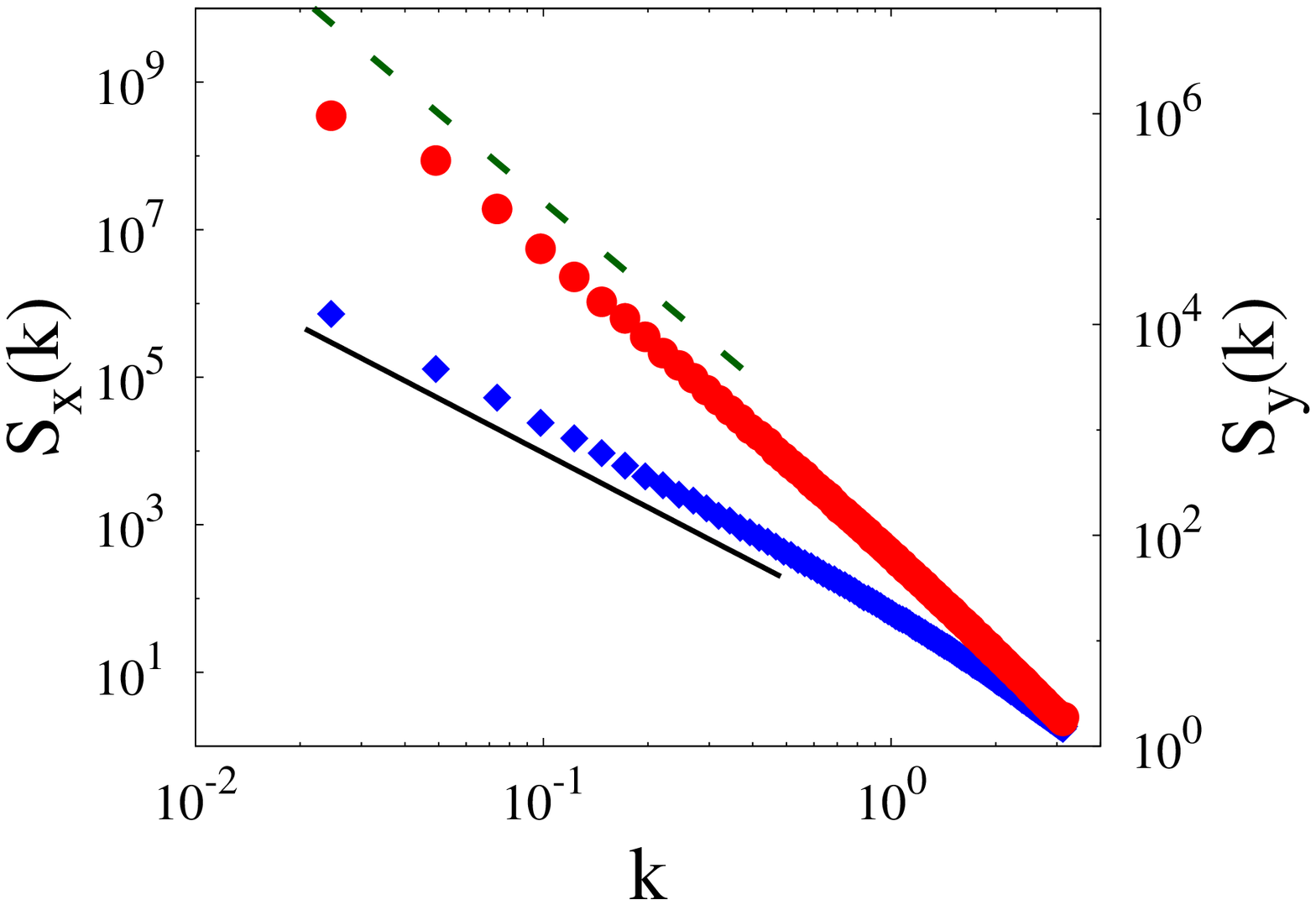}
\end{minipage}
\caption{ Numerical integrations of Eq.\ \eqref{eq:cakpz_3} for parameters $\nu_x = \nu_y = 1$, $D=1$, $\lambda_x=3$, $L=256$, $\Delta x = 1$, and $\Delta t = 0.05$. Left panel: One-dimensional projections $S(k,0)$ (black circles, left axis) and $S(0,k)$ (red diamonds, right axis) of the two-dimensional PSD, averaged over 100 different noise realizations. The solid red line and the dashed blue line are guides for the eye with slopes $-36/11$ and $-4$ respectively. Right panel: PSD of one-dimensional cuts $S_x(k)$ (blue squares, left axis) and $S_y(k)$
(red circles, right axis). The solid black line and the dashed green line are guides for the eye with slope $-27/11$ and $-25/9$ respectively. All units are arbitrary.   }
\label{fig_psd_cakpz_zero}
\end{figure}

Having established the previous result, the only possibility for Eq.\ \eqref{eq:cakpz_2}, and equivalently for the caKPZ equation, to display SA behavior is that one nonlinearity, say $\lambda_y$, is suppressed, but not the other. Hence, we consider equation
\begin{equation}
 \partial_t h = -\nu_x \partial_x^4 h - \nu_y \partial_y^4 h - \frac{\lambda_x}{2}\partial_x^2 (\partial_x h)^2 + \eta .
\label{eq:cakpz_3}
 \end{equation}
In a specific physical situation, this implies a non-generic parameter condition, e.g.\ that the corresponding non-linear contribution to the surface-diffusion current vanishes \cite{Barabasi} due to a special parameter choice. This case seems not to have been considered in \cite{kallabis:1998}. Actually, we can benefit from the DRG analysis performed by Kallabis in order to derive expectations for the critical exponents of Eq.\ \eqref{eq:cakpz_3}: After the coarse-graining step is performed (full details are available in \cite{kallabis_thesis}) as described in Sect.\ \ref{sec_cdyn}, we perform an anisotropic rescaling that restores the original wave-vector cut-off, namely,
\begin{equation}
x \to b x , \;\; y \to b^{\zeta} y , \;\; t \to b^{z_x} t , \;\; h \to b^{\alpha_x} h .
\label{eq:rescaling}
\end{equation}
Using $b=e^{\delta l}$, and taking into account the net modification of the equation parameters after both the coarse-graining and the rescaling transformations, the DRG parameter flow for $\nu_y, \lambda_x$, and $D$ reads particularly simple, namely,
\begin{eqnarray}
\dfrac{d \nu_{y}}{dl} & = & \nu_{y}(z_x - 4\zeta), \label{nuy_ca} \\[5pt]
\dfrac{d \lambda_x}{dl} & = & \lambda_x (\alpha_x + z_x -4), \label{lx_ca} \\[5pt]
\dfrac{d D}{dl} & = & D (z_x - 2\alpha_x - \zeta -1), \label{D_ca}
\end{eqnarray}
which actually coincides with the result of a mere parameter rescaling \cite{Barabasi}. The reasons behind such a simplicity are: (i) Given that in Eq.\ \eqref{eq:cakpz_3} $\lambda_y=0$ to begin with, parameter renormalization can be only due to the remaining nonlinearity $\lambda_x$, which does not contribute to $k_y^2$ order, hence $\nu_y$ does not renormalize; (ii) at one-loop order there is a vertex cancellation \cite{nicoli:2011} by which $\lambda_x$ does not renormalize either; (iii) as standard for conserved equations with non-conserved noise, since the lowest-order non-linear modification of the noise propagator is $O(k_x^2)$, the variance $D$ is not affected and it does not renormalize either. Finally, the fixed points of the RG flow control the scaling behavior. Thus, setting to zero the right-hand sides of Eqs.\ \eqref{nuy_ca}-\eqref{D_ca} we obtain
\begin{eqnarray}
\label{eq:scaling_rel_cakpz}
 & & z_x = 4\zeta , \\ 
 & & \alpha_x+z_x = 4 , \label{eq:scaling_rel_cakpzb} \\ 
 & & z_x = 2 \alpha_x + \zeta + 1 . \label{eq:scaling_rel_cakpzc} 
\end{eqnarray}
These are three equations for three unknowns, whose solution does correspond to SA behavior, namely,
\begin{equation}
\alpha_x = 8/11,\qquad z_x = 36/11, \qquad \zeta = 9/11.
\label{11s}
\end{equation}
We have performed numerical simulations of Eq.\ \eqref{eq:cakpz_3} in order to verify Eq.\ \eqref{11s}. The results, presented in Fig. \ref{fig_psd_cakpz_zero},
are in good agreement with these SA values of the scaling exponents.

In spite of being strongly anisotropic, the $\zeta$ value obtained for Eq.\ \eqref{eq:cakpz_3} is very close to one, so that effectively scaling behavior is not far from a proper WA case. For practical applications, Eq.\ \eqref{eq:cakpz_3}, and thus the caKPZ equation with $\lambda_y=0$, is not the most clear-cut example of strong anisotropy. However, Eqs.\ \eqref{eq:scaling_rel_cakpz}-\eqref{eq:scaling_rel_cakpzc} give us a way to construct an equation similar to \eqref{eq:cakpz_3}, but with a tunable anisotropy exponent $\zeta$. In wave-vector space, such an equation can be written as
\begin{equation}
\partial_t h = -(\nu_x |k_x|^{2n+2} + \nu_y |k_y|^{2m})h_{\mathbf k} - \frac{\lambda_x}{2} |k_x|^{2n} \mathcal{F}[(\partial_x h)^2] + \eta_{\mathbf k},
 \label{eq:cakpz_nm}
\end{equation}
where $\mathcal{F}[\cdot]$ denotes space Fourier transform, and $n$ and $m$ are {\em real} numbers. Notice, Eq.\ \eqref{eq:cakpz_3} corresponds simply to the particular choice $n=1$ and $m=2$.
In exactly the same form as Eqs.\ \eqref{eq:scaling_rel_cakpz}, it is not difficult to derive the following scaling relations for Eq.\ \eqref{eq:cakpz_nm},
\begin{align}
 \alpha_x + z_x &= 2(n+1), \\
 z_x &= 2 m \zeta, \\
 z_x &= 2\alpha_x + 1 + \zeta,
\end{align}
whose solution provides the following values of the exponents, as functions of $n$ and $m$,
\begin{equation}
 \alpha_x = \frac{4 n m +2 m -2 n -2}{6m -1}, \qquad z_x = 2 m\frac{4n+5 }{6m-1},\qquad \zeta = \frac{4n+5 }{6m-1} .
\label{exp_nm}
\end{equation}
Indeed, Eqs.\ \eqref{exp_nm} reduce to Eqs.\ \eqref{eq:scaling_rel_cakpz}-\eqref{eq:scaling_rel_cakpzc} for $n=1$ and $m=2$. The advantage is that now we can make different choices for $(n,m)$ in such a way that $\zeta$ is far from unity and SA behavior is enhanced. Note, a result such as Eq.\ \eqref{exp_nm} is remarkable, as it provides the solution for the scaling exponents of a two-parameter family of non-linear equations. An analogous result was obtained for the HK equation in \cite{HK:1989,HK:1992}, where it was argued to hold at any order in the DRG loop expansion. Indeed, it is due to the symmetries of the system as discussed above, leading to the three scaling relations among exponents. In the case of the HK equation proper this even allows to approximate it by a {\em linear} equation with the exact same scaling exponents \cite{us}.

\begin{figure}[t!]
\centering
\begin{minipage}[t]{0.45\textwidth}
 \includegraphics[width=\textwidth]{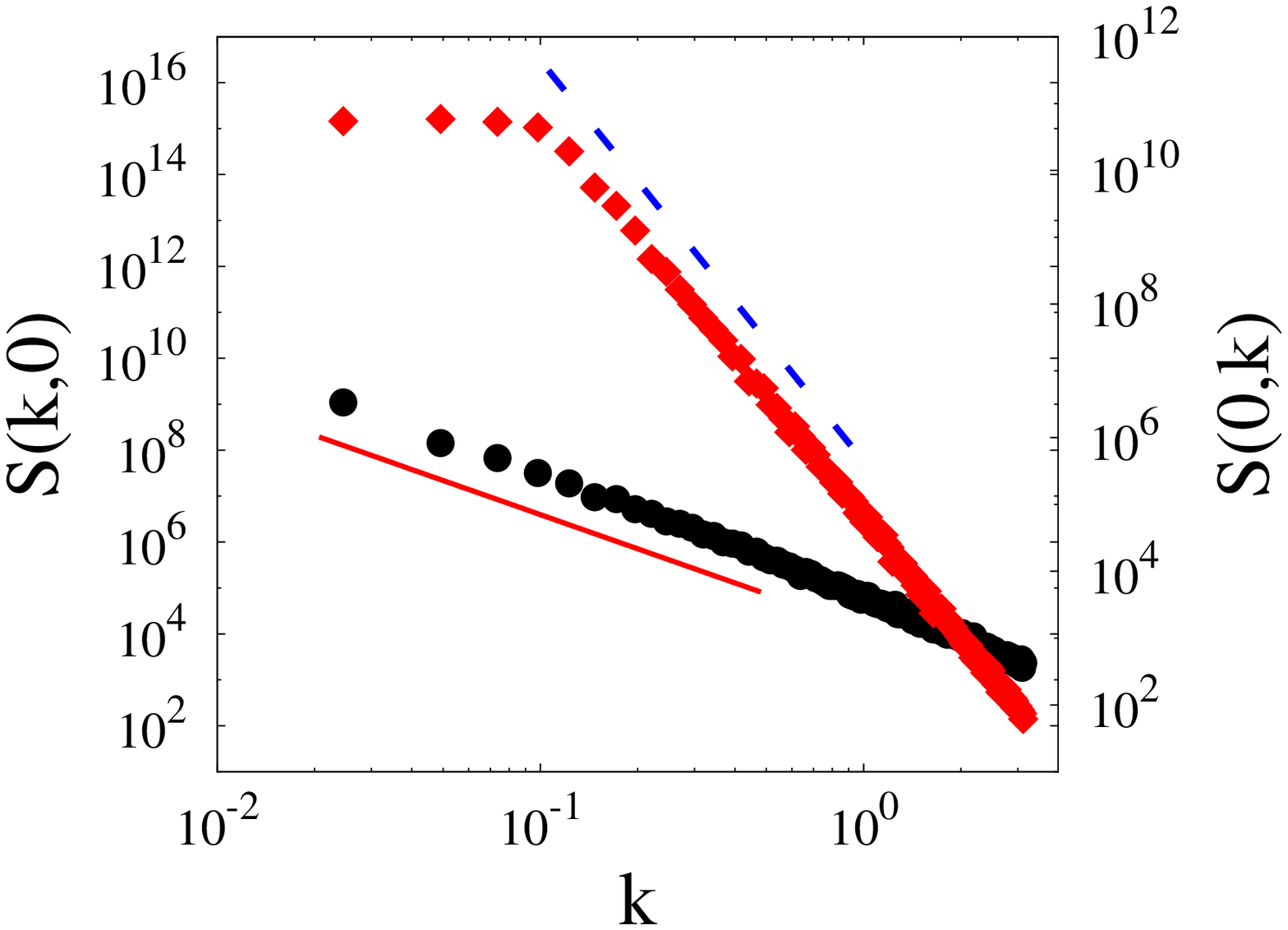}
\end{minipage}
\quad
\begin{minipage}[t]{0.45\textwidth}
\includegraphics[width=\textwidth]{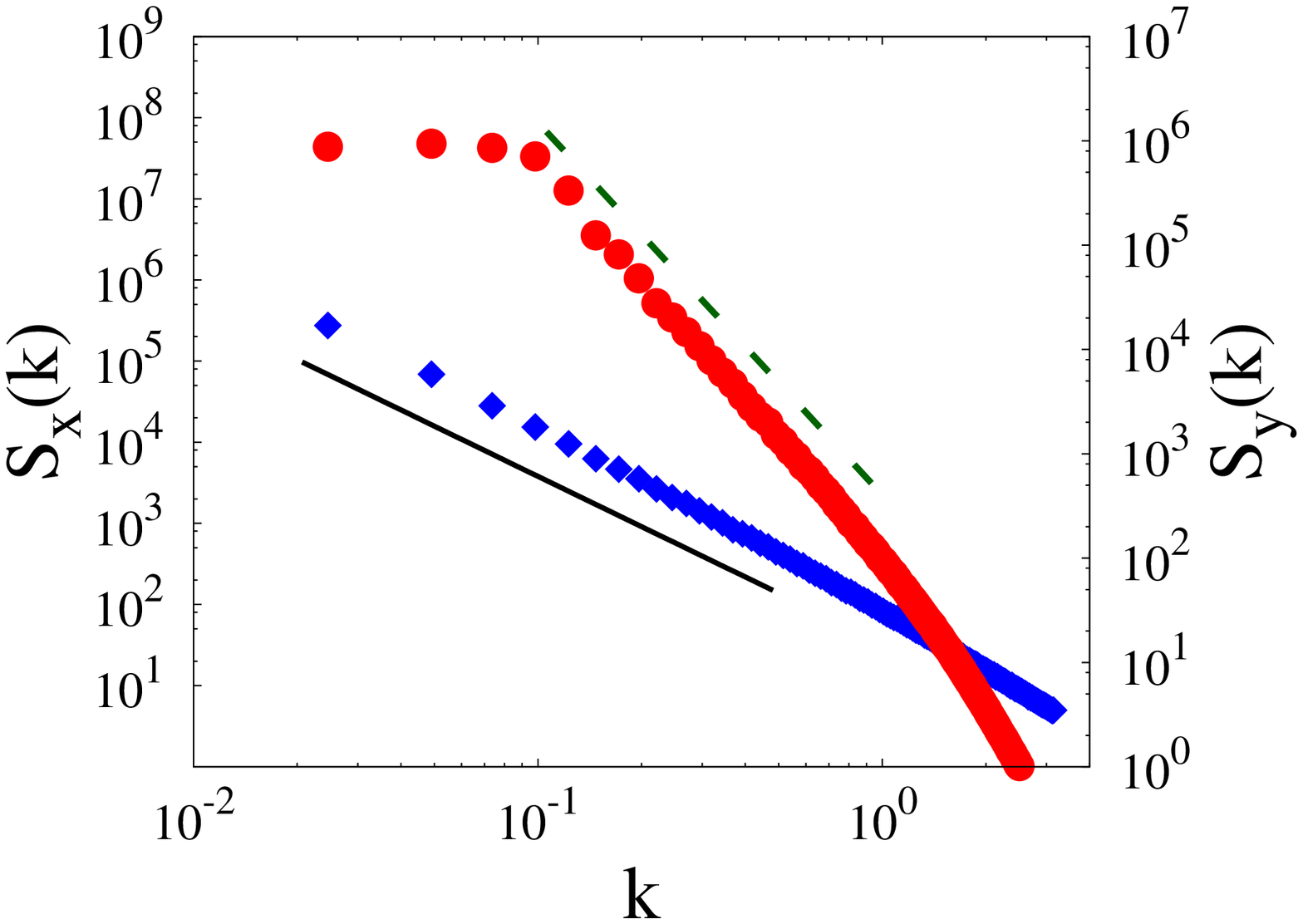}
\end{minipage}
\caption{ Numerical integrations of Eq.\ \eqref{eq:cakpz_3} for $n = 1/2$, $m = 3$, and parameters $\nu_x = \nu_y = 1$, $D=1$, $\lambda_x=2$, $L=256$, $\Delta x = 1$, and $\Delta t = 0.05$. Left panel: One-dimensional projections $S(k,0)$ (black circles, left axis) and $S(0,k)$ (red diamonds, right axis) of the two-dimensional PSD, averaged over 100 different noise realizations. The solid red line and the dashed blue line are guides for the eye with slopes $-42/17$ and $-6$ respectively. Right panel: PSD of one-dimensional cuts $S_x(k)$ (blue squares, left axis) and $S_y(k)$ (red circles, right axis). The solid black line and the dashed green line are guides for the eye with slope $-35/17$ and $-25/7$ respectively. All units are arbitrary.
}
\label{fig_psd_cakpznm}
\end{figure}

As an specific example, we have performed numerical simulations of Eq.\ \eqref{eq:cakpz_nm} with $n=1/2$ and $m=3$ in order to compare with the expected scaling exponents, which are
\begin{equation}
 \alpha_x = \frac{9}{17}, \qquad z_x = \frac{42}{17}, \qquad \zeta = \frac{7}{17}.
\end{equation}
The results, presented in Fig.\ \ref{fig_psd_cakpznm}
indeed agree with these values. Notice in this case full saturation of correlations along the $y$ direction has not been achieved for our longest simulation times, hence the $k_y$-independent behavior of $S(0,k_y)$ and $S_y(k_y)$ at small arguments.

As a summary of the results in this section, we conclude that SA is indeed feasible for conserved equations with non-conserved noise which are invariant under global shifts of the height field. However, this requires the suppression of nonlinearities along one of the substrate directions, which is a non-generic parameter condition. Notice, under such a constraint the equation cannot possibly be brought into isotropic form by any simple combination of coordinate rotations and/or rescalings in the substrate plane.

\subsection{Systems without shift invariance: generalized HK equation}

As described in the Introduction, in our previous work \cite{us} we considered the Hwa-Kardar equation, which was originally proposed to describe the interface dynamics of a running-sandpile model, in the context of SOC \cite{HK:1989,HK:1992}. For a two-dimensional substrate such as we are currently considering, this equation reads
\begin{equation}
\label{eq_hk2}
\partial_t h = \nu_x \partial_x^2 h + \nu_y \partial_y^2 h - \dfrac{\lambda_x}{2}\partial_x h^2 + \eta .
\end{equation}
In the original formulation \cite{HK:1989,HK:1992}, the linear terms model the relaxation of the height of the sand-pile through diffusive transport, whereas the nonlinearity accounts for the lack of inversion symmetry in the $x$ direction, being related to the presence of the external driving provided by the influx of sand. This is assumed to occur along the $x$ axis, which is an example of a non-generic condition for nonlinearities in the context of discussed in the previous section. The noise term $\eta$ in Eq.\ \eqref{eq_hk2} mimics the random addition of sand particles from outside the system, thus being non-conserved. This leads to GSI properties, in spite of the fact that the HK equation depends explicitly on $h$, and not only on its derivatives \cite{grinstein:1995}. In particular, SA occurs, scaling exponents having been obtained, analytically in \cite{HK:1989,HK:1992} through a DRG approach, and numerically in \cite{us}, which are
\begin{equation}
\label{eq_hk_2exp}
\alpha_x^{{\rm HK}} = -\dfrac{1}{5}, \quad z_x^{{\rm HK}} = \dfrac{6}{5}, \quad \zeta^{{\rm HK}} = \dfrac{3}{5}.
\end{equation}
Note that the negative values of $\alpha_{x,y}$ actually imply subdominant (logarithmic) behavior for observables in real space, such as e.g.\ the surface roughness \cite{Barabasi}. As discussed in detail in \cite{us}, it also leads to slow convergence even for observables in Fourier space, but which are integrals of the 2D PSD function, such as $S_{x,y}(k_{x,y})$. We will meet again this type of behavior in some specific examples to be discussed below.

In view of the results of the previous section, a natural question is whether the different behavior of the HK under global shifts of the height, as compared to e.g.\ the caKPZ equation, could allow for the occurrence of SA even for more generic parameter conditions such that, e.g., the non-linear part of the equation could be brought into an isotropic form via appropriate coordinate transformations in the substrate plane. In order to elucidate this possibility, we generalize the HK equation into
\begin{equation}
\frac{\partial h}{\partial t} = \nu_x \partial^2_x h + \nu_y \partial^2_y h + \nu_{xy} \partial_x \partial_y h-
\frac{\lambda_x}{2} \partial_x h^2 - \frac{\lambda_y}{2} \partial_y h^2 + \eta,
\label{chk}
\end{equation}
which will be henceforth referred to as the gHK equation. Indeed, the original HK equation simply corresponds to the particular case of Eq.\ \eqref{chk} in which $\lambda_x \neq 0$ while $\lambda_y=\nu_{xy}=0$. The term proportional to $\nu_{xy}$ has been introduced for technical reasons, as will become clear next.

In order to derive analytical insight into the critical behavior of the gHK equation, we apply to it the DRG procedure described in Sec.\ \ref{sec_cdyn}. The flow equations for the renormalization of the parameters of the gHK equation read
\begin{align}
\dfrac{d\nu_x}{dl} &= \nu_x \left( z_x -2 - \Sigma_{\nu_x}\right), \quad
\dfrac{d\nu_{xy}}{dl} = \nu_{xy}(z_x - \zeta -1 - \Sigma_{\nu_{xy}}), \quad
\dfrac{d\nu_y}{dl} = \nu_y \left( z_x -2\zeta - \Sigma_{\nu_y}\right),
\label{flux_nus}\\[5pt] 
\dfrac{d\lambda_x}{dl} &= \lambda_x (\alpha_x + z_x -1), \qquad
\dfrac{d\lambda_y}{dl} = \lambda_y (\alpha_x + z_x - \zeta),\label{flux_lambdas}
\\[5pt]
\dfrac{d D}{dl} &= D (z_x - 2\alpha_x - \zeta -1). \label{flux_D}
\end{align}
where $\Sigma_{\nu_x}$, $\Sigma_{\nu_y}$ and $\Sigma_{\nu_{xy}}$ are functions of $\nu_{x,y}$ and $\lambda_{x,y}$ which are provided in Appendix \ref{app_gHK}, together with further details on the derivation of Eqs.\ \eqref{flux_nus}-\eqref{flux_D}. From Eq.\ \eqref{flux_nus} note that for the gHK equation, even if the term with bare parameter $\nu_{xy}$ were initially zero, it is in principle generated by the coarse-graining procedure. This is due to the fact that $\Sigma_{\nu_{xy}}$ has a prefactor of $1/\nu_{xy}$, see Eq.\ \eqref{eq:sigmanuxy}, so that the term $\nu_{xy} \Sigma_{\nu_{xy}}$ in the flux equation for $\nu_{xy}$ will not generically vanish, even when $\nu_{xy} = 0$.
This is the reason why we have incorporated it to the definition of Eq.\ \eqref{chk}, in order to correctly take it into account in the DRG analysis.

We can write down an equivalent DRG flow which does not depend explicitly on $\alpha_z$ and $z_x$ through the identification of natural couplings in the system, such as, e.g.,
\begin{equation}
\label{couplings}
g = \frac{\lambda_x^2 D}{16 \pi^2 \Lambda^2\nu_x^3},
\quad r_{\nu} = \frac{\nu_y}{\nu_x}, \quad f_{\nu} = \frac{\nu_{xy}}{\nu_x},
\quad r_\lambda= \frac{\lambda_y}{\lambda_x} .
\end{equation}
Thus, we get
\begin{align}
& \dfrac{d r_\lambda}{dl}= r_\lambda \, (1-\zeta) , \label{flux_rlam}\\[5pt]
& \dfrac{d f_\nu}{dl} =  f_{\nu} \, \Big[1 - \zeta + \Sigma_{\nu_x} -
\Sigma_{\nu_{xy}}\Big], \label{flux_fnu}\\[5pt]
& \dfrac{d r_\nu}{dl}= r_\nu  \, \Big[ 2\,(1-\zeta) + \Sigma_{\nu_x} -
\Sigma_{\nu_y}\Big], \label{flux_rnu}\\[5pt]
& \dfrac{d g}{dl}=  g \, \Big[ 3 - \zeta + 3\, \Sigma_{\nu_x} \Big].
\label{flux_g}
\end{align}

\subsubsection{HK equation as a particular case}

\begin{figure}[t!]
\begin{center}
\centering
\epsfig{clip=,width=0.445\textwidth,file=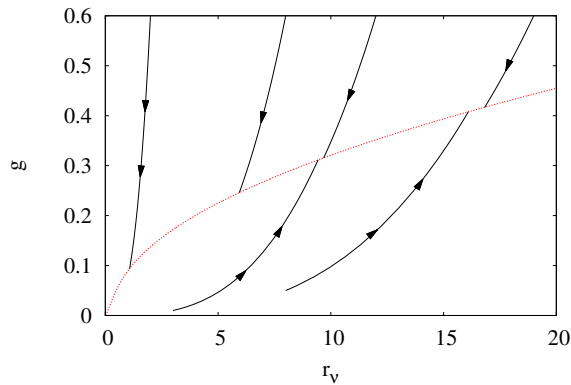}
\caption{(Color online). Numerical integration of the DRG flow for the gHK equation \eqref{chk} in the HK limiting case $r_\lambda = f_\nu = 0$, Eqs.\ \eqref{flux_rnu}-\eqref{flux_g}. The solid black lines are flow trajectories, while the dashed red line is the manifold of fixed points of the flow. All units are arbitrary.}
\label{fig_flux_HK}
\end{center}
\end{figure}

The behavior of the original HK equation, which corresponds to $r_\lambda = f_\nu = 0$, can be readily obtained from the above DRG results, see details in Appendix \ref{appa:hk}. The non-trivial part of the flow reduces in this case to
\begin{align}
& \dfrac{dr_\nu}{dl} = 2r_\nu (1-\zeta) -
g\left[\dfrac{3\zeta+(7+\zeta)r_\nu+5r_\nu^2}{(1+r_\nu)^2}+ 5 \sqrt{r_\nu}\tan^{-1}(\sqrt{r_\nu}) + \dfrac{3\zeta}
{\sqrt{r_\nu}}
		\tan^{-1}\left(\dfrac{1}{\sqrt{r_\nu}}\right)\right], \label{rnu_HK}
\\[5pt]
& \dfrac{dg}{dl} = g(3-\zeta) -
\dfrac{3g^2}{r_\nu}\left[\dfrac{3\zeta+(7+\zeta)r_\nu+5r_\nu^2}{(1+r_\nu)^2}
+ 5 \sqrt{r_\nu}\tan^{-1}(\sqrt{r_\nu}) + \dfrac{3\zeta}
{\sqrt{r_\nu}}
		\tan^{-1}\left(\dfrac{1}{\sqrt{r_\nu}}\right)\right]. \label{g_HK}
\end{align}
The fixed points of Eq.\ \eqref{g_HK} are either $g = 0$ or
\begin{equation}
g = g^* \equiv \frac{3-\zeta}{3} r_\nu \left[\dfrac{3\zeta+(7+\zeta)r_\nu+5r_\nu^2}{(1+r_\nu)^2}
+ 5 \sqrt{r_\nu}\tan^{-1}(\sqrt{r_\nu}) + \dfrac{3\zeta}
{\sqrt{r_\nu}}
		\tan^{-1}\left(\dfrac{1}{\sqrt{r_\nu}}\right)\right]^{-1}.
\label{g*}
\end{equation}
If $g=0$, Eq.\ \eqref{flux_rnu} implies $\zeta=1$, see Eq.\ \eqref{rnu_HK}. In contrast, setting $g=g^*$ requires $\zeta = 3/5$ in order to yield a fixed point for Eqs.\ \eqref{flux_rnu}-\eqref{flux_g} [note Eqs.\ \eqref{flux_rlam}-\eqref{flux_fnu} hold automatically since we have set $\nu_{xy}=\lambda_y=0$]. Moreover, in this case a manifold of fixed points actually exists in $(r_{\nu},g)$ parameter space, described by the equation obtained once we set $\zeta=3/5$ in Eq.\ \eqref{g*}, namely,
\begin{equation}
 g = \frac{4}{5}r_\nu \left[\dfrac{9 + 38 r_\nu + 25 r_\nu^2}{5(1+r_\nu)^2}
+ 5 \sqrt{r_\nu}\tan^{-1}(\sqrt{r_\nu}) + \dfrac{9}
{5\sqrt{r_\nu}}
		\tan^{-1}\left(\dfrac{1}{\sqrt{r_\nu}}\right)\right]^{-1}.
\end{equation}
In order to explore the stability of this family of fixed points, we have numerically integrated the flux \eqref{flux_rnu}-\eqref{flux_g} for $\nu_{xy}=\lambda_y=0$. The result is shown in Fig. \ref{fig_flux_HK}, where it is clear that all the fixed points on the manifold are attractive, an eloquent statement of GSI, and in stark contrast with the role of RG fixed points in equilibrium critical systems, which are unstable due to the relevance of temperature perturbations. Moreover, each point on the manifold corresponds to the \emph{same} set of scaling exponents, which are obtained by going back to Eqs.\ \eqref{flux_nus}-\eqref{flux_D}, and plugging in the values of $g^*$ and $\zeta$. The resulting exponents have the expected values for the HK equation, namely Eq.\ \eqref{eq_hk_2exp}.

\begin{figure}[b!]
\centering
\begin{minipage}[b]{0.425\textwidth}
 \includegraphics[width=\textwidth]{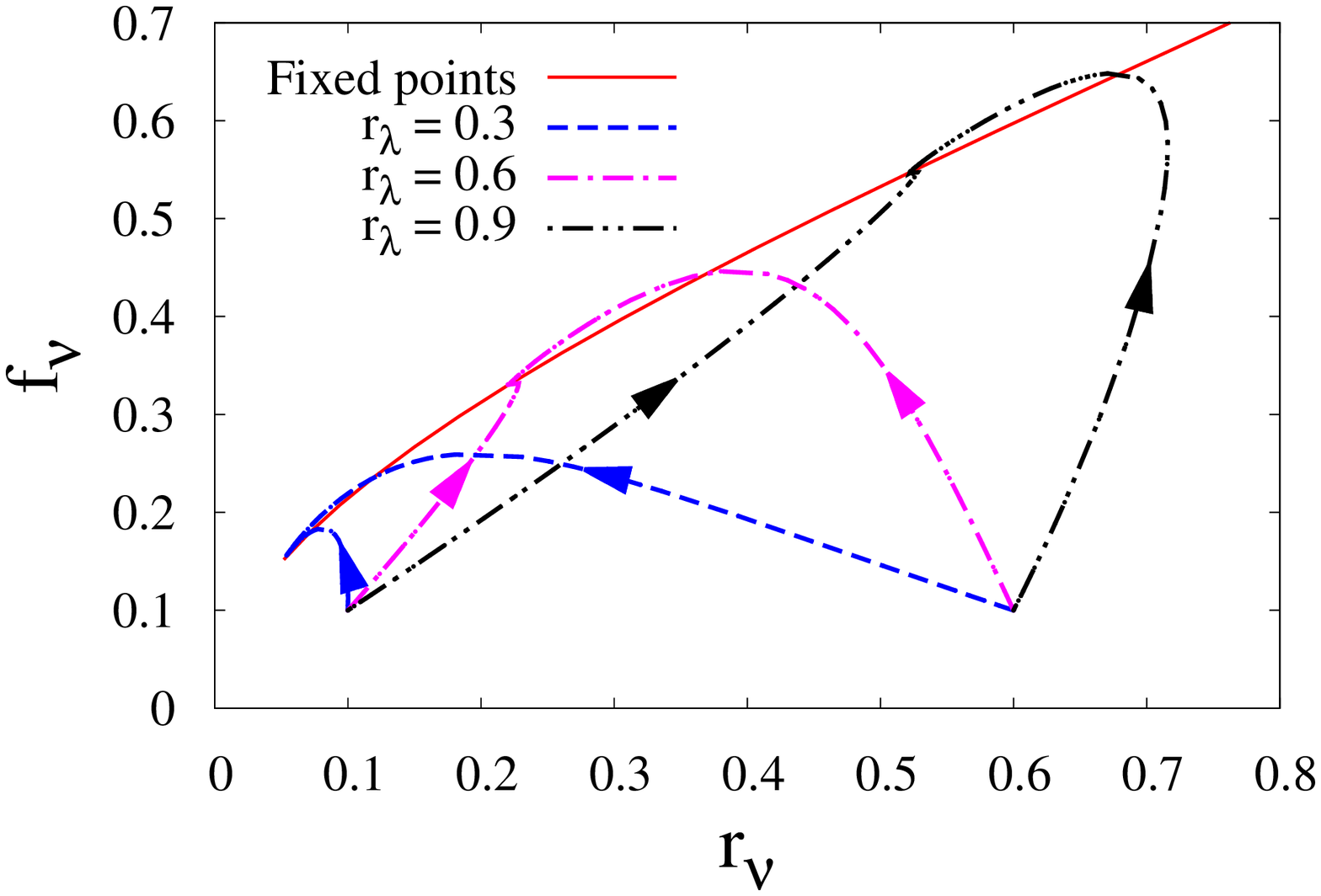}
\end{minipage}
\quad
\begin{minipage}[b]{0.425\textwidth}
 \includegraphics[width=\textwidth]{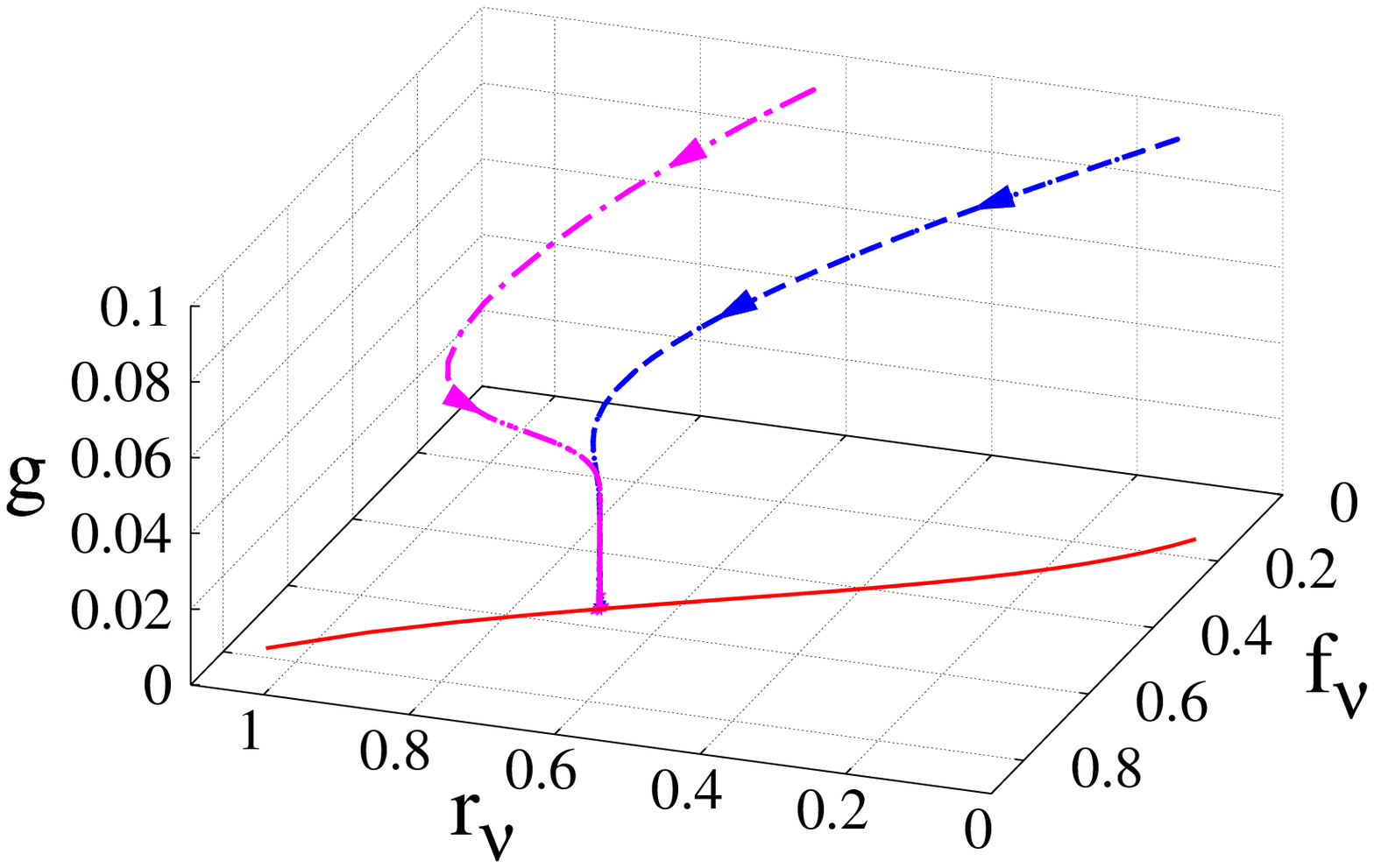}
\end{minipage}
\caption{ (Color online). Numerical integration of the DRG flow for the gHK equation \eqref{chk} in the case $\zeta = 1$ [Eqs. \eqref{flux_fnu}-\eqref{flux_g}]. Left panel: Projection on the $(r_{\nu},f_{\nu})$ plane for two initial conditions, $(r_\nu, f_\nu, g) = (0.1, 0.1,0.1), (0.6, 0.1,0.1)$, and several values of $r_\lambda$ in each case, as indicated in the legend. The solid red line is the manifold of fixed points, parametrized by $r_\lambda$. Right panel: three-dimensional view of two flow trajectories for $r_\lambda = 1$ and the same two initial conditions as in the left panel. Again, the solid red line is the manifold of fixed points. All units are arbitrary. }
\label{fig_flux_gHK}
\end{figure}

\subsubsection{Full generalized HK equation}

In the case of the full gHK equation, it is clear from Eq.\ \eqref{flux_rlam} that if $r_\lambda \neq 0$, then $\zeta=1$ at the RG fixed point, leading to {\em isotropic asymptotic behavior}. In Appendix \ref{app_gHK_iso}, we explicitly provide the three remaining DRG flow equations, Eqs.\ \eqref{flux_fnu}-\eqref{flux_g} in this case. By numerically exploring the parameter space, for $0.3 \leq r_\lambda \leq 1.2$ we have found a non-trivial manifold of fixed points, which can actually be seen as a line parametrized by the value of $r_\lambda$. All points on this manifold share the scaling exponents values
\begin{equation}
\alpha_x = -\frac{1}{3}, \qquad z_x = \frac{4}{3},\qquad \zeta = 1.
\label{az_ghk}
\end{equation}
In Fig.\ \ref{fig_flux_gHK} we show the numerical integration of the DRG flow for $r_{\lambda}$ within this range. Similar considerations can be made as those provided for Fig.\ \ref{fig_flux_HK}.

\begin{figure}[t!]
\centering
\begin{minipage}[t]{0.45\textwidth}
 \includegraphics[width=\textwidth]{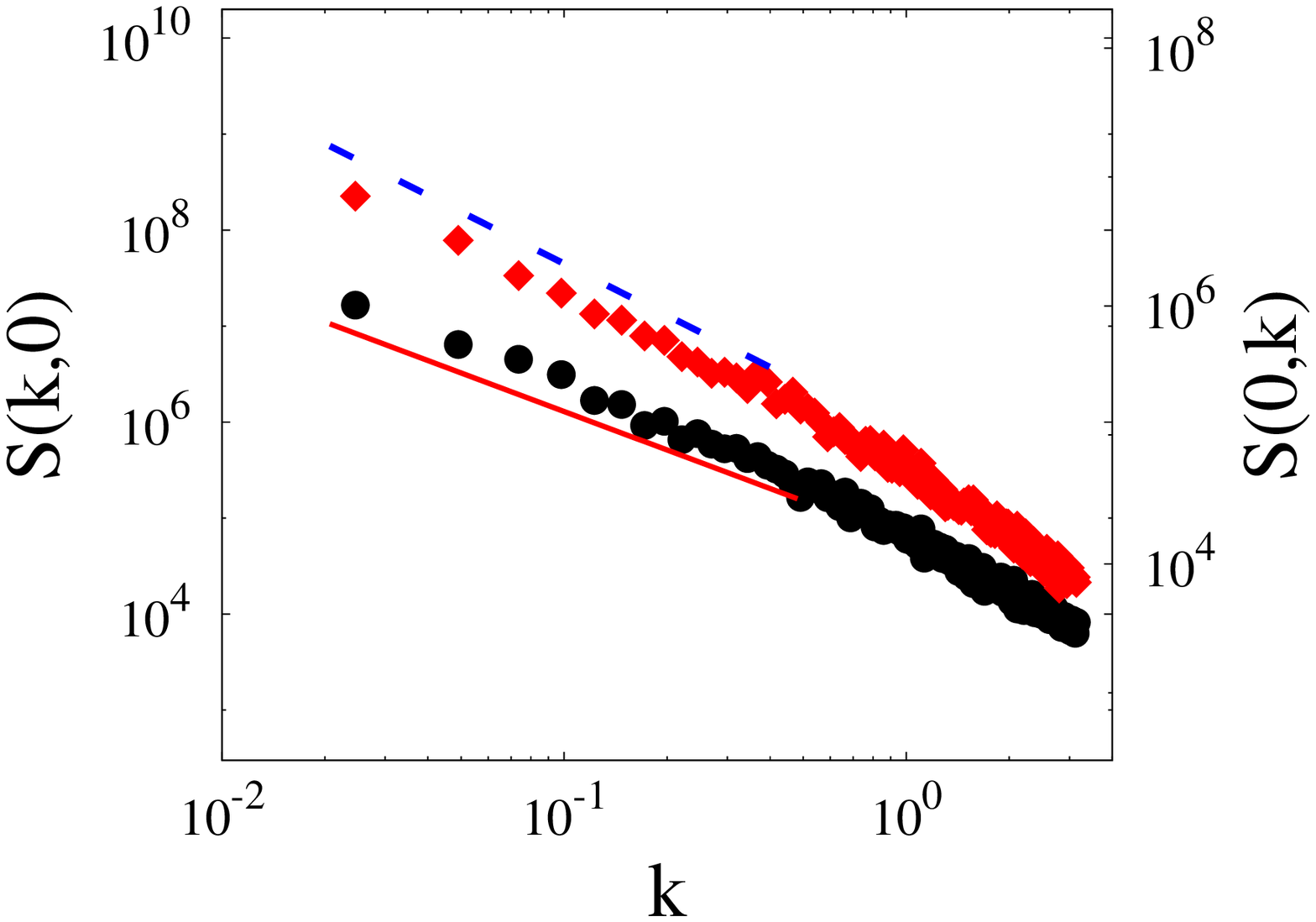}
\end{minipage}
\quad
\begin{minipage}[t]{0.45\textwidth}
\includegraphics[width=\textwidth]{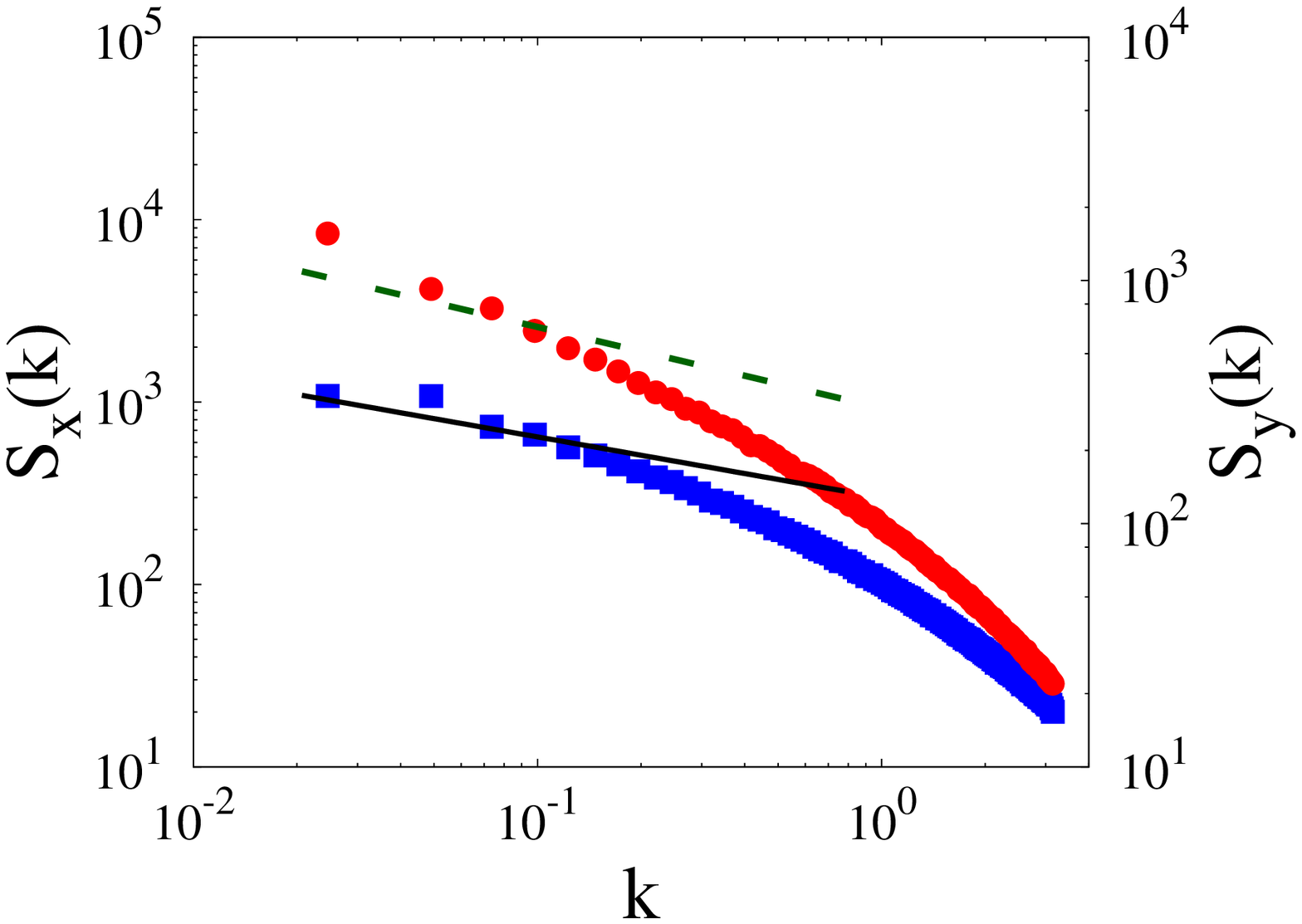}
\end{minipage}
\caption{(Color online) Numerical integrations of the gHK equation, Eq.\ \eqref{chk}, for parameters $\nu_x = \nu_y = 1$, $\nu_{xy}=0$, $D=1$, $\lambda_x=1$, $\lambda_y=-2$ (so that $r_\lambda < 0$), $L=256$, $\Delta x = 1$, and $\Delta t = 0.01$. Left panel: One-dimensional projections $S(k,0)$ (black circles, left axis) and $S(0,k)$
(red diamonds, right axis), averaged over 50 different noise realizations. Both the solid red line and the dashed blue line are guides for the eye with slope $-4/3$. Right panel: PSD of one-dimensional cuts $S_x(k)$ (blue squares, left axis) and $S_y(k)$ (red circles, right axis). Both the solid black line and the dashed green line are guides for the eye with slope $-1/3$. All units are arbitrary.}
\label{fig_psd_gHK_negative}
\end{figure}

\begin{figure}
\centering
\begin{minipage}[t]{0.45\textwidth}
 \includegraphics[width=\textwidth]{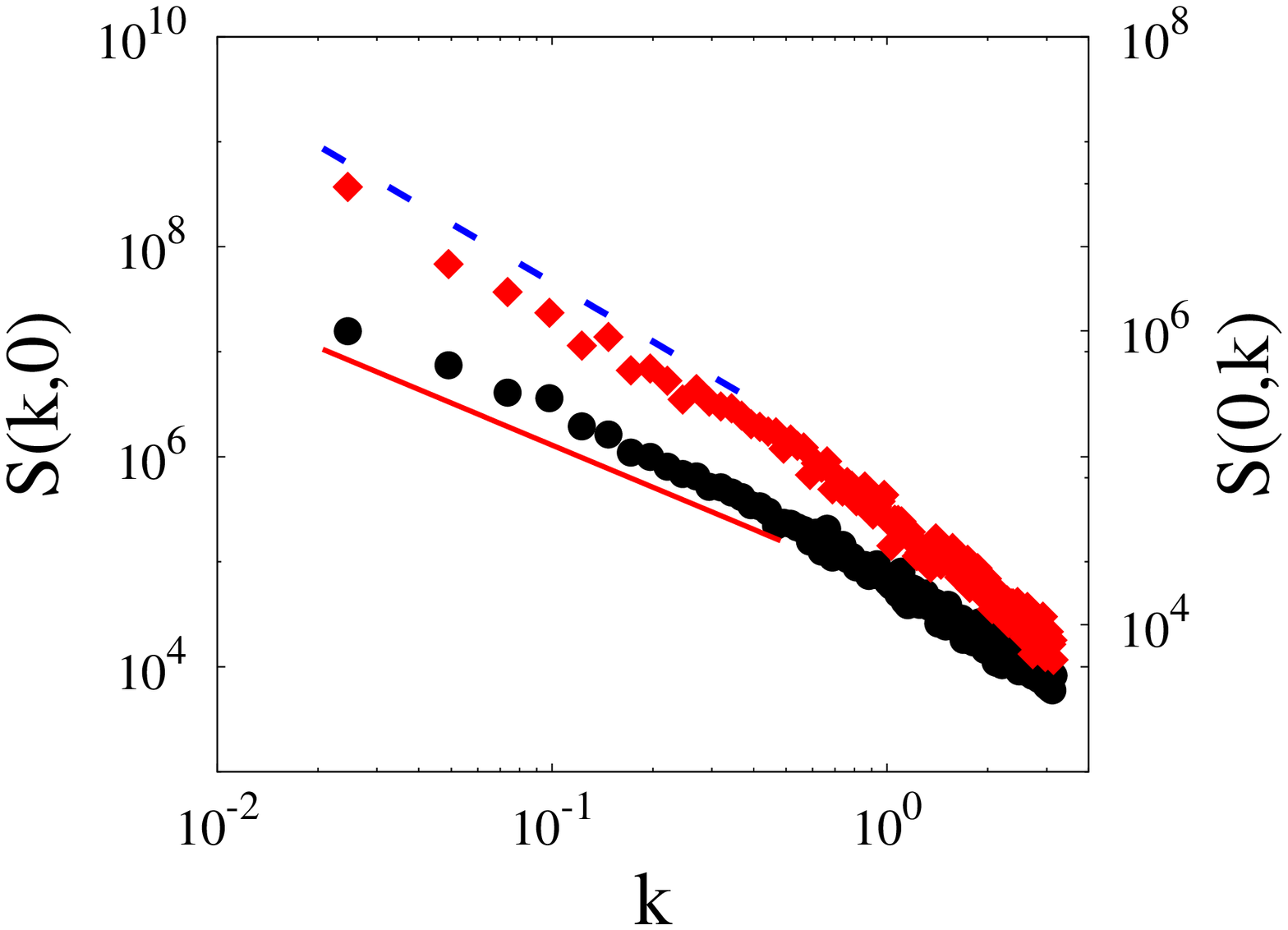}
\end{minipage}
\quad
\begin{minipage}[t]{0.45\textwidth}
\includegraphics[width=\textwidth]{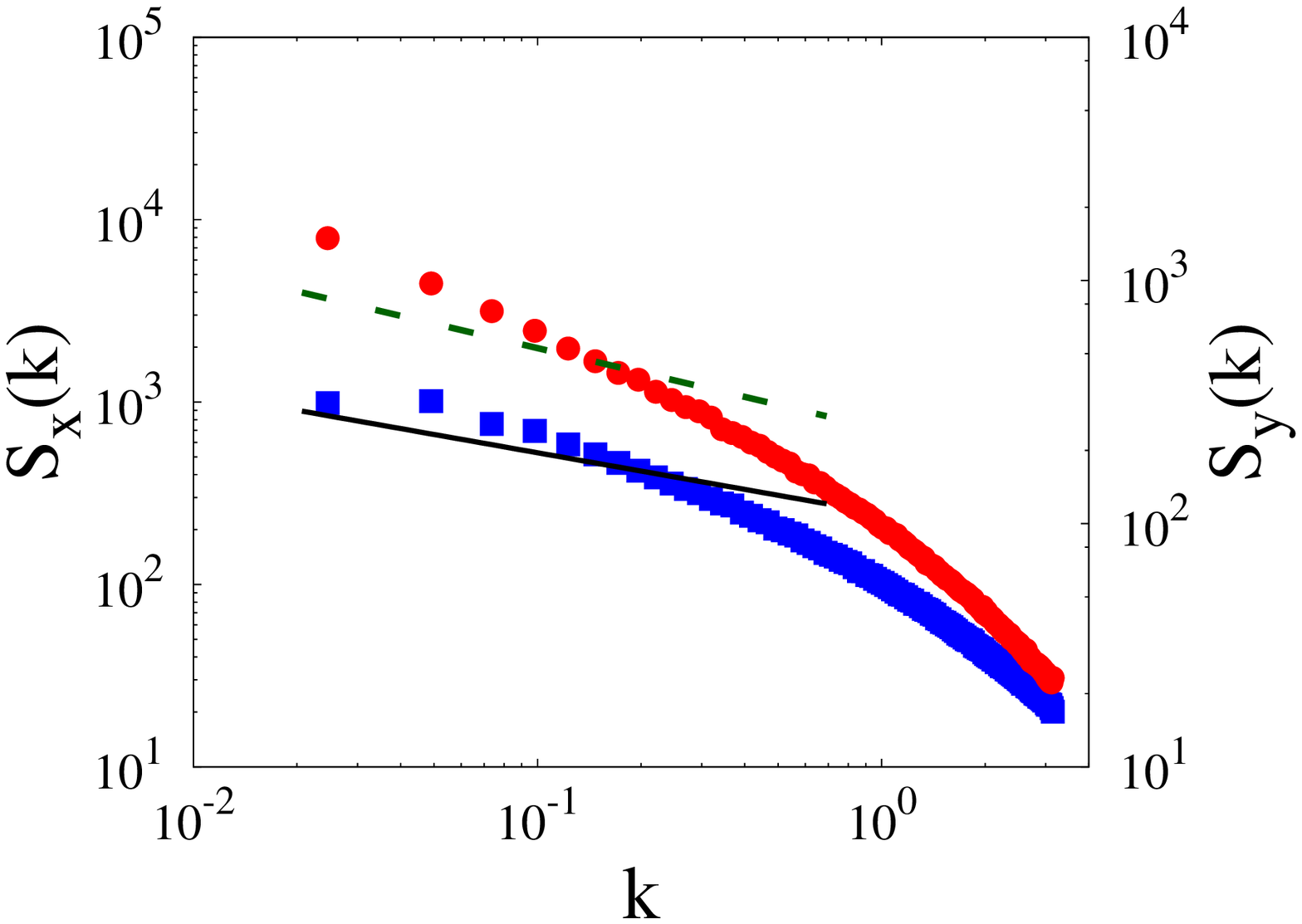}
\end{minipage}
\caption{(Color online) Numerical integrations of the gHK equation, Eq.\ \eqref{chk}, for parameters $\nu_x = \nu_y = 1$, $\nu_{xy}=0$, $D=1$, $\lambda_x=1$, $\lambda_y=2$ (so that $r_\lambda > 0$), $L=256$, $\Delta x = 1$, and $\Delta t = 0.01$. Left panel: One-dimensional projections $S(k,0)$ (black circles, left axis) and $S(0,k)$
(red diamonds, right axis) of the two-dimensional PSD, averaged over 50 different noise realizations. Both the solid red line and the dashed blue line are guides for the eye with slope $-4/3$. Right panel: PSD of one-dimensional cuts $S_x(k)$ (blue squares, left axis) and $S_y(k)$ (red circles, right axis). Both the solid black line and the dashed green line are guides for the eye with slope $-1/3$. All units are arbitrary.}
 \label{fig_psd_gHK_positive}
\end{figure}

However, we have not been able to find a similar set of fixed points for other values of $r_{\lambda}$. Due to the strong non-linear character of the equations that one needs to solve (see Appendix \ref{app_gHK}), it is uncertain whether this is due to lack of convergence of our numerical scheme or to an artifact of the approximations made within our DRG approach. Nevertheless, one would expect such fixed points to also exist and correspond to exponent values as given in Eq.\ \eqref{az_ghk}. In order to verify this conjecture,
%
%
%
we have performed direct numerical simulations of the full gHK equation using the same pseudo-spectral scheme as above. We have paid particular attention to a potential change of scaling behavior due to a relative change in the signs of the nonlinearities $\lambda_x$ and $\lambda_y$. As is clear from the left panels of Figs.\ \ref{fig_psd_gHK_negative} and \ref{fig_psd_gHK_positive}, in the hydrodynamic limit the equation displays the expected isotropic exponent values, Eq.\ \eqref{az_ghk}, irrespective of such a relative sign, analogous in this sense to the caKPZ equation.

However, as in the HK equation proper, a negative value of the roughness exponent induces slow convergence in observables such as the 1D PSD functions of 1D transverse cuts of the two-dimensional interface, to the extent that agreement with the behavior implied by Eq.\ \eqref{az_ghk} is much worse than for the 2D PSD function. Thus, the right panels of  Figs.\ \ref{fig_psd_gHK_negative} and \ref{fig_psd_gHK_positive} indeed show strong finite-size effects for $S_{x,y}(k_{x,y})$, in analogy to the case of the HK equation, analyzed in \cite{us}. Probably related with the value of $\lambda_y$, which is larger in absolute value than $\lambda_x$ in both cases, it is worth mentioning that such a lack of convergence seems more pronounced for the PSD of cuts along the $y$ direction than for cuts along the $x$ direction.

\section{Non-conserved dynamics}


After the previous results, it is natural to ponder whether strongly anisotropic behavior can actually occur for GSI systems with non-conserved dynamics. The prime representative of them is the aKPZ equation \cite{Wolf}, namely,
\begin{equation}
\partial_t h = \nu_x \partial_x^2 h + \nu_y \partial_y^2 h + \frac{\lambda_x}{2} (\partial_x h)^2
+ \frac{\lambda_y}{2} (\partial_y h)^2 + \eta .
 \label{eq:akpz}
\end{equation}
This equation was studied in detail in the seminal paper \cite{Wolf}. The main result was that the scaling behavior is always isotropic, changing from linear Edwards-Wilkinson (EW) type to non-linear KPZ type as a function of the nonlinearities having opposite or the same signs, respectively. However, the case in which only one of the nonlinearities is zero remained basically unexplored. Our results above suggest that it might lead to SA behavior, and for this reason we
will revisit the DRG analysis in \cite{Wolf}, complementing it with direct numerical simulations of the equation. Moreover, while detailed calculations are available for the caKPZ system \cite{kallabis_thesis}, this is not the case of Eq.\ \eqref{eq:akpz}. For this reason we provide details on our analysis in Appendix \ref{app_aKPZ}.

\subsection{DRG analysis of the anisotropic KPZ equation}

In the case of Eq.\ \eqref{eq:akpz}, the DRG flow equations read in general
\begin{align}
& \frac{d \nu_x}{dl} = \nu_x (z_x -2 - \Sigma_{\nu_x}), &
	\frac{d \nu_y}{dl} =  \nu_y (z_x - 2 \zeta - \Sigma_{\nu_y}), \label{akpz_nus} \\[5pt]
&\frac{d \lambda_x}{dl} = \lambda_x (z_x+\alpha_x -2  ),  &
	\frac{d \lambda_y}{dl} = \lambda_y (z_x+\alpha_x -2\zeta), \label{akpz_ls} \\[5pt]
& \frac{d D}{dl} = D(z_x - 2\alpha_x - \zeta -1 + \Phi_D) , \label{akpz_D} &
\end{align}
where functions $\Sigma_{\nu_x}$, $\Sigma_{\nu_y}$, and $\Phi_D$ are reported in Appendix \ref{app_aKPZ}, together with the main steps in their calculation. From Eq.\ \eqref{akpz_ls} we immediately see that, if both nonlinearities are non-zero, $\lambda_x\lambda_y\neq 0$, a fixed point can be attained only for a weakly anisotropic system, that is $\zeta
=1$. By introducing the following couplings,
\begin{equation}
r_{\nu} = \frac{\nu_y}{\nu_x}, \qquad
r_{\lambda} = \frac{\lambda_y}{\lambda_x}, \qquad
g = \frac{ \lambda_x^2 D}{32\pi^2\nu_x^3},
\end{equation}
we again obtain a renormalization flow that is independent of $\alpha_z$ and $z_x$, specifically,
\begin{eqnarray}
&&\dfrac{d r_{\nu}}{d l} = 2r_\nu (1-\zeta ) \left( 1 -  g \left[ \left(\frac{3
+ r_{\nu}}{(1+r_{\nu})^2} -
A_{\zeta, r_{\nu}}\right)
 + \dfrac{r_\lambda}{r_\nu} \left[ \frac{4}{1+r_{\nu}} +
 	\frac{r_{\lambda}}{r_{\nu}} \left( A_{\zeta, r_{\nu}} + \frac{1+
3r_{\nu}}{(1+r_{\nu})^2} \right)\right]  \right] \right),
		\label{rnu_akpz} \\[5pt]
&&\frac{d r_{\lambda}}{dl} = 2 r_{\lambda}(1 -  \zeta), \label{rl_akpz} \\[5pt]
&&\frac{d g}{d l} = g (1-\zeta)\left(1 -  g  \left[\frac{13
+3r_{\nu}}{(1+r_{\nu})^2} -3A_{\zeta, r_{\nu}}
	+ \frac{r_{\lambda}}{r_{\nu}}\left[8\left( A_{\zeta, r_{\nu}} +
\frac{2r_{\nu}+1}{(1+r_{\nu})^2} \right) +
\frac{r_{\lambda}}{r_{\nu}} \left( 3A_{\zeta, r_{\nu}} + \frac{3+
5r_{\nu}}{(1+r_{\nu})^2} \right)\right] \right] \right),
	\label{g_akpz}
\end{eqnarray}
where we have introduced the auxiliary function (proportional to the one defined in \cite{Wolf})
\begin{equation}
A_{\zeta, r_{\nu}} = \dfrac{\tan^{-1}\left(\sqrt{r_\nu}\right) + \zeta
\tan^{-1}\left(\sqrt{1/r_\nu}\right)}{(\zeta-1) \sqrt{r_\nu}}.
\end{equation}

\begin{figure}[t!]
\begin{center}
\centering
\includegraphics[width=0.5\textwidth]{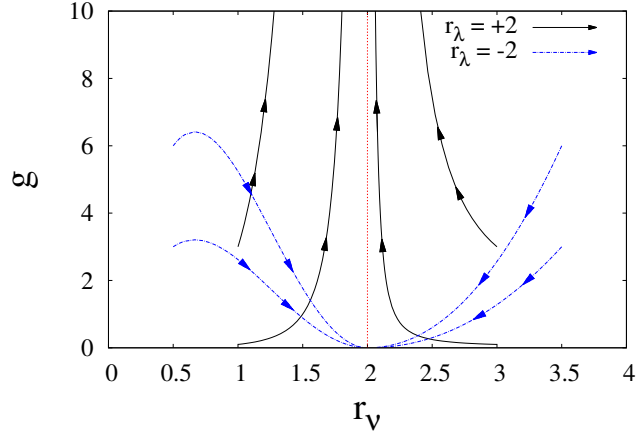}
\caption{(Color online). Numerical intergation of the DRG Flow for the aKPZ equation \eqref{eq:akpz} for the case
$\zeta = 1$ [Eqs.\ \eqref{r_1}-\eqref{g_1}]. The solid black lines correspond to the case of positive $r_\lambda$, while the dashed blue lines correspond to $r_\lambda < 0$. }
\label{fig_wa_aKPZ}
\end{center}
\end{figure}

Without loss of generality, we can consider only the cases $r_\lambda\neq 0 $ and $r_\lambda = 0$ (a zero $\lambda_x$,
i.e.\ $r_\lambda = \infty$, is not taken into account due to the symmetry of the aKPZ equation with respect to
an exchange in the spatial coordinates $x \leftrightarrow y$). For $r_\lambda\neq 0 $, the fixed points of the set of
Eqs.\ \eqref{rnu_akpz}-\eqref{g_akpz} must satisfy the condition $\zeta = 1$ (weak anisotropy), the only terms
different from zero being those proportional to $(1-\zeta)A_{1, r_{\nu}} = - \pi/2\sqrt{r_\nu}$, so that the
non-trivial part of the flow \eqref{rnu_akpz}-\eqref{g_akpz} becomes
\begin{eqnarray}
&&\dfrac{d r_{\nu}}{d l} = g \pi \sqrt{r_\nu}
\left[\left(\dfrac{r_\lambda}{r_\nu}\right)^2 -1 \right], \label{r_1}\\[5pt]
&&\dfrac{d g}{d l} = 4\pi \dfrac{g^2}{\sqrt{r_\nu}} \left(
\frac{r_{\lambda}}{r_{\nu}} +
\dfrac{3}{8}\left[\left(\frac{r_{\lambda}}{r_{\nu}}\right)^2 - 1\right]
\right).
	\label{g_1}
\end{eqnarray}
The fixed points of this set of equations need to belong to the manifold
$(r_\nu^*,r_\lambda^*,g^*,\zeta) = (r_\nu,r_\lambda,0,1)$ with $r_\nu >0$.
Beyond the trivial solution $(0,r_\lambda,0,1)$ (see below), which corresponds to EW behavior,
two submanifolds of $(r_\nu,r_\lambda,0,1)$ provide non-trivial fixed points.
Indeed, by equating \eqref{r_1} to zero we have $r_\lambda = \pm r_\nu$,
and fixed points $(r_\nu,\pm r_\nu,0,1)$, while from Eq.\  \eqref{g_1} we
obtain the two solutions $r_\lambda = r_\nu/3, -3r_\nu$, i.e.\ $(r_\nu, r_\nu/3,0,1)$  and
$(r_\nu, -3 r_\nu,0,1)$. Nevertheless, difficulties arise when we try to compute the stability of these points.
In fact, the stability matrix has three elements equal to zero for $g=0$. Even though a more refined
analysis is possible, the RG flow can be more conveniently studied through numerical integration of Eqs.\ \eqref{r_1}
and \eqref{g_1}. In Fig.\ \ref{fig_wa_aKPZ} we show results of such a study.
If we take a bare parameter condition such that $r_{\lambda} < 0$ (dashed blue lines), the flow is attracted by the
fixed point at the origin, namely, scaling behavior is WA and linear, scaling exponents being those of the EW equation
in 2+1 dimensions, namely, $\alpha_x=\alpha_y=0$ $({\rm log})$ and $z_x=z_y = 2$ \cite{Barabasi}. In contrast, for bare parameter choices such that $r_{\lambda} > 0$ (solid black lines in Fig.\ \ref{fig_wa_aKPZ}) the RG flow lines move towards unbounded values for $g$. This is a manifestation of the occurrence of WA non-linear KPZ scaling, which is well-known not to lead to a finite fixed point in 2+1 dimensions \cite{Barabasi}. Thus, as expected, Wolf's results are recovered through the numerical integration of the RG flow.

Once the well-known results for the aKPZ equation have been retrieved, we focus next in the case of strong anisotropy $\zeta \neq 1$. From Eq.\ \eqref{rl_akpz} we immediately obtain  $r_\lambda = 0$ at the fixed points, so that the DRG flow equations reduce to
\begin{eqnarray}
&&\dfrac{d r_{\nu}}{d l} = 2r_\nu (1-\zeta ) \left[ 1 -  g  \left(\frac{3 +
r_{\nu}}{(1+r_{\nu})^2} -
A_{\zeta, r_{\nu}}\right) \right], \label{flux_akpz_rlambda0_rnu} \\[5pt]
&&\frac{d g}{d l} = g (1-\zeta)\left[1 -  g  \left(\frac{13
+3r_{\nu}}{(1+r_{\nu})^2} -3A_{\zeta, r_{\nu}}\right)  \right].
\label{flux_akpz_rlambda0_g}
\end{eqnarray}
In order to find the fixed points for Eqs.\ \eqref{flux_akpz_rlambda0_rnu}-\eqref{flux_akpz_rlambda0_g}, we need to set
their right hand sides to zero. This gives us several possible solutions, which we proceed to analyze.
Considering Eq.\ \eqref{flux_akpz_rlambda0_rnu}, zeros are obtained setting $r_\nu=r^*_\nu = 0$ or $\displaystyle g=g_1^* = \left[(3 + r_{\nu})(1+r_{\nu})^{-2} - A_{\zeta, r_{\nu}}\right]^{-1}$. However:
 \begin{itemize}
 \item[(a)] When $r_\nu = 0$, Eq. \eqref{flux_akpz_rlambda0_g} cannot be set to zero. This is due to the fact that one of the terms within the equation, namely, $g^2 \left[\tan^{-1}\left(r_\nu^{1/2}\right) + \zeta
 \tan^{-1}\left(r_\nu^{-1/2}\right)\right] r_\nu^{-1/2}$ does not have a well defined limit as a two-variable function for $(r_\nu, g) \rightarrow (0,0)$.
 \item[(b)] Substituting $g_1^*$ into Eq.\ \eqref{flux_akpz_rlambda0_g}, we get
 \begin{equation}\label{eq_A}
A_{\zeta,r_\nu} = \dfrac{5+r_\nu}{(1+r_\nu)^2},
\end{equation}
which gives us the possible values of $r_\nu$ and $\zeta$ corresponding to $g_1^*$. However, it is not difficult to see that Eq.\ \eqref{eq_A} leads to $g_1^* = - (1+r_\nu)^2/2 < 0$, which is not a physically acceptable value.
\end{itemize}
Hence, the formal zeros of Eq.\ \eqref{flux_akpz_rlambda0_rnu} provided by $r_\nu=0$ and $g=g_1^*$ are both to be discarded. On the other hand, if we start out with Eq.\ \eqref{flux_akpz_rlambda0_g}, we obtain zeros for $g=0$ and $\displaystyle g=g_2^* = \left[(13 +3r_{\nu})(1+r_{\nu})^{-2} -3A_{\zeta, r_{\nu}}\right]^{-1}$. Then:
\begin{itemize}
\item[(a)] By substituting $g=0$ into Eq.\ \eqref{flux_akpz_rlambda0_rnu} we get
\begin{equation}
 \dfrac{d r_{\nu}}{d l} = 2r_\nu (1-\zeta ),
\end{equation}
which implies that the line $g=0$, $r_\nu > 0$ is a separatrix for the RG flow. One could then argue that the point $g=0$, $r_\nu=0$ is indeed a fixed point, but strictly speaking this is not true due to the ill-definiteness of the flow at the origin, as discussed above.
\item[(b)] If we substitute $g=g_2^*$ into Eq.\ \eqref{flux_akpz_rlambda0_rnu}, we again obtain Eq.\ \eqref{eq_A}, and therefore no physically acceptable fixed points.
\end{itemize}
\begin{figure}[t!]
\centering
\begin{minipage}[t]{0.45\textwidth}
 \includegraphics[width=\textwidth]{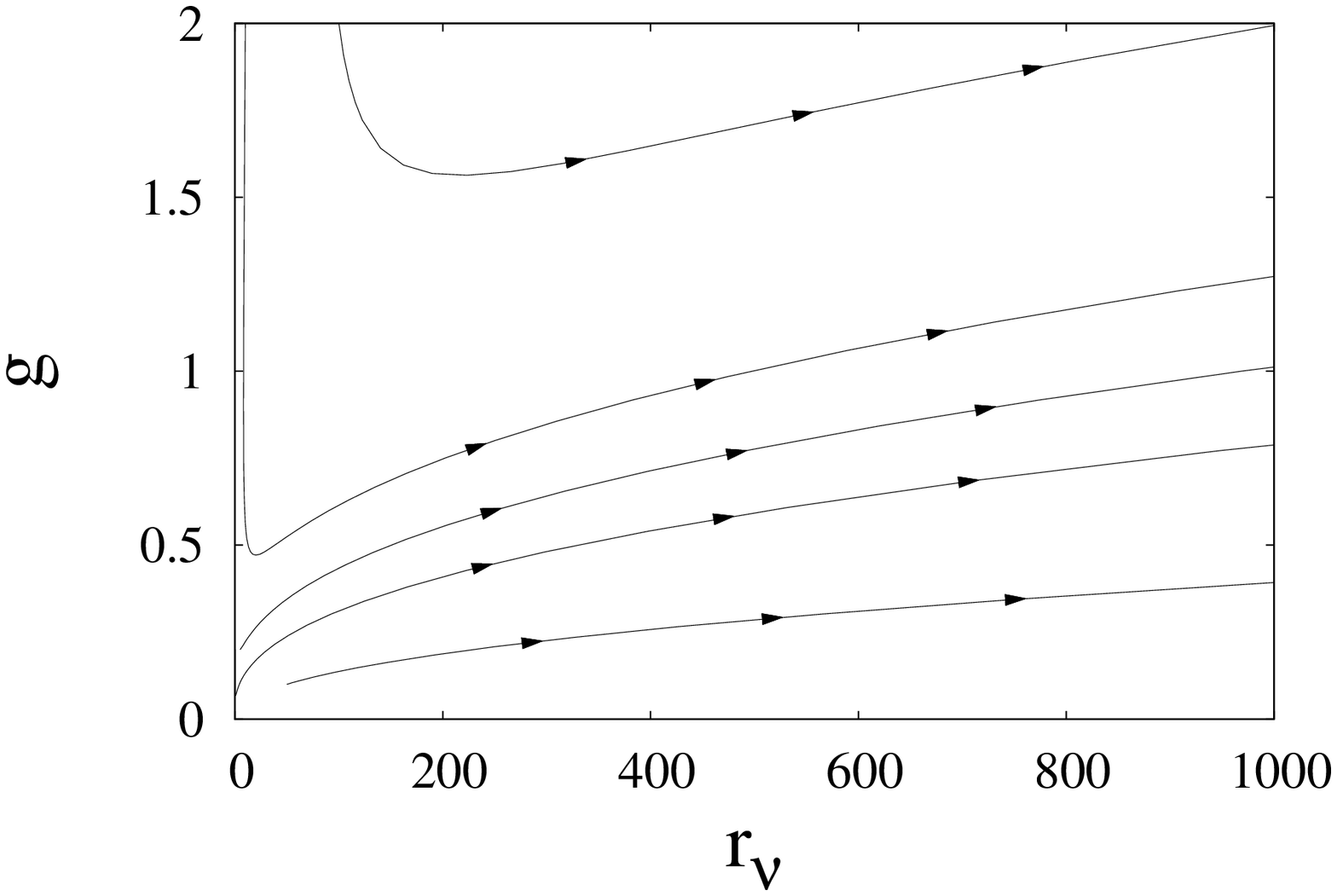}
\end{minipage}
\quad
\begin{minipage}[t]{0.45\textwidth}
 \includegraphics[width=\textwidth]{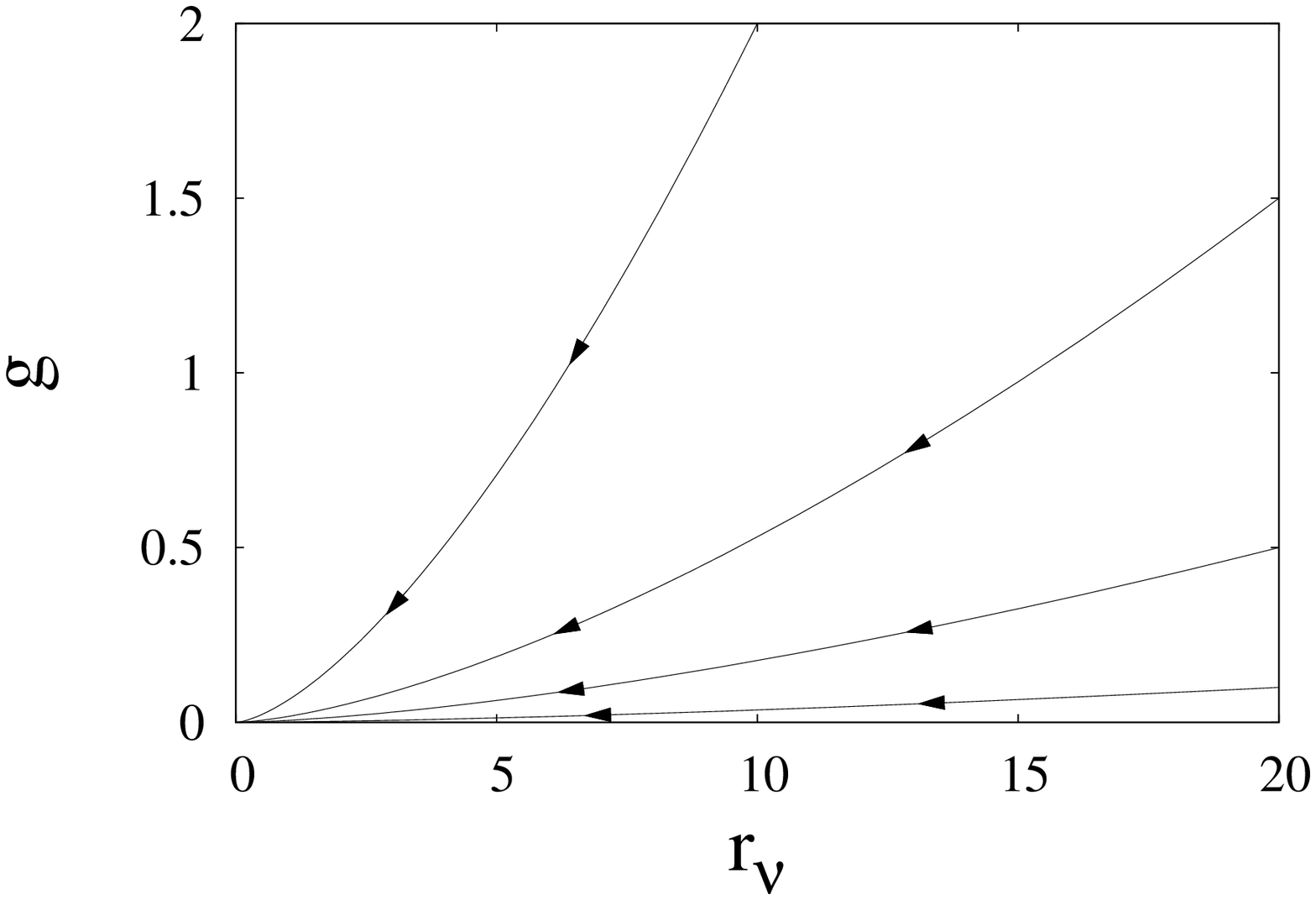}
\end{minipage}
\\
\begin{minipage}[t]{0.45\textwidth}
\includegraphics[width=\textwidth]{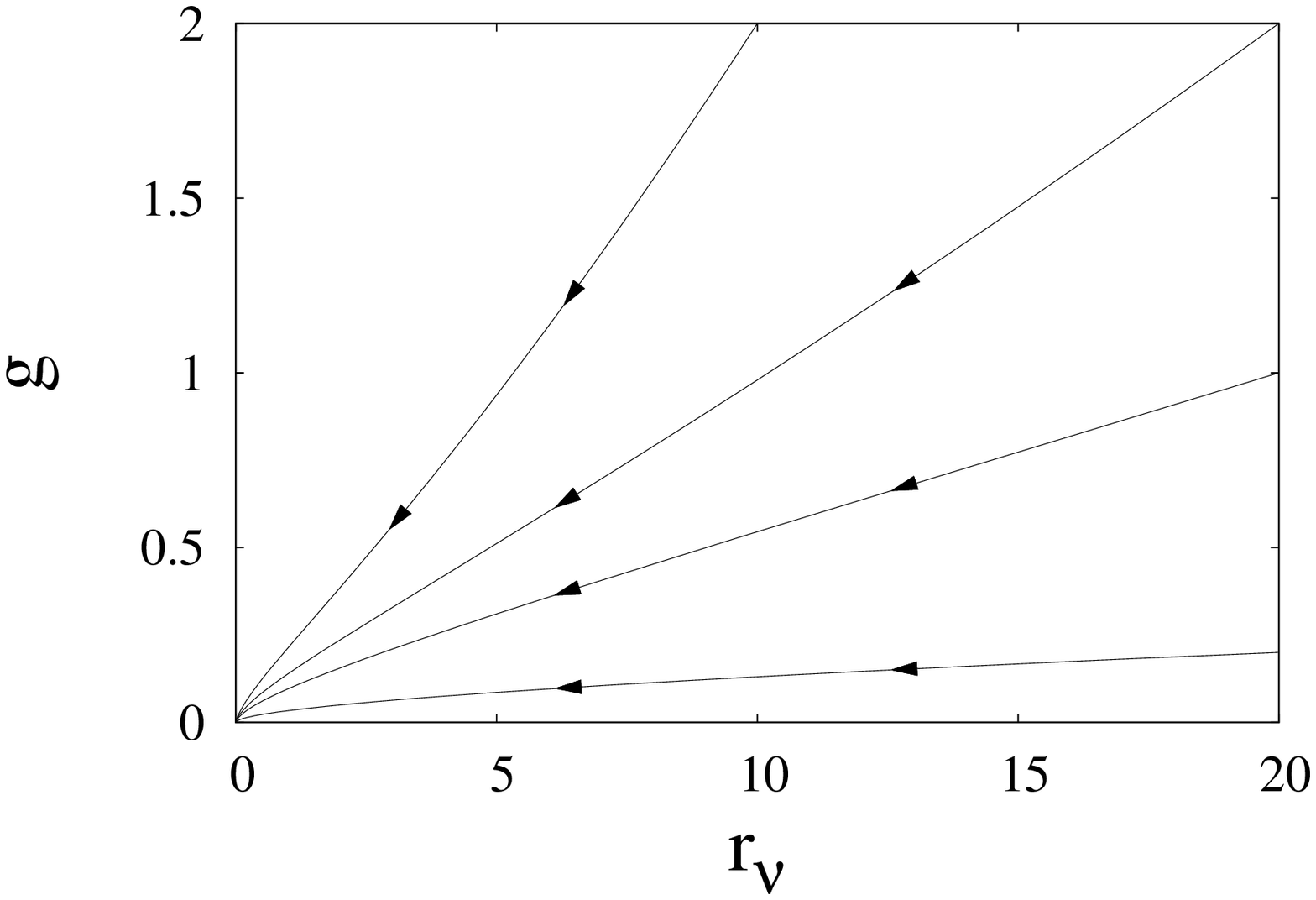}
\end{minipage}
\quad
\begin{minipage}[t]{0.45\textwidth}
\includegraphics[width=\textwidth]{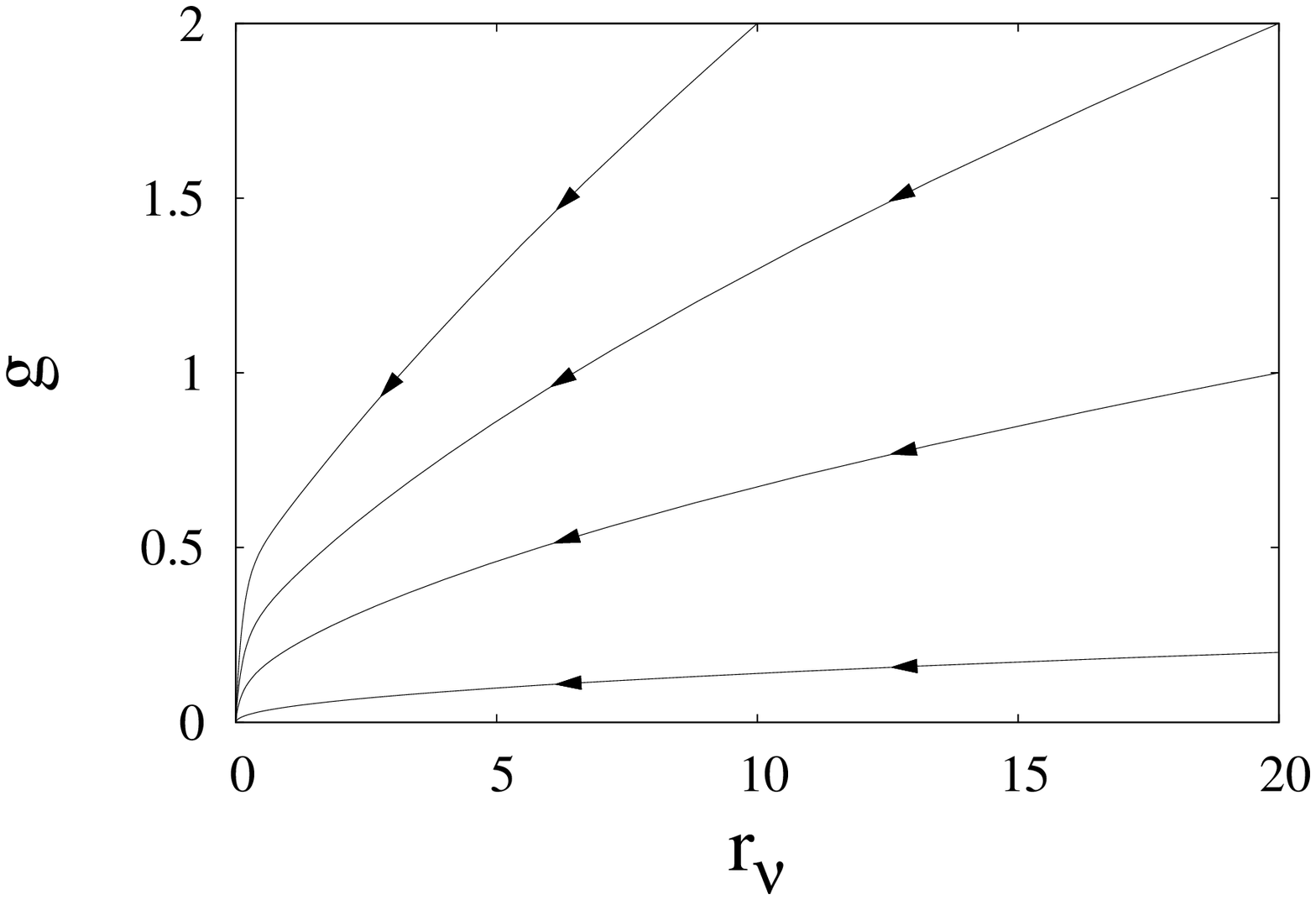}
\end{minipage}
\caption{ (Color online). Numerical integration of the DRG flow for the aKPZ equation \eqref{eq:akpz} for the case $r_\lambda = 0$ [Eqs.\ \eqref{flux_akpz_rlambda0_rnu}-\eqref{flux_akpz_rlambda0_g}], and different values of the anisotropy exponent, $\zeta = 1/2$ (upper left panel), $\zeta=1$  (upper right panel), $\zeta = 3/2$ (bottom left panel), and $\zeta = 5$ (bottom right panel). All units are arbitrary.}
\label{fig_aKPZ_rlambda0}
\end{figure}
A final possibility to find a meaningful fixed point of the flow is to set $\zeta = 1$, which would correspond to isotropic asymptotic behavior. Eqs.\ \eqref{flux_akpz_rlambda0_rnu}-\eqref{flux_akpz_rlambda0_g} then become
\begin{eqnarray}
\label{flux_akpz_rlambda0_rnu_zeta1}
&&\dfrac{d r_{\nu}}{d l} = - \pi g \sqrt{r_\nu},  \\[5pt]
&&\frac{d g}{d l} = -\frac{3 \pi}{2} \frac{g^2}{\sqrt{r_\nu}}.
\label{flux_akpz_rlambda0_g_zeta1}
\end{eqnarray}
As it turns out, these equations can be exactly solved, giving
\begin{eqnarray}
 && r_\nu(l) = \dfrac{r_0^{3/2}}{\sqrt{r_0} + \pi g_0 l}, \label{exact_r} \\[5pt]
 && g(l) = \dfrac{g_0 r_0^{3/4}}{(\sqrt{r_0} + \pi g_0 l)^{3/2}}, \label{exact_g}
\end{eqnarray}
where $r_0 = r_\nu(0)$ and $g_0 = g(0)$ are the initial conditions (bare parameter values). This exact solution tells us that the flow moves towards the point $r_\nu=0$, $g=0$, which is only reached in infinite ``time'', i.e.\ for $l \rightarrow \infty$. Moreover, Eqs.\ \eqref{exact_r}-\eqref{exact_g} can be simply restated as
\begin{equation}
 g(l) = g_0\,\left[\frac{r_\nu(l)}{r_0}\right]^{3/2},
\end{equation}
implying that $g$ vanishes faster than $r_\nu$ in this limit.

\begin{figure}[t!]
\centering
\begin{minipage}[t]{0.45\textwidth}
 \includegraphics[width=\textwidth]{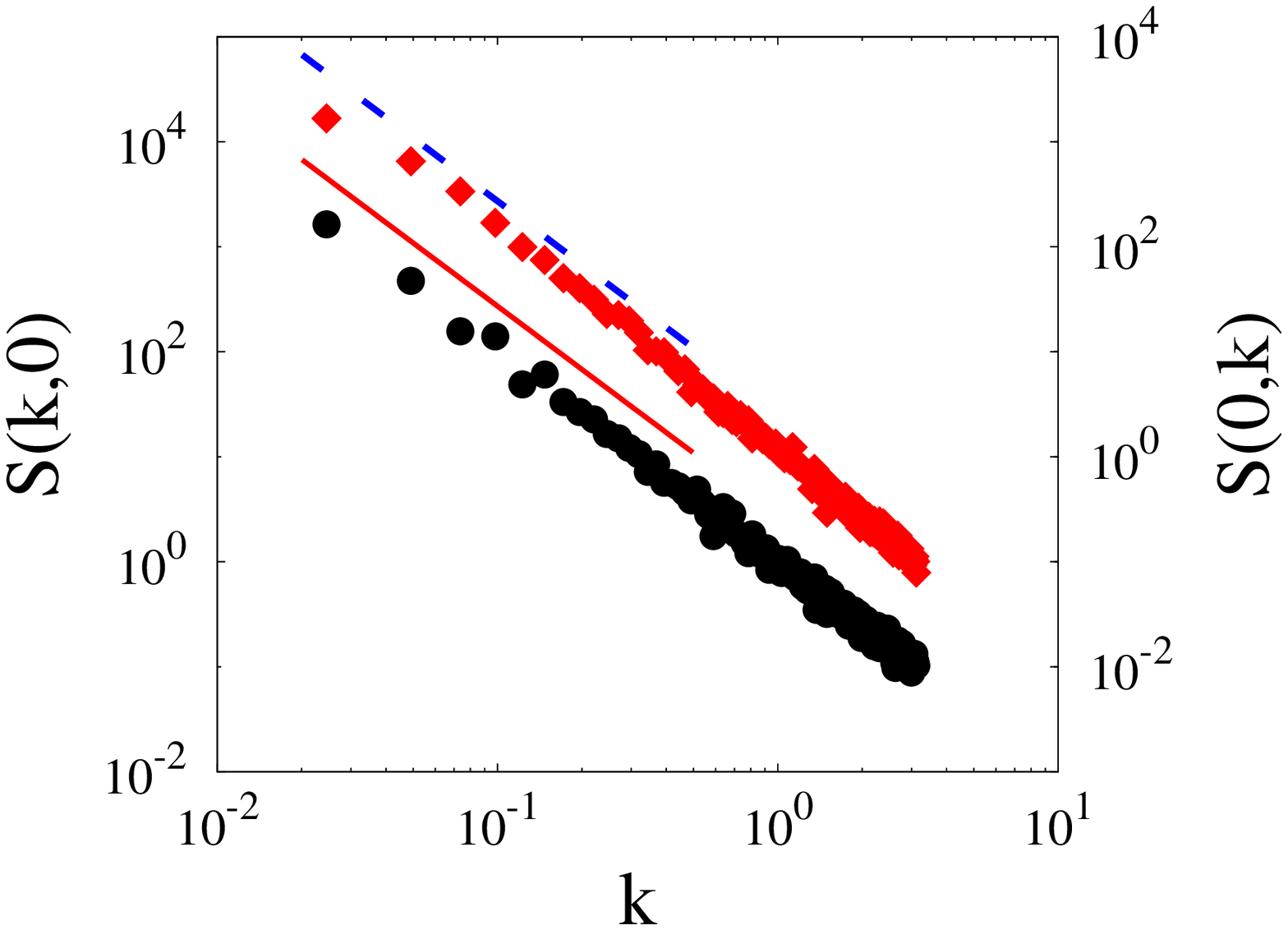}
\end{minipage}
\quad
\begin{minipage}[t]{0.45\textwidth}
\includegraphics[width=\textwidth]{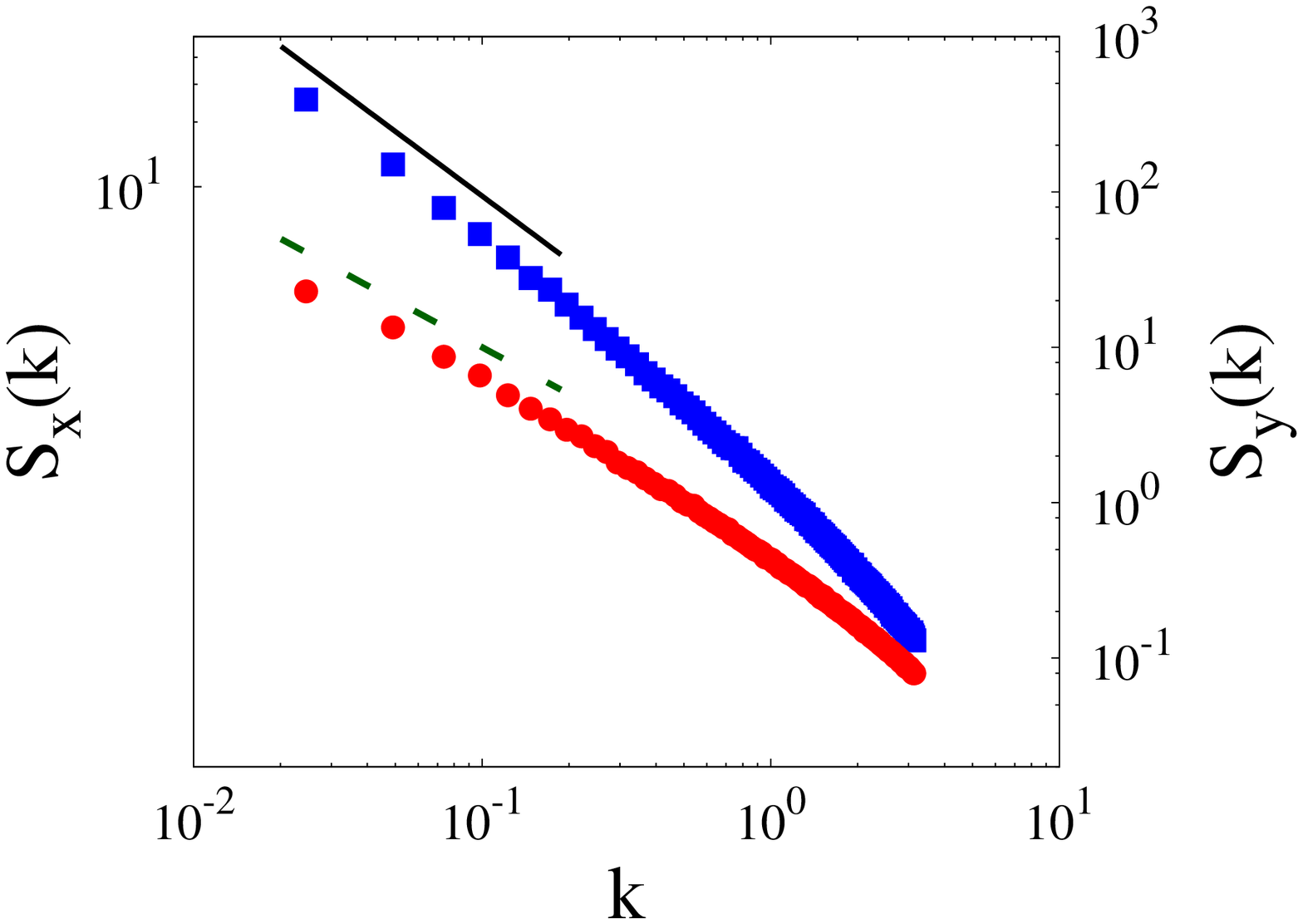}
\end{minipage}
\caption{(Color online). Numerical integrations of the aKPZ equation, Eq.\ \eqref{eq:akpz}, for parameters $\nu_x = \nu_y = 1$, $D=1$, $\lambda_x=3$, $\lambda_y=0$, $L=256$, $\Delta x = 1$, and $\Delta t = 0.01$.
Left panel: One-dimensional projections $S(k,0)$ (black circles, left axis) and $S(0,k)$
(red diamonds, right axis), averaged over 50 different noise realizations. Both the solid red line and the dashed blue line are guides for the eye with slope $-2$. Right panel: PSD of one-dimensional cuts along the $x$ direction, $S_x(k)$ (blue squares, left axis), and along the $y$ direction, $S_y(k)$ (red circles, right axis). Both the solid black line and the dashed green line are guides to the eye with slope $-1$. All units are arbitrary. }
\label{fig_psd_aKPZ}
\end{figure}

The latter result actually allows us to rationalize the behavior of the RG flow, Eqs.\ \eqref{rnu_akpz}-\eqref{g_akpz}, for the anisotropic condition $\zeta\neq 1$, $r_{\lambda}=0$, as obtained through numerical integration of the corresponding
Eqs.\ \eqref{flux_akpz_rlambda0_rnu}-\eqref{flux_akpz_rlambda0_g}. In Fig.\ \ref{fig_aKPZ_rlambda0}
we show such type of results for different values of $\zeta$. Obviously, $\zeta=1$ (see upper right panel) constitutes a natural reference case for which, as we have just seen, the RG flow is both well-defined at, and attracted by, the origin, where scaling behavior is isotropic, EW-type. Even though this point cannot be reached by the RG flow at finite $l$ for other values of $\zeta$, for which no finite fixed points otherwise exist, it still plays an important role. Thus, as can be seen in  Fig.\ \ref{fig_aKPZ_rlambda0} (upper left panel), for $\zeta < 1$ the origin seems to repel the flow lines, which evolve towards arbitrarily large values of $(r_{\nu},g)$. This behavior may be an artifact of the approximations made in the DRG analysis, as suggested by further results. Namely, $\zeta > 1$ is seen in the bottom panels of Fig.\
\ref{fig_aKPZ_rlambda0} to reverse the stability of the origin.
Now it attracts the RG trajectories, which flow into it for infinite $l$, indicating asymptotic isotropic EW behavior. We have checked that it is the latter behavior, rather than the unbounded growth of $r_{\nu}$ and $g$ obtained for $\zeta < 1$, which seems to actually occur for the aKPZ equation under the present type of conditions. Specifically, we have performed direct numerical simulations of the aKPZ equation, Eq.\ \eqref{eq:akpz}, for a case in which one of the nonlinearities is ``suppressed'', $\lambda_y=0$, see Fig.\ \ref{fig_psd_aKPZ}
\cite{note:num_akpz}. As can be seen in the figures, the behavior of correlation functions is well-reproduced by isotropic EW exponents, namely, $\alpha=0$ (log) and $z=2$. This is consistent with the effective $g$ coupling renormalizing to zero much faster than $r_{\nu}$, so that at large length scales the system is effectively behaving as an anisotropic EW equation, for which the scaling is well-known to be of the WA type \cite{us}.

%

\section{Conclusions and Outlook}

The previous sections have allowed us to assess the non-genericity of strong anisotropy for surfaces displaying GSI
and non-linear effects. Thus, for non-conserved dynamics SA simply does not occur, even for special conditions under
which only one of the nonlinearities is suppressed. On the other hand, for systems with conserved dynamics SA can be obtained, and even whole families of equations can be formulated which display this property, such as Eq.\ \eqref{eq:cakpz_nm}. However, both in the presence and in the absence of the shift symmetry $h \to h + {\rm const.}$,
this seems only possible for ``incomplete'' equations in which only one of the nonlinearities is suppressed.

Overall for the type of systems that we have studied here and in \cite{us}, one can conclude that, if the part of the
interface equation which is most relevant for the scaling behavior (e.g.\ non-linear vs linear terms, or surface
tension vs surface diffusion, etc.) can be rewritten in an isotropic form using coordinate transformations, such as rotations or a mere rescaling (in which rescaling factors are {\em positive}), then the system will display {\em weak anisotropy}. Actually, this is a sufficient condition for weak anisotropy, but is not necessary: One also obtains WA
for example in the aKPZ equation when the coefficients of the nonlinearities have different signs. Note that a
rescaling in such a situation still preserves the difference in sign between the two nonlinear terms.

But in order to obtain strong anisotropy, one further needs conserved dynamics, combined with special parameter cancellations such that, e.g.\ $\lambda_x \neq 0$ while $\lambda_y=0$. In general, conditions of this type depend
critically on details of the dynamics that is being described, acting as special constraints, and are in
this sense non-generic in parameter space. Hence, they cannot be obtained from simple-minded derivations of the
equations of motion based on symmetries and conservation laws.

Naturally, there are formulations of the interface equation, such as the original one by HK, in which this type of
special conditions becomes natural, as imposed by the geometry of the external driving fields and/or relaxation
mechanisms (e.g.\ the direction of sand transport, etc.). Beyond driven diffusive systems or models of self-organized criticality, such type of constraints also appear for instance in solidification fronts \cite{golovin:1998}, the dynamics
of localized structures \cite{toh:1989} in plasmas \cite{kuznetsov:1986} and in fluid propagation \cite{demekhin:2010},
the evolution of driven flux lines in superconductors \cite{hwa:1992}, or the effect of shear on interface
fluctuations \cite{bray:2001_2}. Still, such type of constraints leading to ``incomplete'' equations are not to be expected in many other systems. Consider for instance epitaxial growth systems \cite{Wolf,rost:1995} in which lattice
anisotropies are expected to lead to different values of, say, $\lambda_x$ and $\lambda_y$. In general, the physics
is limited to inducing different values in the equation parameters, but not necessarily to exact cancellations of
specific ones.

Additionally, we have to note an additional (implicit) assumption that we have made in our analysis. This is the fact
that the interface equation is morphologically unstable, in the sense that the deterministic terms tend to smooth out surface inhomogeneities. However, many natural contexts for the occurrence of spatial anisotropies are actually systems
in which patterns emerge (convection rolls, ripples under IBS, etc.), some of which correspond to references just quoted \cite{golovin:1998,toh:1989,kuznetsov:1986,demekhin:2010,rost:1995}. Formation of this type of structures requires
morphological instabilities to occur, which suggests pattern-forming systems as a potential context for non-trivial SA behavior. Note, pattern-forming behavior (i.e.\ the emergence of a spatial structure from an homogeneous system) is to some extent the converse interfacial property to GSI, since the former is characterized by the predominance of a characteristic length scale (namely, the pattern periodicity), which is absent in the latter. Nevertheless, studies are already available \cite{keller:2011} in which a highly non-trivial interplay occurs between instability and anisotropy, and
in which the difference between scale invariance (kinetic roughening, or surface GSI) and its opposite property
(pattern formation) is a matter of space and time scales \cite{karma:1993,nicoli:2010}. The anisotropic Kuramoto-Sivashinsky equation \cite{cuerno:1995,rost:1995,keller:2011} is a natural example, albeit itself being possibly confined to WA.
Thus, we believe an interesting avenue for further studies of the occurrence of SA in GSI systems is related with anisotropic models of pattern formation that are compatible with kinetic roughening at the appropriate scales.

\newpage

\appendix
\section{Dynamic Renormalization Group analysis of the generalized Hwa-Kardar
equation}
\label{app_gHK}

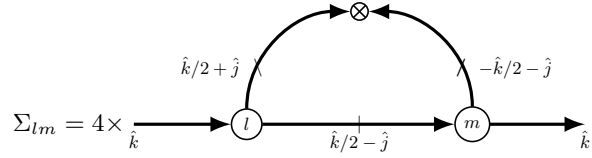
\begin{figure}[t!]
\begin{center}
\scalebox{1.5}{\begin{tikzpicture}
\node [draw = black, shape=circle, scale = 0.5, name = n1] at (1,0) {$l$};
\node [draw = black, shape=circle, scale = 0.5, name = n2] at ($(n1)+(2,0)$) {$m$};
\node [fill = white, shape=circle, draw=black, scale = 0.5, name = n3] at ($(n1) + (1,1)$) {};

\draw (n3.north west)--(n3.south east);
\draw (n3.north east)--(n3.south west);

\node [name=nbar1, scale=0.5] at ($(n1)+(0.1,0.5)$) {$\bf\backslash$};
\node [name=nbar2, scale=0.5] at ($(n1)+(1,0)$) {$\bf |$};
\node [name=nbar3, scale=0.5] at ($(n2)+(-0.1,0.5)$) {\textbf /};

\node [name = text0, scale = 0.7, left] at (0,0) {$\Sigma_{lm} = 4\times$};
\node [name = text1, scale = 0.5, below] at (0,0) {$\hat k$};
\node [name = text2, scale = 0.5, below] at ($(nbar2)$) {$\hat k/2 - \hat j$};
\node [name = text3, scale = 0.5, below] at ($(n2)+(1,0)$) {$\hat k$};
\node [name = text4, scale = 0.5, left] at ($(n1)+(0,0.5)$) {$\hat k/2 + \hat j$};
\node [name = text5, scale = 0.5, right] at ($(n2)+(0,0.5)$) {$-\hat k/2 - \hat j$};

\draw [thick, arrows=-latex] (0,0) -- (n1);
\draw [thick, arrows=-latex] (n1) -- (n2);
\draw [thick, arrows=-latex] (n2) -- ($(n2)+(1,0)$);

\draw [thick, arrows=-latex] (n1) to [out = 90, in = 180] (n3);
\draw [thick, arrows=-latex] (n2) to [out = 90, in = 0] (n3);
\end{tikzpicture}}
\caption{Generic diagrammatic representation of the four different contributions $\Sigma_{xx}$, $\Sigma_{xy}$, $\Sigma_{yx}$, and $\Sigma_{yy}$ to the renormalization of the propagator $G({\bf k},\omega)$ of the gHK and the aKPZ equations. For each equation the exact meaning of the solid lines differ, see Eq.\ \eqref{ghk_sigma_lm} for the former and Eq.\ \eqref{sigma_lm} for the latter.}
\label{fig_prop_akpz}
\end{center}
\end{figure}

The diagrammatic expansion of the integrals that contribute to the renormalization of the bare propagator of the gHK equation is sketched in Fig.\ \ref{fig_prop_akpz}, where we use standard notation for the nonlinearities involved \cite{McComb}. Note, there being two different vertices with couplings $\lambda_{x,y}$, both indices $l,m$ take as values the two spatial variables $x,y$, leading to four different contributions, $\Sigma_{xx}, \Sigma_{xy}, \Sigma_{yx}$, and $\Sigma_{yy}$. After the usual symmetrization  of  the integration variables
$({\bf q},\Omega)\to ({\bf j}+{\bf k}/2,\Omega+\omega/2)$,
we get
\begin{equation}
\Sigma_{lm}({\bf k},\omega) = -2 \lambda_l \lambda_m D \int^>  \dfrac{d{\bf
j}}{(2\pi)^2} \int \dfrac{d\Omega}{2\pi} \,
	 k_l \left(\frac{k_m}{2} - j_m\right) \,
	\left|G_0\left(\frac{\hat k}{2}+\hat j\right)\right|^2
G_0\left(\frac{\hat k}{2}-\hat j\right),
\label{ghk_sigma_lm}
\end{equation}
where $G_0\big(\hat k\big)$ is short-hand notation for the bare propagator
\begin{equation}
G_0\big(\hat k\big) = \left[\nu_x k_x^2 + 2 \nu_{xy} k_x k_y + \nu_y k_y^2 -i
\omega\right]^{-1}.
\end{equation}
An expansion to first order in $k_x/j_x,k_y/j_y\ll 1$ leads to
\begin{eqnarray}
&& \lim_{\omega \rightarrow 0} \left| G_0\left(\hat k / 2+\hat j\right)\right|^2
\sim
	\frac{1}{\Delta^2 + \Omega^2} \left[1 - \frac{2\Delta}{\Delta^2 +
\Omega^2}(\nu_x j_x k_x + \nu_y j_y k_y + \nu_{xy} j_y k_x + \nu_{xy} j_x k_y)
\right],\\[5pt]
&&\lim_{\omega \rightarrow 0}	G_0\left(\hat k / 2-\hat j\right) \sim
\frac{1}{\Delta + i \Omega}
	\left[1 + \frac{1}{\Delta + i\Omega}(\nu_x j_x k_x + \nu_y j_y k_y+
\nu_{xy} j_y k_x + \nu_{xy} j_x k_y) \right].
\end{eqnarray}
where $\Delta = \nu_x j_x^2 + 2 \nu_{xy} j_x j_y + \nu_y j_y^2$. Using these
results in Eq.\ \eqref{ghk_sigma_lm} and after integration
over the frequency variable $\Omega$, we retain terms up to second order in the
components of ${\bf k}$, to get
\begin{equation}
\Sigma_{lm}({\bf k},0) = -\frac{\lambda_l \lambda_m D}{16 \pi^2} \int^{>}
\frac{d {\bf j}}{\Delta^2}
	\left[k_l k_m + \frac{2}{\Delta} k_l j_m \left( \nu_x  k_x  j_x +\nu_y
k_y   j_y + \nu_{xy} j_y k_x + \nu_{xy} j_x k_y\right) \right].
\label{S_general}
\end{equation}
Considering all possible combinations for $l,m=x,y$, we obtain the coarse-grained propagator, $\Sigma =
\Sigma_{xx}+\Sigma_{xy}+\Sigma_{yx}+\Sigma_{yy}$, where
\begin{eqnarray}
\Sigma_{xx}({\bf k},0) &=& -\dfrac{\lambda_x^2  D}{16 \pi^2} \int^{>} \frac{d
{\bf j}}{\Delta^2}
	\left\{ \left[ 1 + \frac{2}{\Delta}(\nu_x j_x^2 + \nu_{xy} j_x j_y)
\right] k_x^2 +
	\frac{2}{\Delta}(\nu_y j_x j_y + \nu_{xy} j_x^2) k_x k_y \right\},
\label{Sigma_xx}\\[5pt]
\Sigma_{xy}({\bf k},0) &=& -\dfrac{\lambda_x\lambda_y  D}{16 \pi^2} \int^{>}
\frac{d {\bf j}}{\Delta^2} \left\{
	\frac{2}{\Delta}(\nu_x j_x j_y + \nu_{xy} j_y^2) k_x^2 + 
	\left[ 1 + \frac{2}{\Delta} (\nu_y j_y^2 + \nu_{xy} j_x j_y)\right] k_x
k_y	\right\}, \\[5pt]
\Sigma_{yx}({\bf k},0) &=& -\dfrac{\lambda_x\lambda_y  D}{16 \pi^2} \int^{>}
\frac{d {\bf j}}{\Delta^2}
	\left\{ \frac{2}{\Delta}(\nu_y j_x j_y + \nu_{xy} j_x^2) k_y^2 +
	\left[1 + \frac{2}{\Delta}(\nu_x j_x^2 + \nu_{xy} j_x j_y) \right] k_x
k_y \right\}, \\[5pt]
\Sigma_{yy}({\bf k},0) &=& -\dfrac{\lambda_y^2  D}{16 \pi^2} \int^{>} \frac{d
{\bf j}}{\Delta^2}
	\left\{ \left[1 + \frac{2}{\Delta}(\nu_y j_y^2 + \nu_{xy} j_x j_y)
\right] k_y^2 + 
	\frac{2}{\Delta}(\nu_x j_x j_y + \nu_{xy} j_y^2) k_x k_y
\right\}. \label{Sigma_yy}
\end{eqnarray}
The next step is to calculate the $\mathbf{k}$-contributions to these integrals induced by the dependence on the
wave-vector components of the integration limits that define the rectangular momentum-shell which is being integrated out.
Due to the lack of symmetry of the function $\Delta$ with respect to $j_x$ and $j_y$, we cannot use only the first
quadrant of the momentum shell to find them. Rather, it is convenient to expand Eqs.\ \eqref{Sigma_xx}-\eqref{Sigma_yy} in the limit $\delta l\to 0$. This allows us to rewrite each contribution $\Sigma_{lm}$ to the renormalization
of the propagator in a simpler form. In fact, for any function $f(j_x,j_y)$ appearing in the integrand of Eq.\ \eqref{S_general}, its integral decomposes into four terms, namely,
\begin{equation}
\int^{>} d {\bf j}\, f(j_x,j_y) = \int_{\Lambda/b}^\Lambda dj_x f_x(j_x) +
\int_{-\Lambda}^{-\Lambda/b} dj_x f_x(j_x)
	+ \int_{\Lambda/b^\zeta}^\Lambda dj_y f_y(j_y) +
\int_{-\Lambda}^{-\Lambda/b^\zeta} dj_y f_y(j_y),
\label{int_shell_cHK}
\end{equation}
where the associated single-variable functions $f_x$ and $f_y$ are simply given by
\begin{align}
& f_x(j_x) = \int_{-\Lambda}^\Lambda ds\, f(j_x,s), \\[5pt]
& f_y(j_y) = \int_{-\Lambda}^\Lambda ds\, f(s,j_y).
\end{align}
By expanding perturbatively Eq.\ \eqref{int_shell_cHK} for $b= e^{\delta l} \to 1$ we
obtain the general result
\begin{equation}
\int^{>} d {\bf j}\, f(j_x,j_y) \sim \left[f_x(\Lambda) +  f_x(-\Lambda) +\zeta
f_y(\Lambda) +\zeta f_y(-\Lambda) \right] \Lambda\,  \delta l.
\end{equation}
For the specific functions appearing in Eq.\ \eqref{S_general}, it is easy to verify
that $f_{x,y}(\Lambda) = f_{x,y}(-\Lambda)$, so that, in this particular case,
\begin{equation}
\int^{>} d {\bf j}\, f(j_x,j_y) \sim 2 \left[f_x(\Lambda)  +\zeta f_y(\Lambda)
\right] \Lambda\,  \delta l.
\end{equation}
Then it is convenient to express the general contribution to the coarse-grained propagator
in the following way
\begin{equation}
\Sigma_{lm}({\bf k},0) = -\frac{\lambda_l \lambda_m D}{8 \pi^2}
	\left[\int_{-\Lambda}^\Lambda ds\,
\dfrac{N_{lm}(\Lambda,s)}{\Delta_y^3(s)}+
		\zeta \int_{-\Lambda}^\Lambda ds\,
\dfrac{N_{lm}(s,\Lambda)}{\Delta_x^3(s)}\right] \Lambda \delta l,
\label{S_lm_2}
\end{equation}
where
\begin{align}
& \Delta_x(s) = \Delta(s,\Lambda) = \nu_x s^2 + 2\nu_{xy}\Lambda s + \nu_y \Lambda^2, \\[5pt]
&\Delta_y(s) = \Delta(\Lambda,s) = \nu_x \Lambda^2 + 2\nu_{xy}\Lambda s + \nu_y s^2, \\[5pt]
& N_{lm}(j_x,j_y) = \left(\nu_x j_x^2 + 2\nu_{xy} j_x j_y + \nu_y j_y^2\right)
k_l k_m +
	2\left(\nu_x j_x j_m + \nu_{xy} j_y j_m \right) k_l k_x + 2\left(\nu_y
j_y j_m + \nu_{xy} j_x j_m \right) k_l k_y.
\end{align}
Only six integrals need to be evaluated in order to cast Eq.\ \eqref{S_lm_2} into a form that can be used in our further analysis, namely,
\begin{align}
& J^x_i = \int_{-\Lambda}^\Lambda ds\, s^i / \Delta_x^3(s), \\[5pt]
& J^y_i = \int_{-\Lambda}^\Lambda ds\, s^i / \Delta_y^3(s), \label{hk_integr_Jxy}
\end{align}
for $i=0,1,2$. At this stage of the calculation it is practical to leave them unspecified, hence
\begin{alignat}{3}
& \Sigma_{xx}({\bf k},0)=   -\frac{\lambda_x^2  D\Lambda}{8 \pi^2} &  \bigg\{
	\Big[3\nu_x\left(\zeta J_2^x + \Lambda^2 J^y_0 \right)+
4\nu_{xy}\Lambda\left(\zeta J_1^x + J_1^y\right)+
		\nu_y\left(\zeta\Lambda^2 J_0^x + J_2^y \right)\Big] & k_x^2&&
\nonumber\\[5pt]
&& 	+ 2\Big[\nu_{xy}\left(\zeta J_2^x + \Lambda^2 J_0^y \right) + \nu_y
\Lambda\left(\zeta J_1^x + J_1^y \right)\Big]& k_x k_y &
		\bigg\}\delta l, &\\[5pt]
& \Sigma_{xy}({\bf k},0)=  -\frac{\lambda_x\lambda_y  D\Lambda}{8 \pi^2} &
\bigg\{
	\Big[ \nu_x\left(\zeta J_2^x + \Lambda J_0^y \right) +
4\nu_{xy}\Lambda\left(\zeta J_1^x + J_1^y \right) + 						
	3\nu_y \left(\zeta\Lambda^2 J_0^x + J_2^y \right) \Big] & k_x k_y &&
\nonumber \\[5pt]
&&	+ 2\Big[\nu_{xy}\left(\zeta\Lambda^2 J_0^x + J_2^y \right)  +
\nu_x\Lambda\left(\zeta J_1^x + J_1^y \right)  \Big] &  k_x^2 & \bigg\}\delta
l, &\\[5pt]
& \Sigma_{yx}({\bf k},0)=  -\frac{\lambda_x\lambda_y  D\Lambda}{8 \pi^2} &
\bigg\{
	\Big[ 3\nu_x\left(\zeta J_2^x + \Lambda^2 J_0^y\right) +
4\nu_{xy}\Lambda \left(\zeta J_1^x + J_1^y \right) +
				\nu_y\left(\zeta \Lambda^2 J_0^x + J_2^y \right)
\Big] & k_x k_y && \nonumber  \\[5pt]
&& + 2\Big[\nu_{xy}\left(\zeta J_2^x + \Lambda^2 J_0^y \right) + \nu_y\Lambda
\left(\zeta J_1^x + J_1^y \right)  \Big] & k_y^2 &  \bigg\}\delta l, & \\[5pt]
& \Sigma_{yy}({\bf k},0)=   -\frac{\lambda_y^2  D\Lambda}{8 \pi^2} &  \bigg\{
	\Big[\nu_x\left(\zeta J_2^x + \Lambda^2 J^y_0 \right)+
4\nu_{xy}\Lambda\left(\zeta J_1^x + J_1^y\right)+
		3\nu_y\left(\zeta\Lambda^2 J_0^x + J_2^y \right)\Big] & k_y^2&&
\nonumber\\[5pt]
&& 	+ 2\Big[\nu_{xy}\left(\zeta \Lambda^2 J_0^x +  J_2^y \right) + \nu_x
\Lambda\left(\zeta J_1^x + J_1^y \right)\Big]& k_x k_y &
		\bigg\}\delta l. &
\end{alignat}
Using the definitions of the couplings $g,f_{\nu},r_{\nu},r_{\lambda}$ provided in Eq.\ \eqref{couplings} of the main text and performing a change of variables, the integrals in Eq.\ \eqref{hk_integr_Jxy} can be expressed as
\begin{align}
& J^x_i = \frac{1}{\nu_x^3 \Lambda^2} J_i(1, r_{\nu}), \\[5pt]
& J^y_i = \frac{1}{\nu_x^3 \Lambda^2} J_i(r_{\nu},1),
\end{align}
where
\begin{equation}
 J_i(a, b) = \int_{-1}^{+1} \frac{s^i\, ds }{(a s^2 + 2 f_{\nu}s + b)^3}.
\label{jiab}
\end{equation}
After some algebra, we can make these integrals explicit in term of the couplings,
\begin{align*}
& J_0(a,b) = \dfrac{1}{8(ab - f^2_\nu)^2}\Bigg\{(a+f_\nu)
\dfrac{2(ab-f_\nu^2)+3a(a+2f_\nu+b)}{(a+2f_\nu+b)^2}
	+(a-f_\nu) \dfrac{2(ab-f_\nu^2)+3a(a-2f_\nu+b)}{(a-2f_\nu+b)^2}
\nonumber \\[5pt]
&  \hskip 6cm + \dfrac{3a^2}{\sqrt{ab-f_\nu^2}} \left[ \tan^{-1}\left(
\dfrac{a+f_\nu}{\sqrt{ab-f_\nu^2}}\right) + \tan^{-1}\left(
\dfrac{a-f_\nu}{\sqrt{ab-f_\nu^2}}\right)	\right] \Bigg\}, \\[7pt]
& J_1(a,b) = -\dfrac{1}{8(ab -
f^2_\nu)^2}\Bigg\{\dfrac{2(ab-f_\nu^2)(f_\nu+b)}{(a+2f_\nu+b)^2}  +
\dfrac{3f_\nu(a+f_\nu)}{a+2f_\nu+b}
	+\dfrac{2(ab-f_\nu^2)(f_\nu-b)}{(a-2f_\nu+b)^2}+\dfrac{3f_\nu(a-f_\nu)}{
a-2f_\nu+b} \nonumber \\[5pt]
&  \hskip 6cm + \dfrac{3a f_\nu}{\sqrt{ab-f_\nu^2}}\left[ \tan^{-1}\left(
\dfrac{a+f_\nu}{\sqrt{ab-f_\nu^2}}\right) + \tan^{-1}\left(
\dfrac{a-f_\nu}{\sqrt{ab-f_\nu^2}}\right)	\right]	\Bigg\}, \\[7pt]
& J_2(a,b) = \dfrac{1}{8(ab -
f^2_\nu)^2}\Bigg\{ \dfrac{2(a-b)\left[
ab(a+b)^2+2(a^2+4ab+b^2)f_\nu^2-16f_\nu^4\right]}									
{\left[(a+b)^2-4f_\nu^2 \right]^2}
\nonumber \\[5pt] 		
& \hskip 6cm + \dfrac{2 f_\nu^2+ab}{\sqrt{ab-f_\nu^2}}\left[ \tan^{-1}\left(
\dfrac{a+f_\nu}{\sqrt{ab-f_\nu^2}}\right) + \tan^{-1}\left(
\dfrac{a-f_\nu}{\sqrt{ab-f_\nu^2}}\right)	\right]	\Bigg\}.
\end{align*}
It is now convenient to write the contributions to the coarse-grained propagator
by gathering together the various terms, according to which parameter they renormalize in the original gHK equation,
\begin{equation}
 \Sigma = \Sigma_{xx} + \Sigma_{xy} + \Sigma_{yx} + \Sigma_{yy} \approx
(\Sigma_{\nu_x} \nu_x k_x^2 + 2 \Sigma_{\nu_{xy}}\nu_{xy} k_x k_y +
\Sigma_{\nu_y}\nu_y k_y^2)\delta l.
\label{xyxy}
\end{equation}
Using Eq.\ \eqref{couplings} of the main text, the functions on the right-hand side of Eq.\ \eqref{xyxy} read
\begin{align}
& \Sigma_{\nu_x} = -2 g \bigg\{ 3 \, \Big[ \zeta J_2(1,r_{\nu}) + J_0(r_{\nu},1)
\Big] + 2\, (r_{\nu} + 2 f_{\nu})\Big[\zeta J_1(1,r_{\nu}) + J_1(r_{\nu},1)
\Big] +  \notag \\
& \hskip 8cm+ (r_{\nu} + 2 f_{\nu} r_{\lambda})\Big[\zeta J_0(1,r_{\nu}) +
J_2(r_{\nu},1)\Big] \bigg\}, \label{eq:sigmanux}\\
& \Sigma_{\nu_{xy}} = - \frac{g}{f_{\nu}} \bigg\{ 2 \, (f_{\nu} + 2 r_{\lambda}
) \, \Big[ \zeta J_2(1,r_{\nu}) + J_0(r_{\nu},1) \Big] + (2 r_{\nu} + 8
r_{\lambda} f_{\nu} + r_{\lambda}^2) \Big[\zeta J_1(1,r_{\nu}) + J_1(r_{\nu},1)
\Big] + \notag   \\
 & \hskip 7.7 cm + r_{\lambda} (4 r_{\nu} + r_{\lambda} f_{\nu})\Big[\zeta
J_0(1,r_{\nu}) + J_2(r_{\nu},1)\Big] \bigg\}, \label{eq:sigmanuxy}\\
&\Sigma_{\nu_y} = -2 \frac{g r_{\lambda}}{r_{\nu}} \bigg\{ (r_{\lambda} + 2
f_{\nu}) \, \Big[ \zeta J_2(1,r_{\nu}) + J_0(r_{\nu},1) \Big] + 2\, (r_{\nu} + 2
f_{\nu} r_{\lambda})\Big[\zeta J_1(1,r_{\nu}) + J_1(r_{\nu},1) \Big] +  \notag
\\
 & \hskip 9cm+ 3 r_{\lambda} r_{\nu}\Big[\zeta J_0(1,r_{\nu}) +
J_2(r_{\nu},1)\Big] \bigg\}, \label{eq:sigmanuy}
\end{align}
so that the coarse-grained propagator can be finally written as
\begin{equation}
G_0^<({\bf k},\omega) = \Big[\nu_x \left(1 - \Sigma_{\nu_x}\delta l \right)
k_x^2 + 2\nu_{xy}\left(1 - \Sigma_{\nu_{xy}}\delta l \right)k_x k_y + \nu_y
\left(1 - \Sigma_{\nu_y}\delta l  \right) k_y^2 - i \omega \Big]^{-1} .
\end{equation}
Hence, the coarse-grained surface tension parameters are
$\nu_x^< = \nu_x(1 - \Sigma_{\nu_x} \delta l )$, $\nu_{xy}^< = \nu_{xy}(1 -
\Sigma_{\nu_{xy}}\delta l)$, and \mbox{$\nu_y^< = \nu_y(1 - \Sigma_{\nu_y} \delta l )$}.
After rescaling as in Eq.\ \eqref{eq:rescaling}, the corresponding flow equations become Eq.\ \eqref{flux_nus}
of the main text.

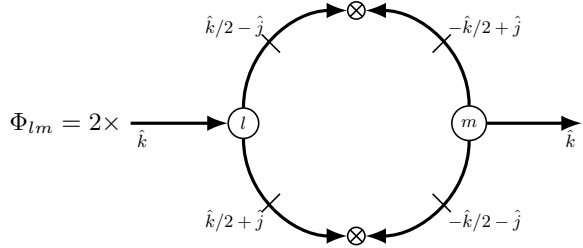
\begin{figure}[h!]
\begin{center}
\scalebox{1.5}{\begin{tikzpicture}

\node [draw=black,fill = white, shape=circle, scale = 0.5, name = n1] at (1,0) {$l$};
\node [draw = black, fill = white, shape=circle, scale = 0.5, name = n2] at ($(n1)+(2,0)$) {$m$};
\node [fill=white, shape=circle, draw=black, scale = 0.5, name = n3] at ($(n1) + (1,1)$) {};
\node [fill=white, shape=circle, draw=black, scale = 0.5, name = n4] at ($(n1) + (1,-1)$) {};
\node [name=center, scale=0.5] at ($(n1)+(1,0)$) {};

\draw (n3.north west)--(n3.south east);
\draw (n3.north east)--(n3.south west);
\draw (n4.north west)--(n4.south east);
\draw (n4.north east)--(n4.south west);

\node [name=nbar1, scale=0.7] at ($(center)+({-1/sqrt(2)-0.05},{1/sqrt(2)})$) {};
\node [name=nbar2, scale=0.7] at ($(center)+ ({-1/sqrt(2)-0.05},{-1/sqrt(2)})$) {};
\node [name=nbar3, scale=0.7] at ($(center)+({1/sqrt(2)+0.05},{1/sqrt(2)})$) {};
\node [name=nbar4, scale=0.7] at ($(center)+({1/sqrt(2)+0.05},{-1/sqrt(2)})$) {};

\node [name = text0, scale = 0.7, left] at (0,0) {$\Phi_{lm} = 2\times$};
\node [name=text1, scale=0.5, above left] at ($(center)+({-1/sqrt(2)-0.05},{1/sqrt(2)})$) {$\hat k/2 - \hat j$};
\node [name=text2, scale=0.5, below left] at ($(center)+ ({-1/sqrt(2)-0.05},{-1/sqrt(2)})$) {$\hat k/2 + \hat j$};
\node [name=text3, scale=0.5, above right] at ($(center)+({1/sqrt(2)+0.05},{1/sqrt(2)})$) {$-\hat k/2 + \hat j$};
\node [name=text4, scale=0.5, below right] at ($(center)+({1/sqrt(2)+0.05},{-1/sqrt(2)})$) {$-\hat k/2 - \hat j$};
\node [name=text5, scale=0.5, below right] at (0,0) {$\hat k$};
\node [name=text6, scale=0.5, below left] at ($(n2) + (1,0)$) {$\hat k$};

\draw (nbar1.north west)--(nbar1.south east);
\draw (nbar2.south west)--(nbar2.north east);
\draw (nbar3.south west)--(nbar3.north east);
\draw (nbar4.north west)--(nbar4.south east);


\draw [thick, arrows=-latex] (0,0) -- (n1);
\draw [thick, arrows=-latex] (n1)  to [out = 90, in = 180] (n3);
\draw [thick, arrows=-latex] (n1)  to [out = -90, in = 180] (n4);
\draw [thick, arrows=-latex] (n2)  to [out = 90, in = 0] (n3);
\draw [thick, arrows=-latex] (n2)  to [out = -90, in = 0] (n4);
\draw [thick, arrows=-latex] (n2) -- ($(n2)+(1,0)$);

\end{tikzpicture}}
\caption{Generic diagrammatic representation of the four different contributions $\Phi_{xx}$, $\Phi_{xy}$, $\Phi_{yx}$, and $\Phi_{yy}$  to the renormalization of the noise variance $2D$ for the gHK and the aKPZ equations. For each equation the exact meaning of the solid lines differ, see Eq.\ \eqref{noise_lm_ghk} for the former and Eq.\ \eqref{noise_lm} for the latter.}
\label{fig_noise_akpz}
\end{center}
\end{figure}
The renormalization of the noise variance is calculated from the standard diagram shown in Fig.\ \ref{fig_noise_akpz}.
Due to the existence of two different vertices, four different contributions occur, analogously to the renormalization of the propagator. In the symmetric momentum variable ${\bf j}$ they read, to leading order,
\begin{equation}
\Phi_{lm}({\bf k},\omega) = \lambda_l \lambda_m D^2 k_l k_m\int^>  \dfrac{d{\bf
j}}{(2\pi)^2} \int \dfrac{d\Omega}{2\pi}
	 	\left|G_0\left(\frac{\hat k}{2}+\hat j\right)\right|^2
\left|G_0\left(\frac{\hat k}{2}-\hat j\right)\right|^2.
\label{noise_lm_ghk}
\end{equation}
Since all contributions given by Eq.\ \eqref{noise_lm_ghk} are proportional to $k_l k_m$, they can be
neglected in the limit $k_{x,y} \rightarrow 0$. Hence, the coefficient $D$ is not renormalized to all orders of the perturbation series, and its flow equation is simply given by Eq.\ \eqref{flux_D} of the main text.
Finally, the one-loop contributions to the renormalization of the
nonlinearities $\lambda_{x,y}$ cancel out \cite{nicoli:2011}, giving rise to Eq.\
\eqref{flux_lambdas} of the main text, thus completing the DRG flow for the gHK equation.

%

\subsection{Standard HK equation}
\label{appa:hk}

The RG flow for the standard HK equation is retrieved from the one derived for the gHK equation by setting $f_\nu=0$ and $r_\lambda=0$. In this case the $J_i$ integrals, Eq.\ \eqref{jiab}, reduce to
\begin{align}
& J_0(a,b) =
\dfrac{1}{8b^2}\left[\dfrac{6a+10b}{(a+b)^2} + \dfrac{6}{\sqrt{ab}}\tan^{-1}
\left(\sqrt{\dfrac{a}{b}} \right)\right], \\[5pt]
& J_2(a,b) =
\dfrac{1}{8ab}\left[\dfrac{2(a-b)}{(a+b)^2} + \dfrac{2}{\sqrt{ab}}\tan^{-1}
\left(\sqrt{\dfrac{a}{b}} \right)\right],
\end{align}
whereas $J_1$ is identically equal to zero. The condition $r_\lambda =0 $
implies $\lambda_y = 0$, so that $\Sigma_{xy}=\Sigma_{yx}=\Sigma_{yy}=0$.
The condition $f_\nu =0$ implies $\nu_{xy} = 0$, so that in this case $\Sigma_{xx}$ does not
generate $\nu_{xy}$ under coarse-graining, provided its bare value is zero.


\subsection{gHK equation for $\zeta=1$}
\label{app_gHK_iso}

If $\zeta=1$, the functions intervening in the RG flow of the gHK equation simplify somewhat. Thus,
\begin{align*}
& J_2(1,r_\nu) + J_0(r_\nu,1) =
\dfrac{1}{8(r_\nu-f_\nu^2)^2}\Bigg\{\dfrac{
4f_\nu^2(5r_\nu-1)-2r_\nu(1+4r_\nu+3r_\nu^2)}{4f_\nu^2-(1+r_\nu)^2} \notag \\
&  \hskip 5cm
+ \dfrac{3 r_\nu^2}{\sqrt{r_\nu - f_\nu^2}}\left[\tan^{-1}\left(\dfrac{r_\nu + f_\nu}{\sqrt{r_\nu - f_\nu^2}}\right) + \tan^{-1}\left(\dfrac{r_\nu - f_\nu}{\sqrt{r_\nu - f_\nu^2}}\right)\right] \notag \\
& \hskip 5cm
+ \dfrac{2 f_\nu^2 + r_\nu}{\sqrt{r_\nu - f_\nu^2}}\left[\tan^{-1}\left(\dfrac{1 + f_\nu}{\sqrt{r_\nu - f_\nu^2}}\right) + \tan^{-1}\left(\dfrac{1 - f_\nu}{\sqrt{r_\nu - f_\nu^2}}\right)\right]  \Bigg\}, \\[5pt]
& J_1(1,r_\nu) + J_1(r_\nu,1) =
\dfrac{1}{8(r_\nu-f_\nu^2)^2}\Bigg\{ \dfrac{2f_\nu(3- 8 f_\nu^2 +2r_\nu + 3
r_\nu^2)}{4f_\nu^2-(1+r_\nu)^2}  \notag \\
& \hskip 5cm
- \dfrac{3 f_\nu r_\nu}{\sqrt{r_\nu - f_\nu^2}}\left[\tan^{-1}\left(\dfrac{r_\nu + f_\nu}{\sqrt{r_\nu - f_\nu^2}}\right) + \tan^{-1}\left(\dfrac{r_\nu - f_\nu}{\sqrt{r_\nu - f_\nu^2}}\right)\right] \notag \\
& \hskip 5cm
- \dfrac{3 f_\nu}{\sqrt{r_\nu - f_\nu^2}}\left[\tan^{-1}\left(\dfrac{1 + f_\nu}{\sqrt{r_\nu - f_\nu^2}}\right) + \tan^{-1}\left(\dfrac{1 - f_\nu}{\sqrt{r_\nu - f_\nu^2}}\right)\right]  \Bigg\}, \\[5pt]
& J_0(1,r_\nu) + J_2(r_\nu,1) =
\dfrac{1}{8(r_\nu-f_\nu^2)^2}\Bigg\{\dfrac{4f_\nu^2(5-r_\nu)-2r_\nu(4+r_\nu)-6}{
4f_\nu^2-(1+r_\nu)^2}  \notag \\
& \hskip 5cm
+ \dfrac{2 f_\nu^2 + r_\nu}{\sqrt{r_\nu - f_\nu^2}}\left[\tan^{-1}\left(\dfrac{r_\nu + f_\nu}{\sqrt{r_\nu - f_\nu^2}}\right) + \tan^{-1}\left(\dfrac{r_\nu - f_\nu}{\sqrt{r_\nu - f_\nu^2}}\right)\right] \notag \\
& \hskip 5cm
+ \dfrac{3 }{\sqrt{r_\nu - f_\nu^2}}\left[\tan^{-1}\left(\dfrac{1 + f_\nu}{\sqrt{r_\nu - f_\nu^2}}\right) + \tan^{-1}\left(\dfrac{1 - f_\nu}{\sqrt{r_\nu - f_\nu^2}}\right)\right]  \Bigg\}.
\end{align*}
These formulae can be employed in order to rewrite the flow of the couplings, which reads
\begin{align}
& \dfrac{dr_\nu}{dl} =
-\dfrac{g}{(r_\nu-f_\nu^2)^2}\Bigg\{ \dfrac{p_{r1}(r_\lambda)}{
4f_\nu^2-(1+r_\nu)^2} +
\dfrac{p_{r2}(r_\lambda)}{4 \sqrt{r_\nu - f_\nu^2} }\left[ \tan^{-1}\left(\dfrac{1 + f_\nu}{\sqrt{r_\nu - f_\nu^2}}\right) + \tan^{-1}\left(\dfrac{1 - f_\nu}{\sqrt{r_\nu - f_\nu^2}}\right)\right] \notag \\
& \hskip 5.7cm
+ \dfrac{p_{r3}(r_\lambda)}{4 \sqrt{r_\nu - f_\nu^2} }\left[ \tan^{-1}\left(\dfrac{r_\nu + f_\nu}{\sqrt{r_\nu - f_\nu^2}}\right) + \tan^{-1}\left(\dfrac{r_\nu - f_\nu}{\sqrt{r_\nu - f_\nu^2}}\right)\right] \Bigg\}, \\[5pt]
& \dfrac{df_\nu}{dl} = \dfrac{g}{2 (r_\nu - f_\nu^2)^2}\Bigg\{
 	  \dfrac{p_{f1}(r_\lambda)}{4f_\nu^2 - (1+r_\nu)^2}
 	   -
\dfrac{p_{f2}(r_\lambda)}{4 \sqrt{r_\nu - f_\nu^2} }\left[ \tan^{-1}\left(\dfrac{1 + f_\nu}{\sqrt{r_\nu - f_\nu^2}}\right) + \tan^{-1}\left(\dfrac{1 - f_\nu}{\sqrt{r_\nu - f_\nu^2}}\right)\right] \notag \\
& \hskip 5.6cm - \dfrac{p_{f3}(r_\lambda)}{4 \sqrt{r_\nu - f_\nu^2} }\left[ \tan^{-1}\left(\dfrac{r_\nu + f_\nu}{\sqrt{r_\nu - f_\nu^2}}\right) + \tan^{-1}\left(\dfrac{r_\nu - f_\nu}{\sqrt{r_\nu - f_\nu^2}}\right)\right] \Bigg\}, \\[5pt]
& \dfrac{dg}{dl} =  2 g + \dfrac{3 g^2}{(r_\nu-f_\nu^2)^2}\Bigg\{ \frac{p_{g1}}{4f_\nu^2-(1+r_\nu)^2}  -
\dfrac{p_{g2}(r_\lambda)}{4 \sqrt{r_\nu - f_\nu^2} }\left[ \tan^{-1}\left(\dfrac{1 + f_\nu}{\sqrt{r_\nu - f_\nu^2}}\right) + \tan^{-1}\left(\dfrac{1 - f_\nu}{\sqrt{r_\nu - f_\nu^2}}\right)\right] \notag \\
& \hskip 5.6cm - \dfrac{p_{g3}(r_\lambda)}{4 \sqrt{r_\nu - f_\nu^2} }\left[ \tan^{-1}\left(\dfrac{r_\nu + f_\nu}{\sqrt{r_\nu - f_\nu^2}}\right) + \tan^{-1}\left(\dfrac{r_\nu - f_\nu}{\sqrt{r_\nu - f_\nu^2}}\right)\right] \Bigg\},
\end{align}
with
\begin{align*}
& p_{r1}(r_\lambda) = \left(f_\nu^2-r_\nu\right) \left[16f_\nu^2 -
(r_\nu+1)(3r_\nu+5)\right] r_\lambda^2
	-f_\nu\left[2f_\nu^2(r_\nu^2-4r_\nu -1)+r_\nu^3+2r_\nu^2+5r_\nu \right]
r_\lambda \\[5pt]
&	\hskip 4cm - r_\nu\left[16 f_\nu^4 +8r_\nu f_\nu^3 - (5r_\nu^2+24r_\nu+
3)f_\nu^2 -(3r_\nu^3+2r_\nu^2 + 3r_\nu)f_\nu + 5r_\nu^3  +8r_\nu^2+ 3r_\nu
	 \right],\\[5pt]
& p_{r2} (r_\lambda) = 10 (f_\nu^2 - r_\nu) r_\lambda^2 + 2 f_\nu (5 r_\nu - 2 f_\nu^2) r_\lambda + 6 r_\nu (r_\nu - f_\nu r_\nu - f_\nu^2), \\[5pt]
& p_{r3}(r_\lambda) = 6 r_\nu (f_\nu^2 - r_\nu) r_\lambda^2 + 2 f_\nu r_\nu (r_\nu + 2 f_\nu^2)r_\lambda + 2 r_\nu^2 (5 r_\nu -3 f_\nu r_\nu -5 f_\nu^2), \\[5pt]
& p_{f1}(r_\lambda) = f_\nu (f_\nu^2 - r_\nu)(1-r_\nu)r_\lambda^2
  +  \left[f_\nu^4 (4 r_\nu - 52) + f_\nu^2(10 r_\nu^2 + 56 r_\nu + 14) - 8 r_\nu(r_\nu^2 + 2 r_\nu + 1) \right] r_\lambda \\[5pt]
& \hskip 4cm + 2 f_\nu \left[ 8 f_\nu^3 (r_\nu + 2 f_\nu) - f_\nu^2(5r_\nu^2 + 23 r_\nu + 4) - f_\nu r_\nu (3 r_\nu^2 + 2r_\nu + 3) + r_\nu (5r_\nu^2 + 7 r_\nu + 4)\right],\\[5pt]
& p_{f2}(r_\lambda) = 4 (7 f_\nu^2 -4 r_\nu) r_\lambda + 4 f_\nu (4 r_\nu - 3 f_\nu r_\nu -4 f_\nu^2), \\[5pt]
& p_{f3}(r_\lambda) = 2 f_\nu (r_\nu - f_\nu^2)r_\lambda^2
 + 4 (-4 r_\nu^2 + 5 f_\nu^2 r_\nu + 2 f_\nu^4) r_\lambda
 + 4 f_\nu r_\nu (5 r_\nu - 3 f_\nu r_\nu -5 f_\nu^2), \\[5pt]
& p_{g1}(r_\lambda) = r_\lambda \left[2 f_\nu^3(r_\nu-5) + f_\nu(r_\nu^2+4r_\nu+3)\right]+ 8f_\nu^3(r_\nu+2f_\nu)
	- f_\nu^2(5r_\nu^2 + 24r_\nu + 3) - f_\nu r_\nu(3r_\nu^2 + 2r_\nu + 3) \\[5pt]
& \hskip 4cm    + r_\nu(5r_\nu^2 + 8 r_\nu + 3),\\[5pt]
& p_{g2}(r_\lambda) = 6 f_\nu r_\lambda + 6 (r_\nu - f_\nu r_\nu - f_\nu^2), \\[5pt]
& p_{g3}(r_\lambda) = 2 f_\nu (2 f_\nu^2 + r_\nu) r_\lambda + 2 r_\nu (5 r_\nu - 3 f_\nu r _\nu - 5 f_\nu^2).
\end{align*}

\section{Dynamic Renormalization Group analysis of the anisotropic
Kardar-Parisi-Zhang equation}
\label{app_aKPZ}

For the aKPZ equation, the diagrammatic expansion of the integrals that contribute to the
renormalization of the bare propagator can be also sketched using general notation as shown in Fig.\ \ref{fig_prop_akpz}. Again $l,m=x,y$ in all possible combinations, leading to four different contributions, which will be denoted
$\Sigma_{xx}, \Sigma_{xy}, \Sigma_{yx}$, and $\Sigma_{yy}$, as in the gHK case. Naturally, the values of these differ for each equation; we hope the context will hinder any potential ambiguity, as we are providing separate discussions for the two equations.
After the usual symmetrization of the integration variables
$({\bf q},\Omega)\to ({\bf j}+{\bf k}/2,\Omega+\omega/2)$, these contributions read
\begin{equation}
\Sigma_{lm}({\bf k},\omega) = 2 \lambda_l \lambda_m D \int^>  \dfrac{d{\bf
j}}{(2\pi)^2} \int \dfrac{d\Omega}{2\pi}
	 \left(j^2_l - \frac{k^2_l}{4}\right)  \left(\frac{k_m}{2} +
j_m\right)k_m \,
	\left|G_0\left(\frac{\hat k}{2}+\hat j\right)\right|^2
G_0\left(\frac{\hat k}{2}-\hat j\right),
\label{sigma_lm}
\end{equation}
where again $G_0\big(\hat k\big)$ is short-hand notation for the bare propagator, which now reads
\begin{equation}
G_0\big(\hat k\big) = \left[\nu_x k_x^2 + \nu_y k_y^2 -i \omega\right]^{-1}.
\end{equation}
An expansion to first order in $k_x/j_x,k_y/j_y\ll 1$ leads to
\begin{eqnarray}
&& \lim_{\omega \rightarrow 0} \left| G_0\left(\hat k / 2+\hat j\right)\right|^2
\sim
	\frac{1}{\Delta^2 + \Omega^2} \left[1 - \frac{2\Delta}{\Delta^2 +
\Omega^2}(\nu_x j_x k_x + \nu_y j_y k_y) \right],\\[5pt]
&&\lim_{\omega \rightarrow 0}	G_0\left(\hat k / 2-\hat j\right) \sim
\frac{1}{\Delta + i \Omega}
	\left[1 + \frac{1}{\Delta + i\Omega}(\nu_x j_x k_x + \nu_y j_y k_y)
\right].
\end{eqnarray}
where $\Delta = \nu_x j_x^2 + \nu_y j_y^2$. Using these results in Eq.\
\eqref{sigma_lm} and after integration over the frequency variable $\Omega$, to second order in the
components of ${\bf k}$ we get
\begin{equation}
\Sigma_{lm}({\bf k},0) = \frac{\lambda_l \lambda_m D}{16 \pi^2} \int^{>} \frac{d
{\bf j}}{\Delta^2}
	\left[ j_l^2 k_m^2 - \frac{2}{\Delta} k_m  j_l^2  j_m\left( \nu_x  k_x
j_x +\nu_y k_y   j_y\right) \right].
\end{equation}
Considering all possible combinations for $l,m=x,y$, we obtain the coarse-grained propagator, $\Sigma =
\Sigma_{xx}+\Sigma_{xy}+\Sigma_{yx}+\Sigma_{yy}$. We now take into account that the momentum shell is
symmetric with respect to $j_x$ and $j_y$; hence, contributions from odd functions cancel out in these integrals, leading to
\begin{eqnarray}
&& \Sigma_{xx}({\bf k},0) = \frac{\lambda_x^2  D}{16 \pi^2} \int^{>} \frac{d
{\bf j}}{\Delta^2}
	\left( j_x^2  - \frac{2}{\Delta} \nu_x   j_x^4   \right) k_x^2, \\[5pt]
&& \Sigma_{xy}({\bf k},0) = \frac{\lambda_x\lambda_y  D}{16 \pi^2} \int^{>}
\frac{d {\bf j}}{\Delta^2}
	\left( j_x^2  - \frac{2}{\Delta} \nu_y   j_x^2 j_y^2   \right) k_y^2,
\\[5pt]
&& \Sigma_{yx}({\bf k},0) = \frac{\lambda_x\lambda_y  D}{16 \pi^2} \int^{>}
\frac{d {\bf j}}{\Delta^2}
	\left( j_y^2  - \frac{2}{\Delta} \nu_x   j_x^2 j_y^2   \right) k_x^2,
\\[5pt]
&& \Sigma_{yy}({\bf k},0) = \frac{\lambda_y^2  D}{16 \pi^2} \int^{>} \frac{d
{\bf j}}{\Delta^2}
	\left( j_y^2  - \frac{2}{\Delta} \nu_y   j_y^4   \right) k_y^2.
\end{eqnarray}
As in the gHK case, the next step is to calculate the contributions to these integrals induced
by the $\mathbf{k}$-dependence of the integration limits defining the momentum shell. Now we can split the momentum
integrals in only two parts,
\begin{align}
& \int^{>} \frac{d {\bf j}}{4\Delta^2} \left( j_x^2  - \frac{2}{\Delta} \nu_x   j_x^4   \right) =
	 \int_{\Lambda/b}^\Lambda dj_x \int_0^\Lambda dj_y  \left( \dfrac{j_x^2}{\Delta^2}  - 2\nu_x\frac{j_x^4}{\Delta^3}       \right) +
		\int_{\Lambda/b^\zeta}^\Lambda dj_y \int_0^\Lambda dj_x  \left(
	\dfrac{j_x^2}{\Delta^2}  - 2\nu_x\frac{j_x^4}{\Delta^3}   \right) \nonumber \\[5pt]
& \hskip 3.8cm =	 \int_{\Lambda/b}^\Lambda dj_x \,	j_x^2\left( I^y_{02} - 2 \nu_x
j_x^2  I^y_{03}   \right) +
		\int_{\Lambda/b^\zeta}^\Lambda dj_y  \left( I^x_{22} - 2 \nu_x
I^x_{43}   \right), \\[5pt]
& \int^{>} \frac{d {\bf j}}{4\Delta^2}
	\left( j_x^2  - \frac{2}{\Delta} \nu_y   j_x^2 j_y^2   \right) =	
	 \int_{\Lambda/b}^\Lambda dj_x \,	j_x^2\left( I^y_{02} - 2 \nu_y
I^y_{23}   \right) +
		\int_{\Lambda/b^\zeta}^\Lambda dj_y  \left( I^x_{22} - 2 \nu_y
j_y^2  I^x_{23}   \right), \\[5pt]
& \int^{>} \frac{d {\bf j}}{4\Delta^2} \left( j_y^2  -
\frac{2}{\Delta} \nu_x   j_x^2 j_y^2   \right) =
	  \int_{\Lambda/b}^\Lambda dj_x \left( I^y_{22} - 2 \nu_x j_x^2
I^y_{23}   \right) +
		\int_{\Lambda/b^\zeta}^\Lambda dj_y \, j_y^2 \left( I^x_{02} - 2
\nu_x   I^x_{23}   \right), \\[5pt]
& \int^{>} \frac{d {\bf j}}{4\Delta^2} \left( j_y^2  -
\frac{2}{\Delta} \nu_y   j_y^4   \right) =
	\int_{\Lambda/b}^\Lambda dj_x \left( I^y_{22} - 2 \nu_y   I^y_{43}
\right) +
		\int_{\Lambda/b^\zeta}^\Lambda dj_y \, j_y^2 \left( I^x_{02} - 2
\nu_y j_y^2   I^x_{03}   \right),
\end{align}
where the values of the integrals
\begin{eqnarray}
&& I^x_{ij} =\int_0^\Lambda ds\,  s^i (\nu_x s^2 + \nu_y j_y^2)^{-j}, \label{ixij} \\
&& I^y_{ij} =\int_0^\Lambda ds\,  s^i (\nu_x j_x^2 + \nu_y s^2)^{-j}, \label{iyij}
\end{eqnarray}
are provided in Table \ref{tab_Iij}. The remaining integrals are solved
perturbatively for $\delta l \to 0$, using that
$\Lambda/b = \Lambda e^{-\delta l} \sim \Lambda (1- \delta l)$, and
$\Lambda/b^\zeta  \sim \Lambda (1- \zeta \delta l)$. We thus get
\begin{eqnarray}
&& \Sigma_{xx}({\bf k},0) = \dfrac{\lambda^2_x D}{16\pi^2 \nu_x} (\zeta -1)
\left[ \dfrac{3\nu_x + \nu_y}{(\nu_x+\nu_y)^2}
	-B_{\nu_x,\nu_y,\zeta} \right]  k_x^2 \delta l, \\[5pt]
&& \Sigma_{xy}({\bf k},0) = \dfrac{\lambda_x\lambda_y D}{16\pi^2 \nu_x} (\zeta
-1) \left[B_{\nu_x,\nu_y,\zeta}- \dfrac{3\nu_x + \nu_y}{(\nu_x+\nu_y)^2}
	 \right]  k_y^2 \delta l, \\[5pt]
&& \Sigma_{yx}({\bf k},0) = \dfrac{\lambda_x\lambda_y D}{16\pi^2 \nu_y} (\zeta
-1) \left[B_{\nu_x,\nu_y,\zeta} +  \dfrac{\nu_x + 3\nu_y}{(\nu_x+\nu_y)^2}
	 \right]  k_x^2 \delta l, \\[5pt]
&& \Sigma_{yy}({\bf k},0) = \dfrac{\lambda_y^2 D}{16\pi^2 \nu_y} (1-\zeta)
\left[\dfrac{\nu_x + 3\nu_y}{(\nu_x+\nu_y)^2} +B_{\nu_x,\nu_y,\zeta}
	 \right]  k_y^2 \delta l,
\end{eqnarray}
where
\begin{equation}
B_{\nu_x,\nu_y,\zeta} = \dfrac{\tan^{-1}\left(\sqrt{\nu_y / \nu_x} \right) +
\zeta \tan^{-1}\left(\sqrt{\nu_x / \nu_y} \right)}
	{(\zeta-1)\, (\nu_x \nu_y)^{1/2}} .
\end{equation}
At this stage of the calculation it is convenient to gather the factors together,
according to the parameter in the original aKPZ equation which they renormalize.
We thus introduce functions $\Sigma_{\nu_{x,y}}$ through $\Sigma_{\nu_x} \delta l \,\nu_x k_x^2 \equiv
\Sigma_{xx}({\bf k},0) + \Sigma_{yx}({\bf k},0)$ and $\Sigma_{\nu_y} \delta l \, \nu_y k_y^2 \equiv
\Sigma_{xy}({\bf k},0) + \Sigma_{yy}({\bf k},0)$, so that the coarse-grained propagator reads
\begin{equation}
G_0^<({\bf k},\omega) = \Big[\nu_x \left(1 - \Sigma_{\nu_x}\delta l \right)
k_x^2 +\nu_y \left(1 - \Sigma_{\nu_y}\delta l  \right) k_y^2 - i \omega
\Big]^{-1} .
\end{equation}
Hence, the coarse-grained surface tension parameters are $\nu_x^< = \nu_x(1 - \Sigma_{\nu_x} \delta l )$ and $\nu_y^< = \nu_y(1 - \Sigma_{\nu_y} \delta l )$. After rescaling as in Eq.\ \eqref{eq:rescaling}, the corresponding flow equations become Eqs.\ \eqref{akpz_nus} of the main text.

\begin{table}[t!]
\begin{tabular}{|l|l|}
\hline
$I_{02}$ & $\dfrac{1}{2a}\left(\dfrac{\tan^{-1}\left(\Lambda\sqrt{b/a}
\right)}{\sqrt{ab}}+ \dfrac{\Lambda}{a + b\Lambda^2}\right)$  \\[12pt]
\hline
$I_{22}$ & $\dfrac{1}{2b}\left(\dfrac{\tan^{-1}\left(\Lambda\sqrt{b/a}
\right)}{\sqrt{ab}} -\dfrac{\Lambda}{a + b\Lambda^2}\right)$  \\[12pt]
\hline
$I_{03}$ & $\dfrac{1}{8a^2}\left(\dfrac{3\tan^{-1}\left(\Lambda\sqrt{b/a}
\right)}{\sqrt{ab}} +\dfrac{\Lambda(5a+3b\Lambda^2)}{(a + b\Lambda^2)^2}
\right)$  \\[12pt]
\hline
$I_{23}$ & $\dfrac{1}{8ab}\left(\dfrac{3\tan^{-1}\left(\Lambda\sqrt{b/a}
\right)}{\sqrt{ab}} + \dfrac{\Lambda (b\Lambda^2-a)}{(a + b\Lambda^2)^2}\right)$
 \\[12pt]
\hline
$I_{43}$ & $\dfrac{1}{8b^2}\left(\dfrac{3\tan^{-1}\left(\Lambda\sqrt{b/a}
\right)}{\sqrt{ab}}-\dfrac{\Lambda(3a+5b\Lambda^2)}{(a + b\Lambda^2)^2}\right)$
\\[12pt]
\hline
\end{tabular}
\caption{Definite integrals $I_{ij}=I_{ij}^x$, Eq.\ \eqref{ixij}, for $a=\nu_y j_y^2$ and $b=\nu_x$, and
$I_{ij}=I_{ij}^y$, Eq.\ \eqref{iyij}, for $a=\nu_x j_x^2$ and $b=\nu_y$.}
\label{tab_Iij}
\end{table}

The renormalization of the noise variance is again calculated from the standard diagram in Fig.\ \ref{fig_noise_akpz}. Similar considerations apply as in the case of the gHK equation. However, now noise does renormalize non-trivially.
Indeed, in the symmetric momentum variable ${\bf j}$, the contributions to the coarse-grained noise variance read
\begin{equation}
\Phi_{lm}({\bf k},\omega) = \lambda_l \lambda_m D^2 \int^>  \dfrac{d{\bf
j}}{(2\pi)^2} \int \dfrac{d\Omega}{2\pi}
	 \left(\frac{k^2_l}{4}- j_l^2\right)  \left(\frac{k^2_m}{4}-
j_m^2\right) \,
	\left|G_0\left(\frac{\hat k}{2}+\hat j\right)\right|^2
\left|G_0\left(\frac{\hat k}{2}-\hat j\right)\right|^2,
\label{noise_lm}
\end{equation}
where $l,m=x,y$ in all four possible combinations. Taking into account that in the perturbative expansion of $\Phi_{lm}$ we only have to retain the zeroth order contribution in ${\bf k}$ components, and that
\begin{equation}
 \lim_{\omega \rightarrow 0} \left| G_0\left(\hat k / 2-\hat j\right)\right|^2
\sim \dfrac{1}{\Delta^2+\Omega^2},
\end{equation}
after the integration in the frequency variable $\Omega$, we obtain
\begin{equation}
\Phi_{lm}({\bf k},0) \sim \dfrac{\lambda_l \lambda_m D^2}{16\pi^2} \int^>  d{\bf
j}\,
	 \dfrac{j_l^2  j_m^2}{\Delta^3} .
\end{equation}
The four contributions are calculated as
\begin{eqnarray}
\Phi_{xx}({\bf k},0) &=& \dfrac{\lambda_x^2  D^2}{16\pi^2} \int^>  d{\bf j}\,
\dfrac{j_x^4}{\Delta^3} =
	\dfrac{\lambda_x^2  D^2}{4\pi^2}\left[\int_{\Lambda/b}^\Lambda dj_x\,
j_x^4 I^y_{03} + \int_{\Lambda/b^\zeta}^\Lambda dj_y\,  I^x_{43} \right]
	\nonumber \\
	&\sim & \dfrac{\lambda_x^2  D^2}{32\pi^2\nu_x^2}(\zeta-1)\left[
3B_{\nu_x,\nu_y,\zeta} - \dfrac{5\nu_x + 3\nu_y}{(\nu_x+\nu_y)^2}\right] \delta
l, \\
\Phi_{xy}({\bf k},0) &=& \Phi_{yx}({\bf k},0)  = \dfrac{\lambda_x \lambda_y
D^2}{16\pi^2} \int^>  d{\bf j}\, \dfrac{j_x^2 j_y^2}{\Delta^3} =
	\dfrac{\lambda_x \lambda_y  D^2}{4\pi^2}\left[\int_{\Lambda/b}^\Lambda
dj_x\, j_x^2 I^y_{23} + \int_{\Lambda/b^\zeta}^\Lambda dj_y\, j_y^2  I^x_{23}
\right]
	\nonumber \\
	&\sim & \dfrac{\lambda_x \lambda_y
D^2}{32\pi^2\nu_x\nu_y}(\zeta-1)\left[ B_{\nu_x,\nu_y,\zeta}  + \dfrac{\nu_x -
\nu_y}{(\nu_x+\nu_y)^2}\right] \delta l, \\
\Phi_{yy}({\bf k},0) &=& \dfrac{\lambda_y^2  D^2}{16\pi^2} \int^>  d{\bf j}\,
\dfrac{j_y^4}{\Delta^3} =
	\dfrac{\lambda_y^2  D^2}{4\pi^2}\left[\int_{\Lambda/b}^\Lambda dj_x\,
I^y_{43} + \int_{\Lambda/b^\zeta}^\Lambda dj_y\,  j_y^4 I^x_{03} \right]
	\nonumber \\
	&\sim & \dfrac{\lambda_y^2  D^2}{32\pi^2\nu_y^2}(\zeta-1)\left[
3B_{\nu_x,\nu_y,\zeta} + \dfrac{3\nu_x + 5\nu_y}{(\nu_x+\nu_y)^2}\right] \delta
l,
\end{eqnarray}
and finally
\begin{equation}
\begin{split}
\Phi({\bf k},0) = \sum_{l,m=x,y} \Phi_{lm}({\bf k},0) &= \dfrac{D^2}{32\pi^2}(\zeta-1)\Bigg[3B_{\nu_x,\nu_y,\zeta}
\left( \dfrac{\lambda_x^2}{\nu_x^2} + \dfrac{2\lambda_x\lambda_y}{3\nu_x\nu_y}
+
	\dfrac{\lambda_y^2}{\nu_y^2}\right) \\
&+\left( \dfrac{\lambda_y^2(3\nu_x+5\nu_y)}{\nu_y^2(\nu_x+\nu_y)^2} +
	\dfrac{2\lambda_x\lambda_y(\nu_x - \nu_y)}{\nu_x \nu_y(\nu_x+\nu_y)^2} -
\dfrac{\lambda_x^2(5\nu_x+3\nu_y)}{\nu_x^2(\nu_x+\nu_y)^2}\right)
	\Bigg]\delta l.
\end{split}
\end{equation}
Note this function is $\mathbf{k}$-independent, hence it implies a non-trivial effect of coarse-graining in the noise variance for the aKPZ equation. By introducing a function $\Phi_D$ through $\Phi_D D \delta l \equiv \Phi({\bf k},0)$,
the coarse-grained noise variance is $D^< = D(1+\Phi_D\delta l)$. After rescaling as in Eq.\ \eqref{eq:rescaling}, the corresponding flow equation becomes Eq.\ \eqref{akpz_D} of the main text. Finally, the one-loop contributions to the renormalization of the nonlinearities $\lambda_{x,y}$ cancel out \cite{nicoli:2011}, giving rise to Eq.\
\eqref{akpz_ls} of the main text, thus completing the DRG flow for the aKPZ equation.

\begin{acknowledgments}

Partial support for this work has been provided by MINECO (Spain) Grant No.\
FIS2012-38866-C05-01. E.\ V.\ acknowledges support by Universidad Carlos III de Madrid.

\end{acknowledgments}

\bibliography{biblio_teor_ii}

\begin{thebibliography}{10}
\bibitem{note:num_akpz}%
  \BibitemOpen
  \bibinfo {note} {The same numerical simulation scheme has also allowed us to
  also check for the aKPZ equation (not shown), the change of universality
  class as a function of the relative signs of the nonlinearities when both are
  non-zero}%
  \bibAnnoteFile{NoStop}{note:num_akpz}%
\end{thebibliography}

\end{document}